\documentclass[a4paper, twocolumn]{article}
\usepackage[margin=2cm]{geometry}
\usepackage[brazilian, english]{babel}
\usepackage{amsmath, amssymb}
\usepackage{pdfpages, graphicx, subcaption}
\usepackage{array, multirow}
\usepackage{enumitem}
\usepackage{xurl, hyperref}
\usepackage{abstract}

\usepackage[font=small]{caption}

\begin{document}

\title{\bf 
    F\'isica de Part\'iculas no Ensino M\'edio\\
	Parte I: Eletrodin\^amica Qu\^antica \\[4mm] 
	\normalsize Particle Physics in High School \\  Part I: Quantum Electrodynamics
}
\author{ \normalsize 
		{\bf Gl\'auber~Carvalho~Dorsch}$^{1}$\thanks{glauber@fisica.ufmg.br}~, 
		{\bf Thaisa~Carneiro~da Cunha~Guio}$^{2}$\thanks{thaisa.guio@ufes.br}\\[3mm]
		\small \it $^{1}$ Departamento de F\'isica, Universidade Federal de Minas Gerais, 31270-901, Belo Horizonte, MG\\
		\small \it $^{2}$ N\'ucleo Cosmo-UFES, Universidade Federal do Esp\'irito Santo, 29075-910, Vit\'oria, ES
}
\date{}

\twocolumn[
    \maketitle
    \selectlanguage{brazilian}
    \vspace*{-5mm}
    \begin{onecolabstract}
        O presente trabalho \'e o primeiro de uma s\'erie de artigos cuja proposta principal \'e apresentar uma nova sequ\^encia did\'atica para o ensino de F\'isica de Part\'iculas no ensino m\'edio. Propomos uma discuss\~ao sistematizada dessa tem\'atica, englobando n\~ao s\'o a compreens\~ao de seus conceitos-chave, mas tamb\'em da pr\'opria natureza da Ci\^encia, dos fatores que circundam sua pr\'atica e sua rela\c{c}\~ao com tecnologia, sociedade e meio-ambiente, em uma perspectiva voltada \`a alfabetiza\c{c}\~ao cient\'ifica. Neste trabalho apresentamos a primeira parte dessa sequ\^encia did\'atica, com foco na teoria qu\^antica do eletromagnetismo: a Eletrodin\^amica Qu\^antica. Analisamos as potencialidades do material aqui proposto, aliado a uma postura dial\'ogica docente, avaliando a presen\c{c}a de indicadores de alfabetiza\c{c}\~ao cient\'ifica e de engajamento dos estudantes ao longo de interven\c{c}\~oes aplicadas em uma escola p\'ublica estadual em tempo integral do Esp\'irito Santo.\\
        
        \noindent {\bf Palavras-chave:} F\'isica de part\'iculas, ensino m\'edio, alfabetiza\c{c}\~ao cient\'ifica, engajamento.
    \end{onecolabstract}

    \selectlanguage{english}
    \begin{onecolabstract}
     This is the first paper of a series aiming to present a new teaching sequence for Particle Physics in high school. We propose a systematic discussion of the subject, covering not only the understanding of its key concepts, but also of the very nature of Science, of the factors that surround its practice and of its relations to technology, society and the environment, in a framework towards scientific literacy. In this work we present the first part of this teaching sequence, focused on the quantum theory of electromagnetism: Quantum Electrodynamics. We analyze the potentialities of the material proposed here, when applied in a dialogic manner by the teacher, by evaluating the presence of scientific literacy and engagement indicators throughout interventions at a full-time public school in Esp\'irito Santo, Brazil.\\
     
    \end{onecolabstract}

    \vspace*{1cm}
]
\saythanks

\selectlanguage{brazilian}

\section{Introdu\c{c}\~ao}

Em um estudo recente realizado no contexto do projeto internacional ROSE (\textit{The Relevance of Science Education})~\cite{SCHREINER2004,SJOBERG2010} foram consultados, entre os anos de 2010 e 2011, 2365 estudantes da 1${^{\text{a}}}$ s\'erie do ensino m\'edio de 84 escolas de todo o Brasil, perguntados sobre os t\'opicos cient\'ificos que mais teriam interesse em aprender nas aulas~\cite{GOUW2013}. Dentre 108 itens propostos no question\'ario, alguns dos mais favoravelmente votados pelos(as) estudantes dizem respeito diretamente \`a F\'isica de Altas Energias e/ou F\'isica de Part\'iculas, tais como ``Como funciona a bomba at\^omica?'' (29${^{\text{o}}}$ mais bem votado no geral, 5${^{\text{o}}}$ na escolha entre meninos), ``Como funciona uma usina nuclear?'' (43${^{\text{o}}}$ na lista geral, 15${^{\text{o}}}$ entre meninos), enquanto outros convidam a F\'isica de Part\'iculas a uma discuss\~ao multidisciplinar, como ``O c\^ancer, o que sabemos e como podemos trat\'a-lo?'' (2${^{\text{o}}}$ mais bem votado no geral e tamb\'em 2${^{\text{o}}}$ na escolha entre meninas), ``Fen\^omenos que os cientistas ainda n\~ao conseguem explicar'' (19${^{\text{o}}}$ na lista geral, 11${^{\text{o}}}$ entre meninos), ``Como os telefones celulares enviam e recebem mensagens?'' (25${^{\text{o}}}$ geral, 28${^{\text{o}}}$ entre meninos e 27${^{\text{o}}}$ entre meninas), ``Inven\c{c}\~oes e descobrimentos que transformaram o mundo'' (30${^{\text{o}}}$ geral, 20${^{\text{o}}}$ entre meninos), ``Como o raio-X, o ultrassom, etc. s\~ao usados na medicina?'' (36${^{\text{o}}}$ geral, 29${^{\text{o}}}$ entre meninas), entre outros.

Tamb\'em os professores e pesquisadores atuantes em diversas \'areas da F\'isica sugerem a F\'isica de Part\'iculas como um dos principais t\'opicos da F\'isica Moderna e Contempor\^anea (FMC) a ser inserido no ensino médio, segundo estudo baseado em entrevistas com 54 f\'isicos te\'oricos e experimentais de v\'arias especialidades e institui\c{c}\~oes brasileiras, 22 pesquisadores em ensino de F\'isica de diversas Universidades, e 22 professores de F\'isica do ensino m\'edio das regi\~oes Sul, Sudeste e Nordeste~\cite{OSTERMANN2000}.

Existe uma rica literatura discutindo a import\^ancia de se abordar tem\'aticas de FMC j\'a no ensino m\'edio~\cite{OSTERMANN2000a}. Muito al\'em de simplesmente transmitir um aglomerado de conhecimentos factuais, \'e essencial que o ensino de ci\^encias nas escolas fomente uma alfabetiza\c{c}\~ao cient\'ifica, educando os(as) discentes para que compreendam o funcionamento da ci\^encia, os fatores que circundam sua pr\'atica, sua rela\c{c}\~ao com tecnologia, sociedade e meio-ambiente, o m\'etodo emp\'irico, entre outros aspectos~\cite{SasseronCarvalho2008, CHASSOT2000, CHASSOT2003, GILVILCHES2001, SASSERONeCARVALHO2011}. No caso especial da F\'isica de Part\'iculas, por se tratar de uma \'area ainda em pleno desenvolvimento, com abundante atividade em pesquisa, e contendo muitos problemas ainda em aberto, essa tem\'atica promove um solo f\'ertil para discuss\~oes sobre o car\'ater da ci\^encia enquanto constru\c{c}\~ao hist\'orico-social humana e os processos de evolu\c{c}\~ao do conhecimento cient\'ifico, demonstrando concretamente que a ci\^encia n\~ao \'e um conjunto fechado de verdades est\'aticas, mas um processo em constante aprimoramento~\cite{OSTERMANN2000a, OSTERMANN2001, OSTERMANN2000b, TERRAZZAN1992}. Mais ainda, uma abordagem de alfabetiza\c{c}\~ao cient\'ifica apoiada em uma tem\'atica contempor\^anea \'e capaz de motivar o(a) estudante a se enxergar como potencial agente desse desenvolvimento cient\'ifico, ao inv\'es de passivo(a) espectador(a) das gl\'orias atingidas por ``g\^enios'' do passado. O combate ao negacionismo cient\'ifico passa crucialmente por essa compreens\~ao da ci\^encia enquanto atividade coletiva, atual e din\^amica. Qualquer tentativa de substituir ideologias negacionistas por apenas um conjunto de ``verdades cient\'ificas'' que devem ser aceitas autoritativamente estaria apenas lubrificando as engrenagens que movem o dogmatismo e a pseudoci\^encia. \'E o dinamismo da ci\^encia, a sua constante autonega\c{c}\~ao e autoreconstru\c{c}\~ao, que garantem a atualidade e relev\^ancia de seus resultados diante das constantes transforma\c{c}\~oes do objeto de estudo e de nossa organiza\c{c}\~ao social, onde esse conhecimento \'e produzido e ser\'a aplicado. ``Verdades eternas'' envelhecem mal.

Ademais, ressalta-se que a apropria\c{c}\~ao dos conhecimentos cient\'ificos e de suas dimens\~oes epistemol\'ogicas \'e \emph{direito} de todo cidad\~ao~\cite{MARQUES2018}, e proporcionar os meios para esse fim \'e dever do Estado. A n\~ao inclus\~ao da FMC no ensino m\'edio priva diversos estudantes do acesso aos recentes desenvolvimentos da F\'isica, visto que a escola \'e, para muitos, a \'unica oportunidade de terem contato com essas tem\'aticas~\cite{TERRAZZAN1992}. \'E justamente um curr\'iculo escolar antiquado que favorece a mistifica\c{c}\~ao de diversos temas relacionados \`a FMC, tais como a mec\^anica qu\^antica e a teoria da relatividade, e torna os cidad\~aos vulner\'aveis a charlatanices.

A exposi\c{c}\~ao dos(as) estudantes a temas atuais de pesquisa em F\'isica \'e tamb\'em essencial para que aqueles(as) inclinados(as) a optar por essa carreira possam tomar uma decis\~ao informada sobre seu futuro, o que potencialmente reduziria o alto \'indice de abandono em cursos de gradua\c{c}\~ao em F\'isica~\cite{BARROSO2004}.

Apesar de tudo isso, embora haja consenso, pelo menos desde a d\'ecada de 90, quanto \`a import\^ancia de se incorporar o ensino de F\'isica de Part\'iculas j\'a no ensino m\'edio~\cite{OSTERMANN2001,SIQUEIRA2006, SWIBANK1992}, s\~ao escassas as propostas concretas e detalhadas de introdu\c{c}\~ao dessa tem\'atica em sala de aula. Os trabalhos existentes na literatura incluem sequ\^encias t\'opicas~\cite{Siqueira} e/ou voltadas \`a classifica\c{c}\~ao das part\'iculas elementares~\cite{2019Silva, 2010AlvesCosta}, materiais resumidos com orienta\c{c}\~oes gerais voltadas \`a forma\c{c}\~ao de professores nessa tem\'atica~\cite{Ostermann:1999} e propostas de ensino investigativo, incluindo um livreto administr\'avel aos estudantes~\cite{Jonas}. Um material interessante, traduzido e comentado a partir de~\cite{1992Bettelli}, e baseado em imagens de intera\c{c}\~oes, cria\c{c}\~oes e aniquila\c{c}\~oes de part\'iculas em c\^amaras de bolhas, pode ser encontrado em~\cite{BubbleChamber}. Uma proposta que se assemelha \`a nossa perspectiva de apresenta\c{c}\~ao da F\'isica de Part\'iculas tamb\'em enquanto constru\c{c}\~ao hist\'orico-social pode ser encontrada na ref.~\cite{Renan_Milnitsky}. No entanto, o assunto \'e t\~ao rico e abrangente que proporciona uma fonte praticamente inesgot\'avel de abordagens pedag\'ogicas. Al\'em disso, s\~ao ainda mais raros os estudos acompanhados de uma avalia\c{c}\~ao
de indicadores de engajamento e aprendizado consequentes de aplica\c{c}\~oes de sequ\^encias de ensino proposta sobre F\'isica de Part\'iculas.
 
O presente trabalho \'e o primeiro de uma s\'erie de artigos com o objetivo de apresentar uma nova sequ\^encia did\'atica para o ensino de F\'isica de Part\'iculas no ensino m\'edio.
A sequ\^encia aborda a F\'isica de Part\'iculas com \^enfase na alfabetiza\c{c}\~ao cient\'ifica, e frequentemente explorando a integra\c{c}\~ao entre Ci\^encia, Tecnologia, Sociedade e Meio-Ambiente (CTSA). \'E bem sabido que a promo\c{c}\~ao de debates a respeito das implica\c{c}\~oes sociais do saber cient\'ifico, e da \'etica que perpassa o uso social das for\c{c}as naturais, fomenta o desenvolvimento de valores vinculados aos interesses coletivos~\cite{SANTOS2002, BYBEE1987}, e a forma\c{c}\~ao de cidad\~aos conscientes de seu papel na sociedade~\cite{AIKENHEAD2005}. Mais ainda, a sequ\^encia se prop\~oe a apresentar a F\'isica de Part\'iculas n\~ao apenas como um aglomerado de curiosidades sobre part\'iculas com nomes estranhos, mas com uma discuss\~ao sistematizada sobre o assunto, englobando a compreens\~ao de conceitos-chave, o amadurecimento do pensamento cient\'ifico dos(as) educandos(as), e \emph{principalmente} explorando a F\'isica de Part\'iculas como via de desenvolvimento de sua intui\c{c}\~ao f\'isica e capacidade argumentativa.

\`A primeira vista essa \'ultima afirma\c{c}\~ao pode parecer contradit\'oria: como desenvolver intui\c{c}\~ao discutindo part\'iculas microsc\'opicas e fen\^omenos que parecem t\~ao distantes da viv\^encia cotidiana dos estudantes? Comparativamente, outras \'areas da F\'isica, como a mec\^anica Newtoniana, \emph{parecem} se oferecer como muito mais intuitivas, como se permitissem uma rela\c{c}\~ao imediata entre a abstra\c{c}\~ao de conceitos, formula\c{c}\~oes matem\'aticas, e o senso comum. Mas essa impress\~ao \'e question\'avel. Em primeiro lugar, a formula\c{c}\~ao da mec\^anica tamb\'em envolve conceitos contraintuitivos~\cite{LopesCoelho2010}, como os de for\c{c}a, massa e o princ\'ipio da in\'ercia, que causam dificuldades a estudantes em seus primeiros contatos com a disciplina~(Barojas (1988) \textit{apud}~\cite{OSTERMANN2000a}, p.~24). Ainda mais, no contexto da mec\^anica, em que vigoram as ``a\c{c}\~oes \`a dist\^ancia'', o comportamento da for\c{c}a gravitacional, variando com o inverso do quadrado da dist\^ancia, \'e apresentado como um fato, de maneira pouco ou nada elucidativa. \'E somente em uma formula\c{c}\~ao em termos de campos 
que se torna poss\'ivel relacionar esse comportamento \`a geometria do espa\c{c}o, e a F\'isica de Part\'iculas lan\c{c}a ainda outra luz \`a quest\~ao, associando-a ainda \`a maneira como as part\'iculas mediadoras da for\c{c}a interagem entre si. Em suma, pode-se compreender o comportamento $\sim 1/r^2$ como oriundo de uma intera\c{c}\~ao mediada por part\'iculas n\~ao-massivas e que n\~ao interagem consigo pr\'oprias, de modo que n\~ao h\'a cria\c{c}\~ao e aniquila\c{c}\~ao de novos mediadores na regi\~ao entre a fonte e o observador, fazendo com que o fluxo do campo caia com a \'area de uma esfera, como sumariza a lei de Gauss.
 A discuss\~ao do caso oposto, em que as part\'iculas mediadores possuem autointera\c{c}\~ao, leva a uma compreens\~ao intuitiva do fen\^omeno de confinamento dos quarks em h\'adrons, que ser\'a discutida em mais detalhes em publica\c{c}\~ao posterior. Esse \'e apenas um exemplo de como uma rica intui\c{c}\~ao heur\'istica pode, sim, ser elaborada e explorada em um contexto de F\'isica de Part\'iculas, resultando em uma maior compreens\~ao da natureza da mat\'eria e suas intera\c{c}\~oes.

Com esse vi\'es em mente, um dos objetivos do presente trabalho \'e mostrar que a F\'isica de Part\'iculas \emph{n\~ao \'e} um terreno \'arido no qual se \'e \emph{proibido} enveredar sem um ferramental matem\'atico absurdamente elaborado, mas que pode, sim, ser acess\'ivel a estudantes de n\'ivel m\'edio --- embora, inegavelmente, ferramentas e t\'ecnicas apropriadas possibilitam avan\c{c}os mais aprofundados e mais r\'apidos.

Neste artigo, o foco ser\'a voltado \`a primeira parte da sequ\^encia did\'atica elaborada sobre F\'isica de Part\'iculas. Em termos de conte\'udo, ela se relaciona \`a Eletrodin\^amica Qu\^antica (QED\footnote{Da sigla em ingl\^es para \textit{Quantum Electrodynamics}.}), a teoria qu\^antica da intera\c{c}\~ao eletromagn\'etica. Essa parte da sequ\^encia did\'atica, apresentada na se\c{c}\~ao~\ref{sec:momentos}, \'e constru\'ida de maneira tal que percorre um caminho acess\'ivel aos estudantes, e com muitas interse\c{c}\~oes com o conte\'udo usualmente ministrado no ensino m\'edio, de modo que a F\'isica de Part\'iculas pode ser inclu\'ida na ementa sem sacrificar a aprendizagem de elementos de f\'isica cl\'assica. 
Especificamente, nesta primeira parte partiremos da estrutura at\^omica da mat\'eria, passando pela descoberta dos el\'etrons (protagonistas da QED), a intera\c{c}\~ao eletromagn\'etica (comparando-a com a intera\c{c}\~ao gravitacional, mais familiar a estudantes que j\'a tenham feito um curso de mec\^anica), ondas eletromagn\'eticas (e aplica\c{c}\~oes interdisciplinares), a dualidade onda-part\'icula e os f\'otons, culminando na descri\c{c}\~ao fundamental da QED em que a intera\c{c}\~ao entre cargas el\'etricas \'e mediada pela troca de f\'otons. Na se\c{c}\~ao~\ref{sec:resultados} apresentamos uma an\'alise da viabilidade de nossa proposta, bem como sua efetividade e suas potencialidades, considerando um estudo de caso em que as atividades aqui propostas foram aplicadas em uma sala de aula de ensino m\'edio de uma escola p\'ublica do Esp\'irito Santo. Considerando alguns indicadores presentes na literatura para avaliar ind\'icios de alfabetiza\c{c}\~ao cient\'ifica e de engajamento dos estudantes, veremos que a sequ\^encia, quando aliada a uma postura dial\'ogica por parte do(a) docente, pode proporcionar elevados \'indices de engajamento e ind\'icios de alfabetiza\c{c}\~ao cient\'ifica. A se\c{c}\~ao~\ref{sec:conclusoes} \'e dedicada \`as nossas conclus\~oes. No ap\^endice, apresentamos propostas de atividades relacionadas ao conte\'udo sugerido e que envolvem e desenvolvem m\'ultiplas compet\^encias do(a) estudante.

\section{Propostas de momentos did\'aticos}
\label{sec:momentos}

Mencionamos a seguir alguns momentos did\'aticos que podem ser explorados em sala de aula contendo a tem\'atica de Eletrodin\^amica Qu\^antica no contexto da F\'isica de Part\'iculas. O material foi elaborado com intuito de servir como refer\^encia sugestiva ao docente, ficando este livre para adequ\'a-lo a seus prop\'ositos, ao inv\'es de ter que segui-lo como receita impositiva. 

Dessa forma, h\'a v\'arias possibilidades de se adotar a proposta deste trabalho em sala de aula. Por suposto, uma op\c{c}\~ao seria a inser\c{c}\~ao de uma sequ\^encia de aulas sobre F\'isica de Part\'iculas ao final do terceiro ano letivo do ensino m\'edio, ap\'os se ter discutido extensivamente eletromagnetismo e aspectos da teoria qu\^antica. A F\'isica de Part\'iculas seria, assim, um ap\^endice ao conte\'udo de f\'isica moderna. A desvantagem dessa abordagem \'e privar o(a) aluno(a), durante todo o decurso do ensino m\'edio,  do contato com aspectos fascinantes da f\'isica contempor\^anea, exceto por poucas aulas ministradas em um momento em que ele(a) possivelmente j\'a tenha formalizado a escolha do curso que pretende seguir em n\'ivel superior. 
Uma segunda possibilidade seria ministrar essas aulas a grupos de alunos(as) que manifestem interesse pr\'evio em seguir carreira cient\'ifica. Pode-se, ainda, discutir a tem\'atica em disciplinas eletivas em escolas de tempo integral. Note que esse foi o caso da pr\'atica dos presentes autores. 
Ainda uma outra abordagem poss\'ivel consiste na inser\c{c}\~ao de discuss\~oes conectadas \`a F\'isica de Part\'iculas em aulas que versam sobre outras grandes tem\'aticas mais tradicionais do ensino m\'edio. Por exemplo, pode-se complementar uma sequ\^encia did\'atica sobre eletromagnetismo mencionando a interpreta\c{c}\~ao dessa intera\c{c}\~ao sob a perspectiva da Eletrodin\^amica Qu\^antica, ou introduzir tal discuss\~ao em uma aula sobre o f\'oton, o efeito fotoel\'etrico e/ou outros elementos de f\'isica moderna. De fato, o presente trabalho j\'a est\'a estruturado de maneira prop\'icia a essa abordagem, por estar dividido em propostas de momentos did\'aticos male\'aveis, adapt\'aveis aos prop\'ositos do(a) docente. Antecipa-se, assim, uma resposta \`a previs\'ivel cr\'itica de que a restrita carga hor\'aria de F\'isica no ensino m\'edio impossibilita a introdu\c{c}\~ao de tem\'aticas para al\'em das necess\'arias \`a prepara\c{c}\~ao dos(as) estudantes para exames qualificat\'orios, como ENEM e vestibulares. Por fim, h\'a ainda uma outra possibilidade de leitura deste trabalho, ainda mais ousada e mais abrangente, qual seja: utilizar a F\'isica de Altas Energias n\~ao apenas como um conte\'udo particular, mas como \emph{um dos} panos de fundo para o ensino de todo o conte\'udo de F\'isica do ensino m\'edio. 

O conte\'udo das se\c{c}\~oes~\ref{sec:intro} \`a \ref{sec:dualidade} \'e parte da ementa usual do ensino m\'edio, com a qual os(as) docentes certamente est\~ao familiarizados. Nessas se\c{c}\~oes, apenas sugerimos uma poss\'ivel abordagem desses t\'opicos com \^enfase na tem\'atica relacionada \`a F\'isica de Part\'iculas. J\'a as se\c{c}\~oes~\ref{sec:qed} e \ref{sec:positrons} versam sobre elementos de Eletrodin\^amica Qu\^antica, que provavelmente ser\~ao menos familiares a docentes de ensino m\'edio. Por isso, essa parte do texto passa a ser direcionada tamb\'em \`a forma\c{c}\~ao desses professores nos principais elementos desse paradigma te\'orico, mas com uma linguagem acess\'ivel tamb\'em a estudantes. Pode-se julgar que o conte\'udo dessas se\c{c}\~oes \'e mais extenso do que se desejaria discutir em sala de aula. Em conformidade com a proposta neste trabalho, deixamos a crit\'erio do(a) docente decidir quais t\'opicos poderiam ser abordados com os(as) estudantes dentro do contexto das salas de aula em que lecionam.

\subsection{Introdu\c{c}\~ao ao tema}
\label{sec:intro}

O primeiro momento da sequ\^encia propicia a oportunidade de di\'alogo com os(as) estudantes a fim de averiguar sua familiaridade com a grande tem\'atica de F\'isica de Part\'iculas, que aparece frequentemente em livros de divulga\c{c}\~ao cient\'ifica ou mesmo em seriados televisivos, canais de plataformas de compartilhamento de v\'ideos e s\'itios que discutem ci\^encia. Os(as) estudantes conseguem imaginar o qu\^e (e quais) s\~ao essas part\'iculas a que a tem\'atica se refere? J\'a tiveram contato com alguma literatura que falasse sobre quarks, neutrinos, b\'oson de Higgs, etc.? J\'a ouviram falar sobre o LHC (Grande Colisor de H\'adrons), ou outros aceleradores de part\'iculas e colisores? Interessam-se pelo tema? Por que \'e relevante estud\'a-lo? \'E um momento de deixar os(as) alunos(as) se expressarem, seja publicamente ou em pequenos grupos com os(as) colegas, de modo a se divertirem ao se envolverem com a tem\'atica.

Ap\'os essa discuss\~ao, conv\'em apresentar, a t\'itulo de motiva\c{c}\~ao, alguns fen\^omenos que se pode explicar gra\c{c}as \`a nossa compreens\~ao da estrutura da mat\'eria em diversas escalas, incluindo aplica\c{c}\~oes tecnol\'ogicas que envolvem F\'isica de Part\'iculas. Por exemplo,
\begin{itemize}
	\item a descri\c{c}\~ao at\^omica da mat\'eria nos permite explicar a origem das for\c{c}as normais de resist\^encia dos materiais enquanto for\c{c}as eletromagn\'eticas repulsivas entre orbitais moleculares. Ou seja, explica-se fen\^omenos simples e cotidianos, como o fato de n\~ao atravessarmos uma cadeira quando nela sentamos;
	\item a Eletrodin\^amica Qu\^antica (QED) nos d\'a uma descri\c{c}\~ao muito acurada do comportamento da luz, possibilitando-nos manipul\'a-la para o desenvolvimento de lasers desde os mais simples, vendidos comercialmente, aos mais potentes, utilizados em escala industrial para diversos fins;
	\item a mesma QED prev\^e a exist\^encia de part\'iculas chamadas p\'ositrons, que t\^em aplica\c{c}\~ao medicinal nos chamados \emph{PET scans}\footnote{PET \'e sigla em ingl\^es para \emph{Tomografia por Emiss\~ao de P\'ositrons}.};
	\item o brilho do Sol \'e devido a rea\c{c}\~oes relacionadas \`as intera\c{c}\~oes nucleares fraca e forte, que tamb\'em est\~ao relacionadas a decaimentos radioativos, com aplica\c{c}\~oes na data\c{c}\~ao de f\'osseis ou rochas, e em in\'umeros aparatos cotidianos, como detectores de fuma\c{c}a;
	\item o poder preditivo da f\'isica nuclear \'e ilustrado de modo estarrecedor pela capacidade destrutiva dos armamentos nucleares, ou pela capacidade de gera\c{c}\~ao de energia a partir de poucos quilogramas de mat\'eria em usinas nucleares.
\end{itemize} 
Todos esses fen\^omenos e aplica\c{c}\~oes tecnol\'ogicas poder\~ao ser explicados em maiores detalhes em momentos posteriores da sequ\^encia did\'atica, quando o assunto relevante for discutido em maior detalhe.

\subsection{Estrutura at\^omica da mat\'eria}
\label{sec:atom}

O objeto de estudo da F\'isica de Part\'iculas \'e a mat\'eria em seu n\'ivel mais elementar, precedendo a forma\c{c}\~ao de estruturas compostas mais complexas. Ou seja, \'e o estudo dos constituintes fundamentais da mat\'eria e de suas intera\c{c}\~oes m\'utuas. A tem\'atica \'e convidativa, portanto, a uma discuss\~ao que desemboque na teoria at\^omica e na natureza discreta da mat\'eria.

Em um primeiro momento, mais importante do que explicar as nuances dos diversos modelos at\^omicos, \'e preciso que fique claro aos(\`as) discentes a g\^enese do conceito de \'atomo, a diferen\c{c}a entre as concep\c{c}\~oes grega, moderna e contempor\^anea desse termo, e, principalmente, as \textbf{evid\^encias emp\'iricas} em favor do atomismo. N\~ao basta que a ideia de \'atomo seja \emph{aceita}, \'e preciso que o(a) estudante saiba convencer outrem de que a hip\'otese atomista resulta em uma excelente descri\c{c}\~ao do comportamento da mat\'eria. Ademais, mesmo que os(as) alunos(as) saibam sobre atomismo, discutir as evid\^encias que corroboram essa hip\'otese leva a um debate salutar sobre epistemologia e ci\^encia, sobre as condi\c{c}\~oes para que uma hip\'otese seja ``validada'' pela comunidade cient\'ifica. No caso da teoria at\^omica, h\'a o fator explicativo e o poder preditivo da hip\'otese, ilustrado em evid\^encias como:
\begin{itemize}
	\item {\bf As Leis Ponderais da Qu\'imica}: sob uma perspectiva plenista/continuista da mat\'eria, \'e dif\'icil explicar por que rea\c{c}\~oes qu\'imicas ocorrem sempre na mesma propor\c{c}\~ao fixa de massa ou de moles dos reagentes, e por que qualquer excedente al\'em dessa propor\c{c}\~ao fixa permanece sem reagir. Por outro lado, a hip\'otese atomista explica prontamente esse comportamento, bem como a lei de conserva\c{c}\~ao de massas de Lavoisier, ao postular que rea\c{c}\~oes qu\'imicas s\~ao simplesmente um rearranjo de \'atomos. Um experimento ilustrativo, que pode ser invocado durante a aula e at\'e mesmo realizado ao vivo devido \`a sua simplicidade, \'e a eletr\'olise da \'agua, em que o volume de g\'as hidrog\^enio gerado \'e \emph{sempre} o dobro do volume de oxig\^enio. Um v\'ideo ilustrativo desse experimento pode ser encontrado em~\cite{VideoEletrolise}.
	\item {\bf Teoria Cin\'etica dos Gases}: admitindo-se a hip\'otese at\^omica, \'e poss\'ivel \emph{deduzir} a equa\c{c}\~ao de estado dos gases ideais a partir das leis da mec\^anica de Newton. Mais ainda, \'e poss\'ivel aprimorar a descri\c{c}\~ao dos gases reais levando-se em conta, por exemplo, o tamanho das mol\'eculas e suas intera\c{c}\~oes m\'utuas, obtendo uma descri\c{c}\~ao ainda mais precisa de sistemas gasosos. Em \'ultima inst\^ancia, a hip\'otese atomista d\'a origem \`a unifica\c{c}\~ao da mec\^anica e da termodin\^amica na {mec\^anica estat\'istica}.
	\item O {\bf Movimento Browniano} \'e o movimento aleat\'orio de micropart\'iculas suspensas em um l\'iquido. O(a) docente pode mostrar um v\'ideo ou anima\c{c}\~ao ilustrando o fen\^omeno como visto em um microsc\'opio. Em 1905, Einstein foi capaz de explicar esse movimento como devido \`a constante colis\~ao das mol\'eculas do l\'iquido com as micropart\'iculas, e, com essa descri\c{c}\~ao, foi capaz de prever corretamente que a dist\^ancia m\'edia percorrida por uma part\'icula \'e proporcional \`a raiz quadrada do tempo decorrido, $d\sim \sqrt{t}$, um comportamento t\'ipico do car\'ater estat\'istico do fen\^omeno.
\end{itemize}
Outros exemplos podem ser apresentados. O importante \'e enfatizar que n\~ao \'e necess\'ario enxergarmos os \'atomos para saber que existem (embora isso j\'a seja poss\'ivel com microsc\'opios eletr\^onicos). Ao supormos que a mat\'eria \'e assim constitu\'ida, podemos deduzir suas propriedades e prever corretamente seu comportamento, cumprindo, assim, uma das fun\c{c}\~oes da Ci\^encia enquanto atividade humana: nosso empoderamento diante das for\c{c}as naturais.

\subsection{El\'etrons}
\label{sec:eletrons}

Uma vez que se tenha consolidado a concep\c{c}\~ao de estrutura at\^omica da mat\'eria, pode-se partir para o rompimento da ideia de que o \'atomo \'e indivis\'ivel. Aqui come\c{c}a a F\'isica de Part\'iculas de fato, introduzindo a primeira part\'icula elementar, protagonista da Eletrodin\^amica Qu\^antica: o el\'etron.

\begin{figure}[h!]
	\centering
	\includegraphics[width=.5\textwidth]{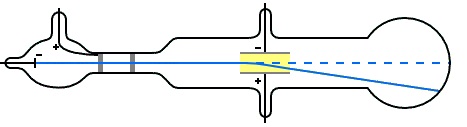}
	\caption{Tubo de raios cat\'odicos usado no experimento de Thomson~\cite{Thomson:1897cm}. Um feixe de part\'iculas \'e emitido do filamento met\'alico conectado ao p\'olo negativo da bateria (\`a esquerda da figura), e se propaga em um tubo a v\'acuo at\'e colidir com a outra extremidade, no bulbo \`a direita. O feixe \'e desviado por um campo el\'etrico inserido no meio do tubo (regi\~ao amarela entre as placas horizontais), indicando que as part\'iculas emitidas possuem carga. Com aux\'ilio de um campo magn\'etico (n\~ao simbolizado na imagem), pode-se calcular a raz\~ao entre a carga e a massa dessas part\'iculas. Fonte: Wikimedia Commons/Dom\'inio P\'ubico.} 
	\label{fig:CRT}
\end{figure}

Uma maneira natural de se introduzir part\'iculas subat\^omicas \'e discutir o experimento de J. J. Thomson com tubos de raios cat\'odicos, ilustrado na figura~\ref{fig:CRT}. Para uma sequ\^encia did\'atica investigativa explorando essa tem\'atica vide ref.~\cite{Dayane}. Tais tubos s\~ao os elementos constitutivos b\'asicos de televisores antigos (as chamadas ``TVs de tubo''), e o(a) docente pode usar isso como motiva\c{c}\~ao inicial \`a aula, propondo uma discuss\~ao sobre como esses aparelhos funcionam. A discuss\~ao sobre esses tubos \'e tamb\'em prop\'icia ao conte\'udo da sequ\^encia, pois constituem exemplos simples de \emph{aceleradores de part\'iculas}. Essencialmente, dois filamentos met\'alicos s\~ao inseridos em um tubo a v\'acuo e conectados a uma fonte de energia (uma bateria, no caso do experimento de Thomson, ou a tomada dom\'estica no caso de um televisor), gerando os p\'olos - e + na figura. Ao faz\^e-lo, el\'etrons s\~ao acelerados do c\'atodo ao \^anodo e colidem com o bulbo \`a direita da figura, causando o aparecimento de um ponto luminoso, que gera a imagem em aparelhos televisores. 

Quando realizou seu experimento, Thomson n\~ao sabia sobre a exist\^encia de el\'etrons, e seu prop\'osito era justamente desvendar a natureza desses raios que geravam o ponto luminoso. Submetendo esse feixe a campos el\'etricos e magn\'eticos, chegou a duas conclus\~oes principais, que merecem \^enfase no contexto dessa sequ\^encia did\'atica:
\begin{itemize}
	\item O feixe \'e constitu\'ido de part\'iculas cujas massas s\~ao cerca de 2000 vezes menor do que a massa do \'atomo mais leve (o hidrog\^enio)\footnote{Mais precisamente, o experimento de Thomson de 1897~\cite{Thomson:1897cm} propiciou a medi\c{c}\~ao apenas da raz\~ao carga/massa do el\'etron, obtendo a mencionada diferen\c{c}a frente ao valor conhecido dessa raz\~ao para um \'ion de hidrog\^enio (i.e. um pr\'oton). No entanto, Thomson inferiu que a discrep\^ancia fosse devido \`a menor massa do el\'etron com base em outras propriedades dos raios cat\'odicos, como sua maior capacidade de penetra\c{c}\~ao ao propagar-se em um g\'as. Foi apenas em outro experimento, realizado em 1899~\cite{Thomson:1899}, que Thomson conseguiu mostrar que a carga do el\'etron \'e igual \`a do pr\'oton, e que portanto a diferen\c{c}a entre as raz\~oes carga/massa deve-se unicamente a uma diferen\c{c}a entre suas massas.}. Ou seja, essas part\'iculas n\~ao poderiam ser constitu\'idas de quaisquer \'atomos conhecidos at\'e ent\~ao.
	\item Ap\'os repetir o experimento com c\'atodos de v\'arios materiais (alum\'inio, chumbo, estanho, cobre e ferro) e obter sempre os mesmos resultados, Thomson concluiu que essas part\'iculas devem estar contidas no interior de todos os \'atomos.
\end{itemize}
Thomson descobriu, assim, o el\'etron: a primeira part\'icula subat\^omica a ser identificada.

Outra discuss\~ao interessante a se explorar, no contexto dessa aula, \'e  o uso de tubo de raios cat\'odicos em televisores e monitores como um dos muitos exemplos de uma pesquisa a princ\'ipio puramente acad\^emica, mas que deu origem a uma aplica\c{c}\~ao tecnol\'ogica que, poucas d\'ecadas depois, veio a se tornar parte da vida cotidiana em nossa sociedade. Outro exemplo famoso \'e o nascimento da \emph{World Wide Web} no CERN\footnote{Antigo acr\^onimo da Organiza\c{c}\~ao Europeia para a Pesquisa Nuclear (\textit{Conseil Europ\'een pour la Recherche Nucl\'eaire} em franc\^es),  o maior laborat\'orio de F\'isica de Part\'iculas do mundo, localizado em Genebra, Su\'i\c{c}a, em sua fronteira com a Fran\c{c}a.}, rede criada inicialmente para compartilhar dados entre os cientistas trabalhando no laborat\'orio e espalhados no mundo, e que hoje faz parte da experi\^encia cotidiana de todos que navegam na \emph{internet}.
 
\subsection{A intera\c{c}\~ao eletromagn\'etica}
\label{sec:em}
 
A F\'isica de Part\'iculas versa n\~ao somente sobre os constituintes fundamentais da mat\'eria, mas tamb\'em sobre o modo como essas part\'iculas interagem entre si, e como essas intera\c{c}\~oes microsc\'opicas d\~ao origem \`a f\'isica macrosc\'opica que observamos cotidianamente.

Dentre as intera\c{c}\~oes fundamentais da mat\'eria, a \emph{intera\c{c}\~ao eletromagn\'etica} se destaca por ser \emph{a mais simples de todas} (veremos a justificativa para essa afirma\c{c}\~ao na se\c{c}\~ao~\ref{sec:emgrav}), e tamb\'em pela sua ubiquidade em nosso cotidiano: al\'em das in\'umeras aplica\c{c}\~oes tecnol\'ogicas, todas as ``for\c{c}as de contato'' que experimentamos t\^em origem eletromagn\'etica.

O eletromagnetismo j\'a \'e um t\'opico pertencente ao curr\'iculo usual de F\'isica no ensino m\'edio, de modo que a discuss\~ao apresentada aqui n\~ao foge \`a tem\'atica que j\'a seria tratada em algum momento pelo(a) docente em sala de aula. Entretanto, a apresenta\c{c}\~ao do t\'opico sob uma perspectiva motivada pela F\'isica de Part\'iculas tem certas vantagens, por possibilitar uma melhor compreens\~ao sobre o comportamento dessa intera\c{c}\~ao, al\'em de induzir uma conex\~ao natural entre f\'isica cl\'assica e moderna.

\subsubsection{Semelhan\c{c}as e diferen\c{c}as entre eletromagnetismo e gravita\c{c}\~ao}
\label{sec:emgrav}

\begin{figure}[h!]
 	\centering
	\includegraphics[page=1]{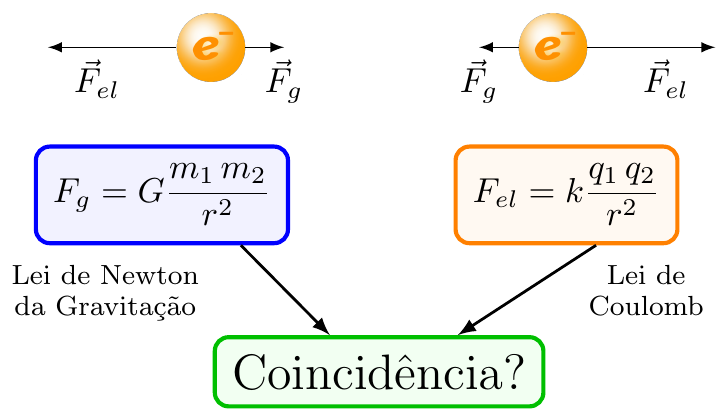}
	\caption{Dois el\'etrons sob a a\c{c}\~ao da atra\c{c}\~ao gravitacional $\vec{F}_g$, devido a suas massas $m_e$, e da repuls\~ao eletrost\'atica $\vec{F}_{el}$ devido a suas cargas $-e$.}
	\label{fig:gel}
\end{figure}

A lei de for\c{c}a da intera\c{c}\~ao entre duas part\'iculas est\'aticas eletricamente carregadas (Lei de Coulomb) tem importantes semelhan\c{c}as e diferen\c{c}as frente \`a lei de atra\c{c}\~ao gravitacional de Newton (vide figura~\ref{fig:gel}), e uma discuss\~ao aprofundada sobre as causas dessas (dis)similitudes abre in\'umeras possibilidades de discuss\~oes em sala de aula, todas ricas em conte\'udo f\'isico, algumas prop\'icias ao desenvolvimento da intui\c{c}\~ao dos estudantes, e ainda outras que levam a quest\~oes profundas que continuam em aberto na F\'isica de Part\'iculas.

A \textbf{semelhan\c{c}a} \'e \'obvia: ambas leis de for\c{c}a seguem o padr\~ao
\begin{equation}
	\text{for\c{c}a} = \text{constante}\times \frac{\text{(produto de cargas)}}{\text{(dist\^ancia)}^2}~.
\end{equation}
Por que isso ocorre? O \emph{produto das cargas} tem que aparecer para que a terceira lei de Newton seja satisfeita: a for\c{c}a agindo sobre A devido a B deve ser proporcional\footnote{No caso da gravita\c{c}\~ao, a depend\^encia da for\c{c}a com a massa inercial \'e devido \`a observa\c{c}\~ao da universalidade da acelera\c{c}\~ao de queda livre. No caso da Lei de Coulomb, trata-se de uma \emph{defini\c{c}\~ao} da carga el\'etrica, que \'e medida justamente atrav\'es da for\c{c}a atuante sobre ela quando posta a interagir com uma carga teste.} \`a carga de A, e deve ter mesma magnitude que a for\c{c}a em B devido a A, que por sua vez \'e proporcional \`a carga de B. A constante aparece devido \`a escolha de unidades. Por exemplo, em unidades Gaussianas, a constante da Lei de Coulomb \'e igual a 1.

A f\'isica mais interessante est\'a no comportamento $\sim 1/r^2$ dessas for\c{c}as, que merece uma discuss\~ao mais aprofundada e cuidadosa com os(as) alunos(as). Em \'ultima inst\^ancia, a Lei de Coulomb tem essa forma porque o campo eletromagn\'etico/o f\'oton/a luz n\~ao possui carga nem massa. Para explicar essa afirma\c{c}\~ao, deve-se primeiramente introduzir o conceito de \emph{campo el\'etrico}, incluindo sua manifesta\c{c}\~ao visual em termos de \emph{linhas de campo} e como as cargas el\'etricas constituem fontes e sumidouros dessas linhas\footnote{Uma simula\c{c}\~ao interativa interessante para esse prop\'osito encontra-se na ref.~\cite{PhET_cargas}.}. A partir da\'i, segue-se que:
\begin{itemize}
	\item Como o pr\'oprio campo n\~ao carrega a carga com a qual ele interage (i.e. o campo eletromagn\'etico n\~ao \'e autointeragente), a \'unica fonte de linhas de campo \'e a mat\'eria, que \'e o enunciado da Lei de Gauss,
\begin{equation}
	\underbrace{\nabla \cdot \vec{E}}_{\text{fonte de campo}} \sim \underbrace{\rho}_{\text{carga}},
\end{equation}
expressa ilustrativamente na figura~\ref{fig:campo} (b). Por isso, o fluxo de linhas de campo atrav\'es de uma superf\'icie fechada englobando a carga \'e sempre constante, pois n\~ao h\'a novas linhas de campo surgindo no espa\c{c}o entre a carga e a superf\'icie. Aqui, uma analogia com um fluido pode ser evocada: se interpretarmos a carga central na figura~\ref{fig:campo} (b) como uma fonte de \'agua em um chafariz, o fluxo de \'agua atravessando as circunfer\^encias pontilhadas em um determinado instante \'e sempre igual ao fluxo de \'agua jorrando da fonte, contanto que n\~ao haja novas fontes ou sumidouros de \'agua no meio do percurso. Portanto a quantidade de \'agua que atravessa as circunfer\^encias pontilhadas, por unidade de comprimento dessa circunfer\^encia, decresce com a dist\^ancia \`a fonte como $\sim 1/2\pi r$. Ou seja, se temos em nossas m\~aos um copo, e queremos ench\^e-lo com a \'agua escorrendo atrav\'es da borda do chafariz, o tempo necess\'ario para encher o copo variar\'a com o inverso da dist\^ancia da borda \`a fonte. Retornando ao caso eletromagn\'etico de interesse, em que as linhas de campo se espalham por todo o espa\c{c}o tridimensional, o fluxo dessas linhas de campo \'e constante sobre qualquer esfera de raio $r$, de modo que a \emph{densidade de linhas de campo} (que corresponde \`a intensidade do campo) decai com a \'area dessa esfera, $E \sim 1/4\pi r^2$. Pelo mesmo argumento, podemos identificar como seria a Lei de Coulomb em espa\c{c}os de $d$-dimens\~oes: $E\sim 1/r^{d-1}$ (o volume de uma esfera $d$-dimensional \'e $\sim r^d$, e sua \'area \'e $\sim r^{d-1}$). 

\begin{figure}[h!]
	\centering
	\begin{subfigure}[t]{.23\textwidth}
	    \includegraphics[page=2, scale=.72]{tikz_figs}
	    \caption{}
	\end{subfigure}
	\begin{subfigure}[t]{.23\textwidth}
    	\includegraphics[scale=.3]{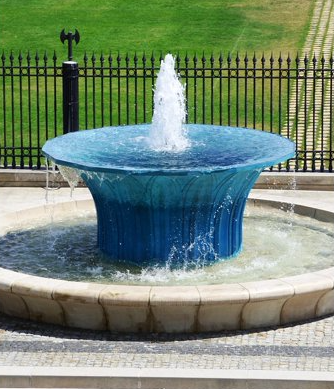}
    	\caption{}
    \end{subfigure}
	\caption{(a) Linhas de campo el\'etrico geradas por uma carga pontual. O n\'umero de linhas atravessando uma superf\'icie fechada \'e sempre constante, o que significa que a densidade de linhas diminui com a dist\^ancia. Para um espa\c{c}o tridimensional, a redu\c{c}\~ao \'e $\sim 1/r^2$. (b) Toda a \'agua que emana da fonte escorre atrav\'es da borda do chafariz. A quantidade de \'agua que atravessa uma unidade de comprimento desse per\'imetro varia com o inverso da dist\^ancia da borda \`a fonte, $\sim 1/r$. Fonte: Dom\'inio P\'ublico.}
	\label{fig:campo}
\end{figure}

Caso o campo eletromagn\'etico portasse carga el\'etrica, as pr\'oprias linhas de campo seriam fonte de novas linhas de campo, e o fluxo n\~ao seria constante, invalidando o argumento acima.

Mais ainda: o fato de o campo eletromagn\'etico n\~ao ser autointeragente \'e tamb\'em a causa da linearidade das equa\c{c}\~oes de Maxwell, e do \emph{princ\'ipio da superposi\c{c}\~ao} em particular. \'E f\'acil entender por que: os campos produzidos por duas cargas se propagam no espa\c{c}o sem interagirem entre si e, portanto, sem que a informa\c{c}\~ao que carregavam inicialmente seja modificada durante o percurso. O campo resultante \'e, portanto, simplesmente a soma dos campos produzidos por cada carga.

Note que, com essa discuss\~ao, o(a) estudante de ensino m\'edio \'e apresentado, de maneira muito clara e intuitiva, a uma das leis de Maxwell do eletromagnetismo e suas implica\c{c}\~oes. Expande-se, assim, o entendimento do(a) estudante sobre fen\^omenos eletromagn\'eticos para al\'em de meras memoriza\c{c}\~oes de leis de for\c{c}as e atividades operacionais puramente mec\^anicas. 

\item A depend\^encia da lei de for\c{c}a com a massa do campo \'e menos trivial, por se tratar de um aspecto qu\^antico, mas possivelmente interessante mencionar (principalmente para aulas posteriores, quando se discutir intera\c{c}\~oes nucleares forte e fraca). Caso os f\'otons tivessem massa $m$, o potencial eletrost\'atico teria a forma de um potencial de Yukawa,
\begin{equation}
	V_{\rm Yukawa} \sim \frac{e^{-m r}}{r},
\end{equation}
correspondendo a uma for\c{c}a que decai muito mais rapidamente do que $1/r^2$, tendo um curto alcance efetivo, $r_{\rm ef} \sim 1/m$. O fato de o f\'oton ser n\~ao-massivo \'e respons\'avel pelo longo alcance da intera\c{c}\~ao eletromagn\'etica, bem como pelo fato de se propagar \`a m\'axima velocidade permitida para a troca de informa\c{c}\~oes na Natureza: a velocidade da luz no v\'acuo.
\end{itemize}

E a gravita\c{c}\~ao? O campo gravitacional tamb\'em n\~ao possui massa --- quer dizer, o \emph{gr\'aviton}, que \'e a part\'icula associada \`a propaga\c{c}\~ao do campo gravitacional, tem massa nula ---, o que explica por que essa intera\c{c}\~ao \'e de longo alcance, efetiva a dist\^ancias em escala cosmol\'ogica (e tamb\'em se propaga no v\'acuo \`a velocidade da luz). Mais ainda, na gravita\c{c}\~ao de Newton apenas as massas agem como fonte de gravita\c{c}\~ao, e portanto o campo gravitacional newtoniano \emph{n\~ao} interage gravitacionalmente (i.e. n\~ao \'e autointeragente), o que implica no comportamento $\sim 1/r^2$, como vimos acima. 

Entretanto, hoje se sabe que a teoria newtoniana n\~ao descreve corretamente todos os fen\^omenos gravitacionais, e \'e, em \'ultima inst\^ancia, suplantada pela teoria da relatividade geral de Einstein. Nessa teoria, n\~ao s\'o a massa dos corpos \'e fonte de campo gravitacional, mas toda e qualquer forma de energia e momento. Logo, a energia contida no campo gravitacional tamb\'em gravita. Por isso \emph{as equa\c{c}\~oes da relatividade geral s\~ao altamente n\~ao-lineares}. Assim, a gravita\c{c}\~ao Newtoniana est\'a contida na relatividade geral no limite em que se negligencia a autointera\c{c}\~ao dos gr\'avitons.

Segue-se, dessas discuss\~oes, que \textbf{a intera\c{c}\~ao eletromagn\'etica \'e a intera\c{c}\~ao mais simples da Natureza}, pois \'e a \'unica linear.

H\'a, ainda, outras \textbf{diferen\c{c}as} entre gravita\c{c}\~ao e eletromagnetismo.
\begin{itemize}
	\item \emph{A intera\c{c}\~ao gravitacional \'e muito mais fraca que a eletromagn\'etica.} De fato, comparando-se a raz\~ao entre as magnitudes das for\c{c}as eletrost\'atica e gravitacional agindo sobre dois el\'etrons, vem que
	\begin{equation}
		\left|\frac{\vec{F}_{el}}{\vec{F}_g}\right| = 
				\frac{k\,e^2}{G\,m_e^2} \sim 10^{42}.
	\end{equation}
	Esse fato \'e facilmente comprovado notando-se que a for\c{c}a eletrost\'atica entre um bal\~ao de festa e os cabelos de uma pessoa \'e suficiente para superar a atra\c{c}\~ao gravitacional do \emph{planeta inteiro}, como se v\^e na figura~\ref{fig:cabelo}.
\begin{figure}[h!]
	\centering
	\includegraphics[trim=90 170 50 120, clip, width=.3\textwidth]{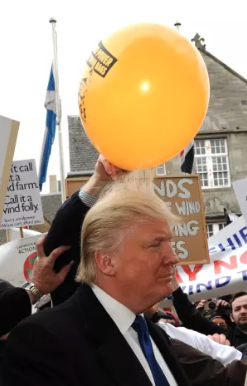}
	\caption{A for\c{c}a eletrost\'atica devido a um bal\~ao de festa \'e suficiente para superar a atra\c{c}\~ao gravitacional de todo o planeta e fazer o cabelo levantar. Fonte: adaptado de SWNS/Splash News.}
	\label{fig:cabelo}
\end{figure}

	A raz\~ao para a exist\^encia dessa hierarquia entre essas intera\c{c}\~oes \'e um dos maiores mist\'erios abertos da F\'isica de Altas Energias, denominado o \textbf{problema da hierarquia}. Uma proposta de solu\c{c}\~ao consiste em supor que existem dimens\~oes espaciais extras (al\'em das tr\^es usuais com as quais estamos familiarizados), e que em \'ultima inst\^ancia todas as intera\c{c}\~oes t\^em a mesma intensidade, mas que a geometria das dimens\~oes extras faz com que a gravita\c{c}\~ao efetiva em tr\^es dimens\~oes se torne muito mais fraca~\cite{Randall:1999ee}. 
	\item A intera\c{c}\~ao eletromagn\'etica pode ser \emph{atrativa} ou \emph{repulsiva}, enquanto a gravita\c{c}\~ao \'e sempre atrativa. Em um sistema de part\'iculas carregadas, as for\c{c}as eletromagn\'eticas dominam e fazem com que elas estejam em constante movimento, sempre se repelindo e se atraindo, at\'e o momento em que a velocidade m\'edia das part\'iculas (i.e. a temperatura do sistema) seja suficientemente baixa para que a atra\c{c}\~ao entre cargas opostas seja capaz de prend\^e-las em \'orbita, formando um sistema ligado neutro. Esse processo ocorreu no Universo primordial, cerca de 380.000 anos ap\'os o Big Bang, numa transi\c{c}\~ao de fase em que os pr\'otons e el\'etrons do plasma primordial se combinaram em \'atomos --- um processo chamado na literatura de \textbf{Recombina\c{c}\~ao}, e que d\'a origem \`a Radia\c{c}\~ao C\'osmica de Fundo~\cite{Mukhanov, Dodelson, Liddle}. Importante se ressaltar esse argumento \`a turma, pois ilustra como a efetiva neutralidade do Universo n\~ao \'e simplesmente uma condi\c{c}\~ao inicial arbitr\'aria, mas decorre de um processo f\'isico no Universo primordial.
		
	Apesar de a intera\c{c}\~ao eletromagn\'etica ser muito mais forte que a gravitacional, seus efeitos de atra\c{c}\~ao e repuls\~ao tendem a se cancelar em grandes escalas, enquanto os efeitos gravitacionais s\~ao sempre cumulativos. Assim, para sistemas formados de grandes quantidades de mat\'eria (formando corpos celestes, gal\'axias, conglomerados e demais estruturas c\'osmicas), a gravita\c{c}\~ao se torna a intera\c{c}\~ao mais relevante. De fato, um dos princ\'ipios b\'asicos da Cosmologia moderna \'e que, em escalas cosmol\'ogicas, \'e necess\'ario levar em conta \emph{apenas} a intera\c{c}\~ao gravitacional entre as estruturas interagentes.
\end{itemize}

\subsubsection{Eletromagnetismo e for\c{c}as de contato}
\label{sec:contato}

\begin{figure*}[h!]
    \centering
    \includegraphics[width=.28\textwidth]{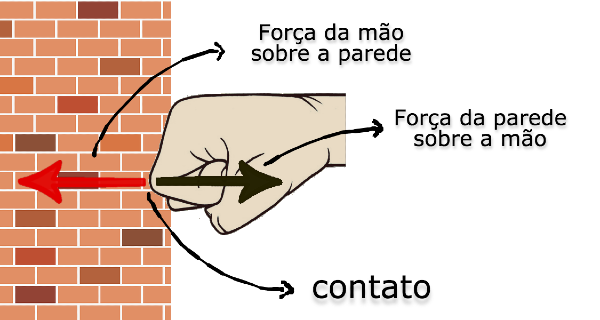}
    \hspace*{-5mm}
	\includegraphics[width=.3\textwidth]{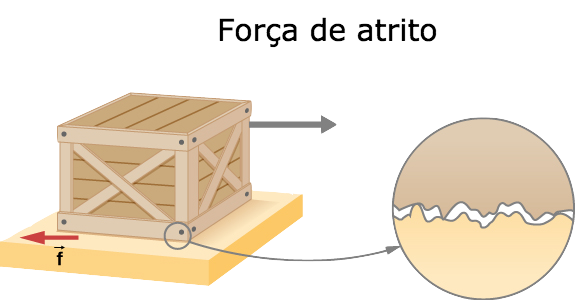}
	\qquad
	\includegraphics[width=.3\textwidth]{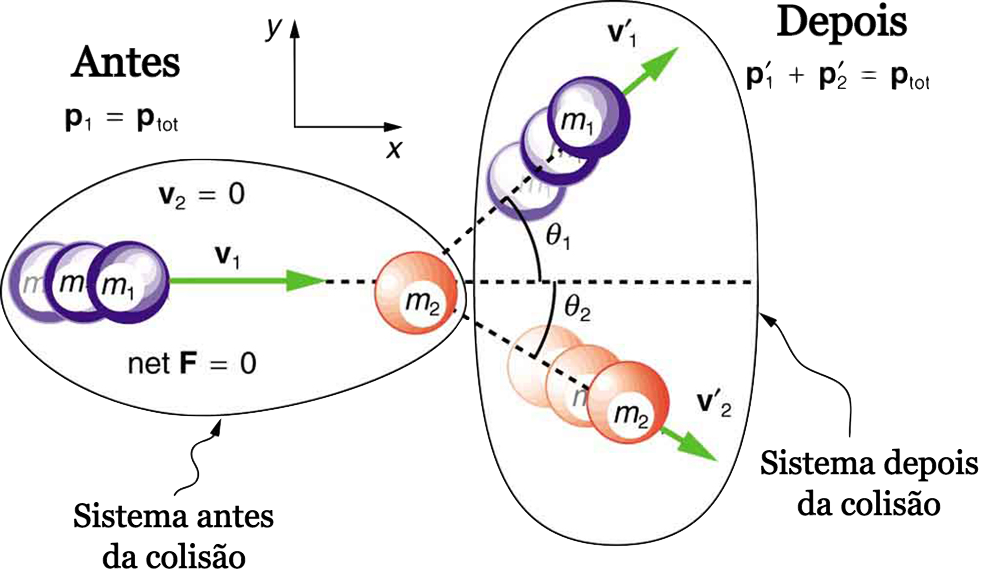}    
	\caption{For\c{c}as de contato. Em \'ultima inst\^ancia, todas t\^em origem na repuls\~ao eletromagn\'etica. Fontes: os autores e adapta\c{c}\~oes de~\cite{Atrito} e~\cite{Collisions}, respectivamente.}
    \label{fig:forcas_contato}
\end{figure*}

Outro ponto de importante men\c{c}\~ao \'e a ubiquidade da intera\c{c}\~ao eletromagn\'etica em fen\^omenos onde n\~ao se os reconhece a princ\'ipio. De seu curso de mec\^anica, o(a) estudante est\'a familiarizado(a) com o conceito de gravita\c{c}\~ao e for\c{c}a peso atuando sobre um corpo, assim como outras formas de for\c{c}a que emergem do \emph{contato} entre corpos (figura~\ref{fig:forcas_contato}), como a for\c{c}a normal, for\c{c}a de atrito, tens\~ao em uma corda, for\c{c}as atuantes em uma colis\~ao, etc.

\begin{figure*}[h!]
	\centering
	\includegraphics[scale=.7, page=10]{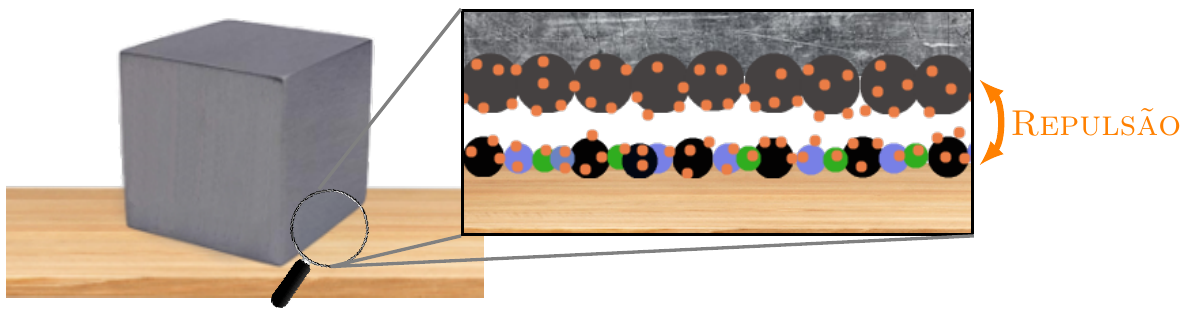}
	\caption{A for\c{c}a normal de uma mesa sobre um bloco \'e devida \`a repuls\~ao entre os el\'etrons dos \'atomos das superf\'icies.}
	\label{fig:contato}
\end{figure*}

Entretanto, em sentido estrito, for\c{c}as de contato inexistem: todas essas for\c{c}as t\^em origem eletromagn\'etica, advindas da repuls\~ao entre os el\'etrons dos \'atomos que constituem os corpos, como ilustrado na figura~\ref{fig:contato}. Apesar de a magnitude da for\c{c}a de repuls\~ao entre el\'etrons ser tipicamente muito menor do que a for\c{c}a peso de um corpo macrosc\'opico, a repuls\~ao aumenta drasticamente \`a medida que as dist\^ancias entre as superf\'icies diminuem, existindo uma dist\^ancia n\~ao-nula para a qual a for\c{c}a peso e a repuls\~ao eletrost\'atica se equilibram.

A revela\c{c}\~ao de que nenhum objeto est\'a em contato com outro, e que estamos sempre ``flutuando'' sobre as superf\'icies onde nos apoiamos, usualmente causa fascina\c{c}\~ao e entusiasmo em estudantes. Mas mais do que isso, \'e importante enfatizar esse ponto porque mostra que todas as for\c{c}as que vivenciamos em nosso cotidiano --- atrito, for\c{c}as normais, empuxo, etc. --- t\^em origem eletromagn\'etica ou gravitacional, o que abre caminho \`a asser\c{c}\~ao futura de que existem poucas\footnote{De acordo com nosso entendimento atual, existem somente quatro intera\c{c}\~oes fundamentais: a gravitacional, a eletromagn\'etica, e as intera\c{c}\~oes nucleares fraca e forte.} intera\c{c}\~oes elementares na natureza. 

\subsection{Ondas eletromagn\'eticas e a natureza da luz}
\label{sec:luz}

O fato de o eletromagnetismo ser uma intera\c{c}\~ao \`a dist\^ancia~\cite{2007Sonia} suscita uma quest\~ao importante: como part\'iculas distantes sabem do comportamento das demais para reagirem de maneira a satisfazer as leis da eletrodin\^amica? Como um el\'etron no receptor de um celular sabe que existem el\'etrons se movendo na antena emissora do sinal? Essa problem\'atica tem rela\c{c}\~ao direta com aplica\c{c}\~oes tecnol\'ogicas que permeiam o cotidiano dos(as) estudantes, que podem (e devem) ser usadas como motiva\c{c}\~ao para essa aula. Perguntas do tipo ``Como os telefones celulares enviam e recebem mensagens?'', ou ``Como funcionam tecnologias \emph{wireless}?'' podem servir como problema inicial em uma aula investigativa.

\begin{figure}[h!]
 \centering
    \includegraphics[scale=.26]{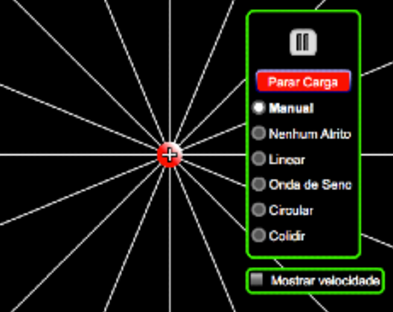}
    \includegraphics[scale=.26]{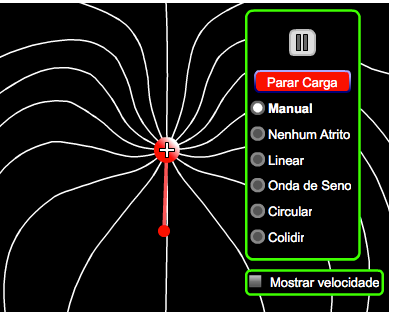}
	\caption{Simula\c{c}\~ao interativa PhET Colorado: Irradiando Carga. Na imagem da esquerda, a carga est\'a em repouso e as linhas de campo s\~ao radiais. \`A direita, a carga foi posta em movimento e as linhas se deformam para acompanh\'a-la. Essa deforma\c{c}\~ao se propaga, constituindo a onda eletromagn\'etica. Fonte:~\cite{PhET_radiacao}.}
	\label{fig:PhET_radiacao}
\end{figure}

\begin{figure}[h!]
 \centering
    \includegraphics[scale=.25]{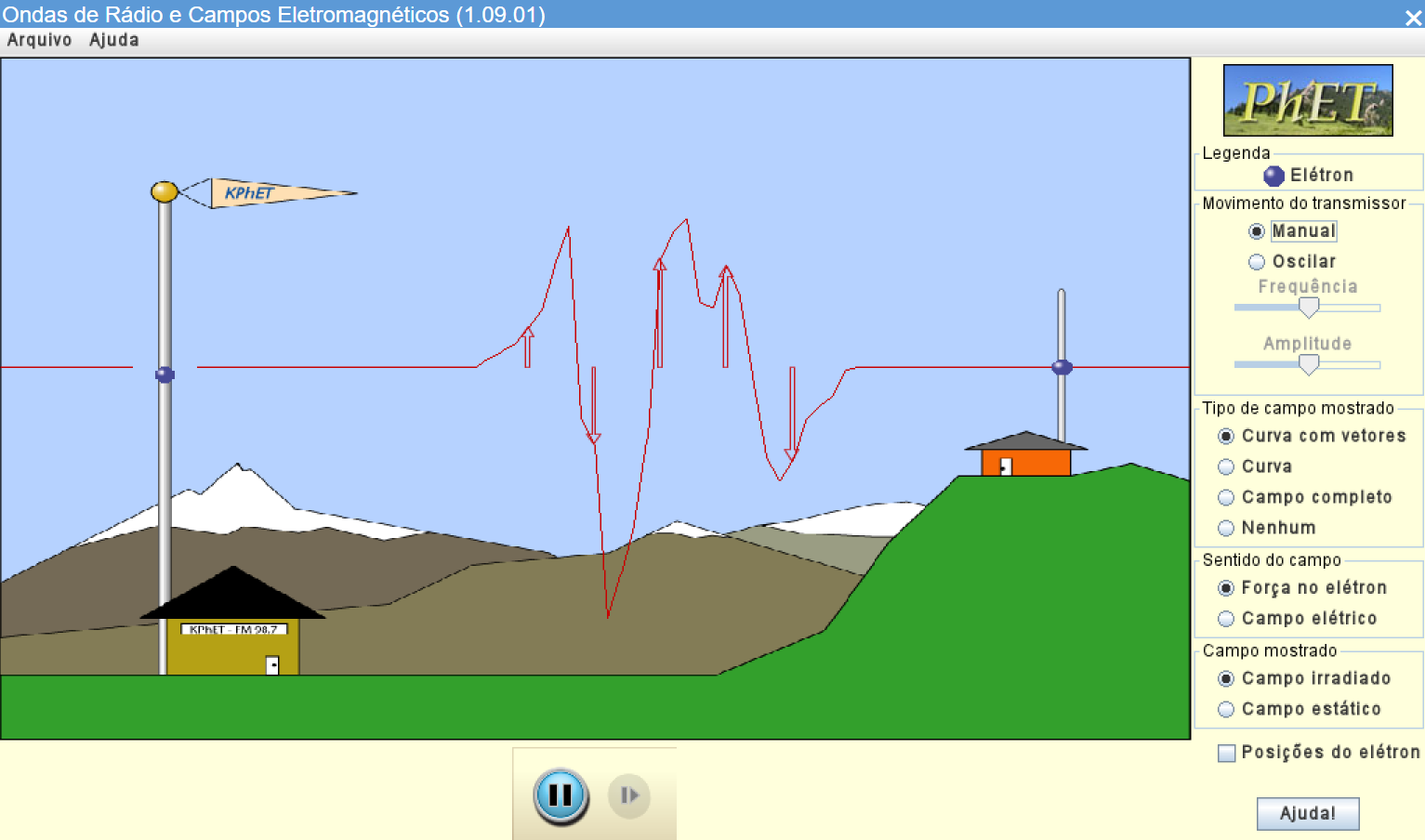}
    \includegraphics[scale=.25]{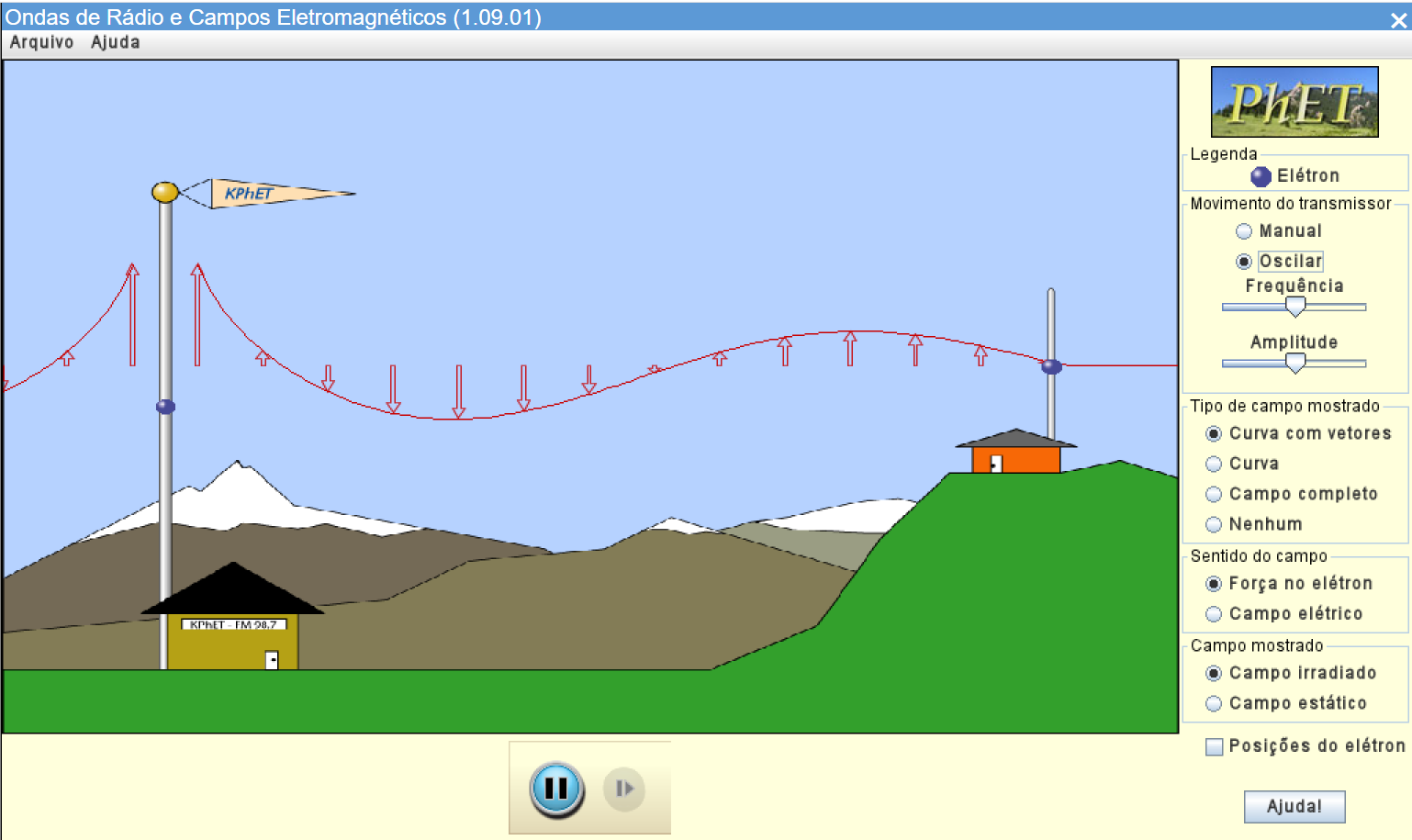}
	\caption{Simula\c{c}\~ao interativa PhET Colorado: Ondas de R\'adio e Campos Eletromagn\'eticos. A simula\c{c}\~ao ilustra como a oscila\c{c}\~ao dos el\'etrons em uma antena transmissora gera uma onda eletromagn\'etica, que se propaga at\'e outra antena receptora, fazendo seus el\'etrons oscilarem \`a mesma frequ\^encia que os da fonte. Assim, o sinal \'e transmitido de um ponto a outro sem interm\'edio de fios. Nessa simula\c{c}\~ao o(a) estudante pode interagir movendo o el\'etron da antena com o \emph{mouse} e produzindo seu pr\'oprio sinal. Fonte:~\cite{PhET_antena}.}
	\label{fig:PhET_antena}
\end{figure}

A discuss\~ao pode ser ent\~ao guiada a uma \^enfase no papel do \emph{campo eletromagn\'etico} enquanto mediador da intera\c{c}\~ao e, mais importante, enquanto ente f\'isico com din\^amica pr\'opria. As conclus\~oes centrais que devem ser atingidas pelos(as) estudantes s\~ao: (i) a exist\^encia de uma carga el\'etrica gera um campo eletromagn\'etico; (ii) caso o estado de movimento da carga seja alterado, o campo \'e modificado em seu entorno imediato, e essa modifica\c{c}\~ao se propaga como uma onda; (iii) quando a onda atinge uma outra carga, transfere-lhe energia e momento, alterando seu estado de movimento, i.e. comunicando-lhe o movimento apropriado. Algumas simula\c{c}\~oes interativas visando auxiliar a compreens\~ao desses efeitos podem ser encontradas em~\cite{PhET_radiacao, PhET_antena} (ver figuras~\ref{fig:PhET_radiacao} e~\ref{fig:PhET_antena}). A conclus\~ao final da investiga\c{c}\~ao \'e que a exist\^encia de radia\c{c}\~ao eletromagn\'etica garante que a causalidade seja preservada pela intera\c{c}\~ao.

O processo de emiss\~ao e recep\c{c}\~ao de ondas eletromagn\'eticas engloba, em uma s\'o tem\'atica, todos os aspectos centrais da teoria eletromagn\'etica. A etapa (i) oferece a oportunidade de se discutir a Lei de Gauss, a etapa (ii) possibilita discutir o fen\^omeno de indu\c{c}\~ao de campos el\'etricos pela varia\c{c}\~ao de campos magn\'eticos (e vice-versa), enquanto (iii) est\'a associado \`a for\c{c}a de Lorentz sobre uma carga. 

Mais especificamente, pode-se apresentar as equa\c{c}\~oes de Maxwell de maneira intuitiva e acess\'ivel a estudantes de ensino m\'edio, substituindo-se as express\~oes matem\'aticas, envolvendo derivadas e integrais, por suas interpreta\c{c}\~oes f\'isicas em forma de senten\c{c}as enunciativas (cf. ref.~\cite{Feynman}, Vol.~II, Cap.~1).

Por exemplo, a Lei de Gauss, que rege a maneira como o campo el\'etrico \'e gerado por uma carga, foi apresentada na se\c{c}\~ao~\ref{sec:emgrav} em termos do fluxo das linhas de campo e como elas se originam e terminam em cargas el\'etricas. Especificamente, o fluxo do campo el\'etrico $\vec{\rm E}$ sobre uma superf\'icie \'e constante, proporcional \`a carga contida em seu interior, 
\begin{equation}
	\parbox{3.5cm}{\centering fluxo de $\vec{\rm E}$\\ atrav\'es de\\ superf\'icie fechada} = 
	\parbox{3.cm}{\centering carga contida\\ no interior\\ da superf\'icie}.
\end{equation}
A partir da\'i, o docente pode estimular os(as) pr\'oprios(as) alunos(as) a enunciar a lei an\'aloga para campos magn\'eticos, bastando dizer-lhes que nunca foram observadas cargas magn\'eticas (i.e. monopolos magn\'eticos). O lado direito deve ser, ent\~ao, igualado a zero, concluindo-se que o fluxo de campo magn\'etico $\vec{\rm B}$ se anula sobre superf\'icies fechadas.

\begin{figure}[h!]
	\centering
	\begin{picture}(200,110)
		\put(0,-40){\includegraphics[angle=45, width=.4\textwidth]{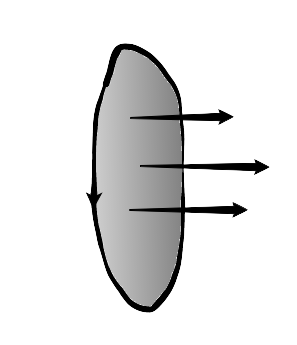}}
		\put(45,50){\scalebox{1.4}{$\mathcal{C}$}}
		\put(115,20){\scalebox{1.4}{$\mathcal{S}$}}
	\end{picture}
	\caption{Uma superf\'icie arbitr\'aria $\mathcal{S}$ delimitada por uma curva fechada $\mathcal{C}$.}
	\label{fig:curva}
\end{figure}

``Qual \'e, ent\~ao, a fonte de campos magn\'eticos?''. A quest\~ao gera a ocasi\~ao para apresentar as outras duas equa\c{c}\~oes de Maxwell,
\begin{equation}\begin{split}
	\parbox{25mm}{\centering circula\c{c}\~ao de $\vec{\rm B}$\\
				em torno de $\mathcal{C}$}
	&=
	\parbox{15mm}{\centering corrente\\ el\'etrica\\ atrav\'es de $\mathcal{S}$}
	+ \parbox{25mm}{\centering varia\c{c}\~ao\\ temporal\\ do fluxo de $\vec{\rm E}$\\ atrav\'es de $\mathcal{S}$},
\end{split}\end{equation}

\begin{equation}
	\parbox{3cm}{\centering circula\c{c}\~ao de $\vec{\rm E}$\\
				em torno de $\mathcal{C}$}
	= \parbox{3cm}{\centering varia\c{c}\~ao temporal\\ do fluxo de $(-\vec{\rm B})$\\ atrav\'es de $\mathcal{S}$},
\end{equation}
onde $\mathcal{S}$ \'e uma superf\'icie arbitr\'aria e $\mathcal{C}$ seu contorno, como ilustrado na figura~\ref{fig:curva}.

\begin{figure*}[h!]
	\centering
	\begin{subfigure}[t]{.45\textwidth}
		\includegraphics[page=12, scale=1]{tikz_figs}
		\caption{}
	\end{subfigure}
	\begin{subfigure}[t]{.45\textwidth}
		\includegraphics[scale=.325]{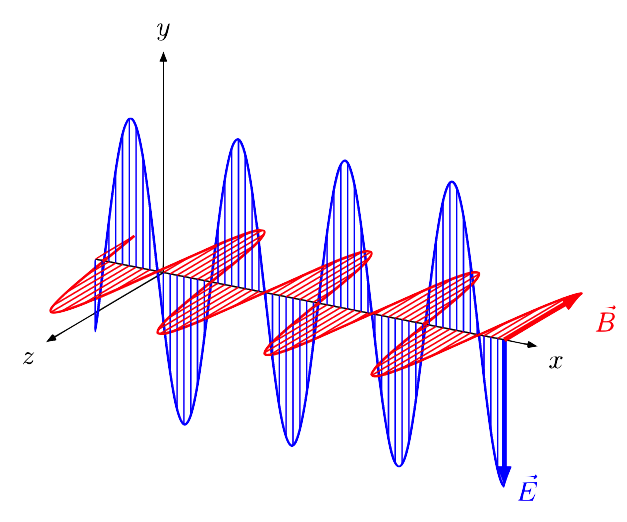}
		\caption{}
	\end{subfigure}
	\caption{(a) Esquema do ciclo retroalimentativo que sustenta a propaga\c{c}\~ao da onda eletromagn\'etica. (b) Uma onda eletromagn\'etica propagando-se na dire\c{c}\~ao $x$ consiste em oscila\c{c}\~oes dos campos el\'etrico e magn\'etico no plano $yz$. No v\'acuo esses campos s\~ao sempre mutuamente perpendiculares. Note que n\~ao h\'a componentes dos campos na dire\c{c}\~ao de propaga\c{c}\~ao: diz-se que a onda eletromagn\'etica \'e uma onda \emph{transversal}. Fonte: Wikimedia Commons/CC BY-SA 4.0.}
	\label{fig:emwave_scheme}
\end{figure*}

Assim, a varia\c{c}\~ao de um campo el\'etrico no tempo (por ex., devido \`a acelera\c{c}\~ao da part\'icula fonte desse campo) gera um campo magn\'etico vari\'avel, que gera um campo el\'etrico vari\'avel, e assim sucessivamente. Com essa argumenta\c{c}\~ao, pode-se mostrar aos estudantes o princ\'ipio b\'asico por tr\'as da propaga\c{c}\~ao das \emph{ondas eletromagn\'eticas}. A figura~\ref{fig:emwave_scheme} mostra um esquema simplificado do processo, mais intuitivo e de mais f\'acil apreens\~ao.

A introdu\c{c}\~ao do conceito de ondas eletromagn\'eticas abre um leque de possibilidades ao docente, que pode abordar tem\'aticas consideradas pelos(as) pr\'oprios(as) estudantes como altamente interessantes~\cite{GOUW2013}, tais como:
\begin{itemize}
	\item o funcionamento de antenas de emiss\~ao e recep\c{c}\~ao, e como ondas eletromagn\'eticas est\~ao presentes em tecnologias atuais de telecomunica\c{c}\~ao (celulares, transmiss\~ao de sinal \emph{wifi}, etc.); 
	\item propriedades fundamentais de ondas gerais, e de ondas eletromagn\'eticas em particular (frequ\^encia, comprimento de onda, amplitude); o espectro eletromagn\'etico e aplica\c{c}\~oes tecnol\'ogicas e medicinais de cada uma de suas bandas (ondas de  TV e r\'adio, microondas, infravermelho, raios solares ultravioleta, raios-X, etc.), ilustradas na figura~\ref{fig:emwaves_uses}. O(A) docente pode, por exemplo, discutir como o microondas funciona enquanto uma \textit{cavidade ressonante}: a onda eletromagn\'etica faz as mol\'eculas de \'agua dos alimentos vibrarem, aumentando sua temperatura. Note que as regi\~oes que se localizam nos n\'os da onda eletromagn\'etica do aparelho n\~ao ser\~ao aquecidas. Por isso, o alimento deve ser posto a girar para que nenhum ponto fique parado no mesmo lugar da cavidade durante todo o per\'iodo do processo de aquecimento do alimento. Pode-se usar a pipoca de microondas como um exemplo divertido para os(as) estudantes;
	\item como atividade, o(a) docente pode pedir que os(as) estudantes pesquisem a banda de frequ\^encia em que funcionam os roteadores \emph{wifi} a que t\^em acesso, e computem o comprimento de onda associado, localizando-o no espectro eletromagn\'etico. Que outros aparelhos dom\'esticos funcionam nessa mesma banda do espectro? O funcionamento de roteadores oferece uma interessante aplica\c{c}\~ao tecnol\'ogica cotidiana da teoria eletromagn\'etica, e tem grande potencial de ser um t\'opico atraente para muitos(as) estudantes interessados nessa \'area de inform\'atica e/ou tecnologia da informa\c{c}\~ao;
	\item interessante destacar, tamb\'em, a natureza da luz enquanto onda eletromagn\'etica, e alguns aspectos hist\'oricos a respeito dessa descoberta. A possibilidade de exist\^encia de tais ondas foi estabelecida por Maxwell em 1864, que tamb\'em calculou a velocidade de propaga\c{c}\~ao, encontrando um valor muito pr\'oximo \`a velocidade da luz (cujo valor j\'a era conhecido na \'epoca), concluindo, ent\~ao, que a luz \'e uma onda eletromagn\'etica. Esse resultado tem um valor hist\'orico inestim\'avel: a partir de grandezas medidas em laborat\'orio com o uso de circuitos el\'etricos, fios, volt\'imetros etc., Maxwell p\^ode extrair uma conclus\~ao sobre a natureza da luz, um fen\^omeno que n\~ao parece ter qualquer rela\c{c}\~ao com aqueles experimentos! A natureza ondulat\'oria da luz j\'a era conhecida desde que Young realizou seu experimento de fenda dupla em 1801. No entanto, n\~ao se conhecia ainda o substrato oscilante nessas ondas, cabendo a Maxwell tal identifica\c{c}\~ao;
	\item pode-se, ainda, mencionar o espectro de corpo negro, ou, mais geralmente, o fato de que todo corpo emite radia\c{c}\~ao eletromagn\'etica devido \`a sua temperatura. Esse espectro cont\'em um pico, e a frequ\^encia no pico de emit\^ancia \'e proporcional \`a temperatura do corpo, o que significa que quanto mais quente for o corpo, mais ``azulada'' \'e a radia\c{c}\~ao emitida por ele. O(a) docente pode propor aos(\`as) estudantes interagirem com a simula\c{c}\~ao encontrada na ref.~\cite{PhET_blackbody}, que mostra como o espectro de emit\^ancia de um corpo varia com a temperatura, como na figura~\ref{fig:phet_blackbody}. Essa depend\^encia do pico de emit\^ancia com a temperatura \'e utilizada, por exemplo, para medir a temperatura de estrelas a partir do espectro de luz que elas emitem. A superf\'icie do Sol est\'a a uma temperatura de aproximadamente 5500 K, e seu espectro de emiss\~ao tem um pico justamente na regi\~ao da luz vis\'ivel, i.e. a maior parte da radia\c{c}\~ao emitida pelo Sol encontra-se na faixa que n\'os conseguimos enxergar. Isso n\~ao \'e uma coincid\^encia, mas produto da sele\c{c}\~ao natural. O ap\^endice~\ref{sec:solar} cont\'em uma proposta de atividade interdisciplinar, a ser realizada com os(as) estudantes e potencialmente em conjunto com o(a) professor(a) da \'area de Biologia, explorando esse fato. A figura~\ref{fig:solar} encontrada naquela se\c{c}\~ao ilustra o espectro de emit\^ancia solar sobreposto ao de um corpo negro ideal \`a temperatura $5500$~K.
	
\begin{figure*}[h!]
    \centering
    \includegraphics[page=13, scale=.9]{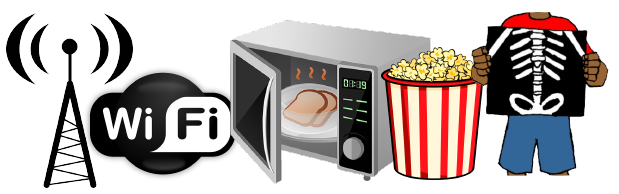}
    \caption{Algumas aplica\c{c}\~oes de diversas bandas do espectro eletromagn\'etico, que podem ser exploradas em sala em um contexto CTSA. Fonte: Dom\'inio P\'ublico.}
    \label{fig:emwaves_uses}
\end{figure*}

\begin{figure}[h!]
    \centering
    \includegraphics[scale=.2]{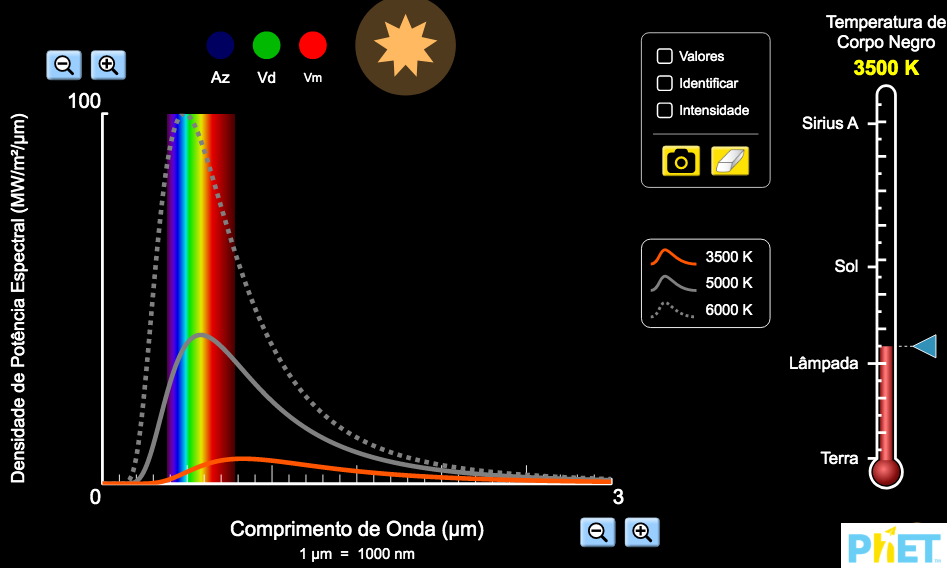}
    \caption{Imagem da simula\c{c}\~ao interativa PhET Colorado: Espectro de Corpo Negro. Fonte:~\cite{PhET_blackbody}.}
    \label{fig:phet_blackbody}
\end{figure}

\end{itemize}

\begin{figure*}
    \centering
    \includegraphics[scale=.5]{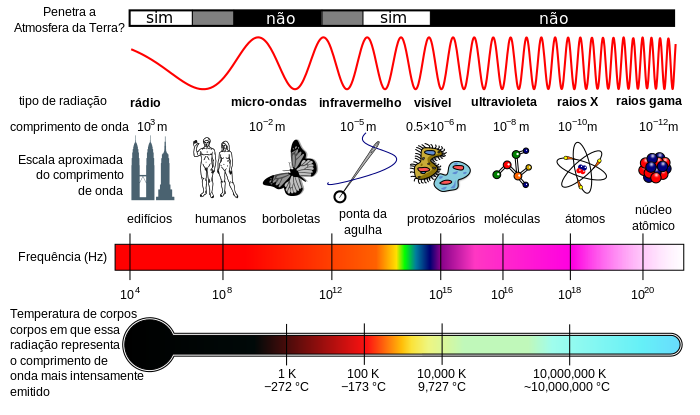}
    \caption{Diversos tipos de radia\c{c}\~ao eletromagn\'etica, caracterizados por uma banda espec\'ifica de frequ\^encia e comprimento de onda do espectro completo. A cada banda, atribui-se tamb\'em um objeto cujo tamanho t\'ipico \'e da mesma ordem de magnitude do comprimento da onda correspondente: tratam-se dos menores objetos que podem ser distinguidos por meio dessas ondas. Por exemplo, microorganismos podem ser vistos sob um microsc\'opio usual, funcionando \`a base de luz vis\'ivel, mas para distinguir mol\'eculas ou \'atomos \'e preciso fazer incidir radia\c{c}\~ao muito mais energ\'etica, de menor comprimento de onda. As frequ\^encias de cada banda s\~ao associadas, tamb\'em, \`as cores da radia\c{c}\~ao correspondente, divididas em tr\^es regi\~oes: o infravermelho, o espectro cont\'inuo da luz vis\'ivel, e o ultravioleta. Por fim, ilustra-se tamb\'em a temperatura de um corpo cujo espectro de emiss\~ao termal \'e predominante na banda em quest\~ao. Fonte: Wikimedia Commons/CC BY-SA 4.0.}
    \label{fig:em_spectrum}
\end{figure*}

A figura~\ref{fig:em_spectrum} ilustra diversos aspectos do espectro eletromagn\'etico, e pode ser extremamente \'util ao(\`a) docente na elabora\c{c}\~ao de sua aula e das atividades a serem propostas relativas a essa tem\'atica.

\subsection{Dualidade onda-part\'icula}
\label{sec:dualidade}

A discuss\~ao da tem\'atica de ondas eletromagn\'eticas abre a possibilidade de se adentrar em temas de f\'isica moderna, como a teoria da relatividade restrita ou a dualidade onda-part\'icula em mec\^anica qu\^antica. Embora ambos sejam de suma import\^ancia para a F\'isica de Part\'iculas, nesta sequ\^encia did\'atica focamos no aspecto qu\^antico visando a introdu\c{c}\~ao do conceito de f\'oton. Para propostas did\'aticas visando o ensino de relatividade no ensino m\'edio vide refs.~\cite{Rel, Capelari, Sa}.

A dualidade onda-part\'icula da luz pode ser introduzida de diversas maneiras.
\begin{itemize}
	\item Invocando-se o experimento de Young da fenda dupla, pode-se, em um primeiro momento da aula,  discutir as expectativas dos alunos quanto \`a imagem formada no aparato de incid\^encia, quando o objeto incidente for uma part\'icula cl\'assica (figura~\ref{fig:ds_particle}) e quando for uma onda (para esse caso pode-se explorar, junto aos(\`as) alunos(as), a simula\c{c}\~ao interativa dispon\'ivel em~\cite{PhET_interferencia}). Em um segundo momento da aula, discute-se (ou realiza-se em sala) o experimento de Young, que detectou franjas de interfer\^encia para a luz, mostrando tratar-se de um fen\^omeno de natureza ondulat\'oria. Posteriormente, discute-se o que ocorre quando se reduz suficientemente a intensidade do feixe luminoso, resultando nos padr\~oes mostrados na figura~\ref{fig:ds_particle_quantum}. As franjas s\~ao reveladas aos poucos, a partir de detec\c{c}\~oes de part\'iculas pontuais, revelando assim o \emph{car\'ater corpuscular} da luz (para imagens e v\'ideo ilustrativo que podem ser usados em sala de aula, consultar ref.~\cite{SIEGEL2019}).
\begin{figure}[h!]
	\centering
	\includegraphics[width=.4\textwidth]{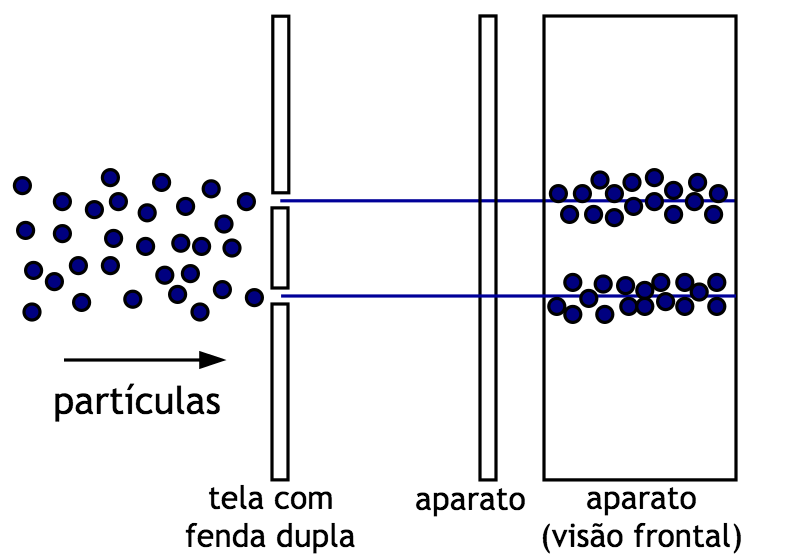}
	\caption{Padr\~ao esperado para um feixe de part\'iculas cl\'assicas lan\c{c}adas contra uma tela com duas fendas. Fonte: adaptado de Wikimedia Commons/Dom\'inio p\'ublico.}
	\label{fig:ds_particle}
\end{figure}
\begin{figure}[h!]
	\centering
	\begin{subfigure}[b]{.3\textwidth}
		\includegraphics[width=\textwidth]{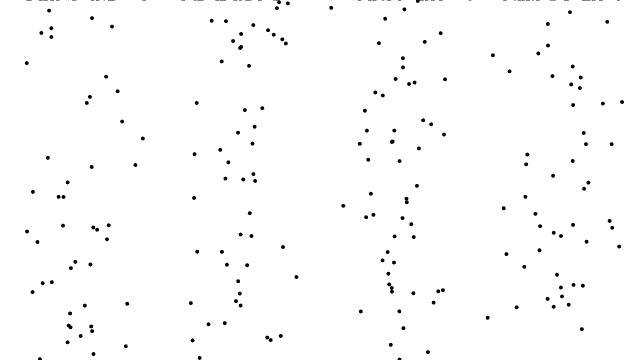}
		\caption{}
	\end{subfigure}\qquad
	\begin{subfigure}[b]{.3\textwidth}
		\includegraphics[width=\textwidth]{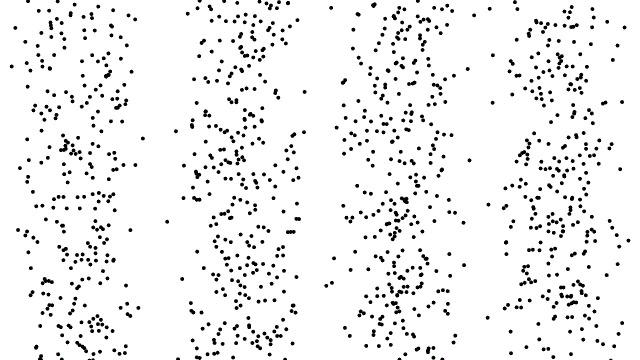}
		\caption{}
	\end{subfigure}\qquad
	\begin{subfigure}[b]{.3\textwidth}
		\includegraphics[width=\textwidth]{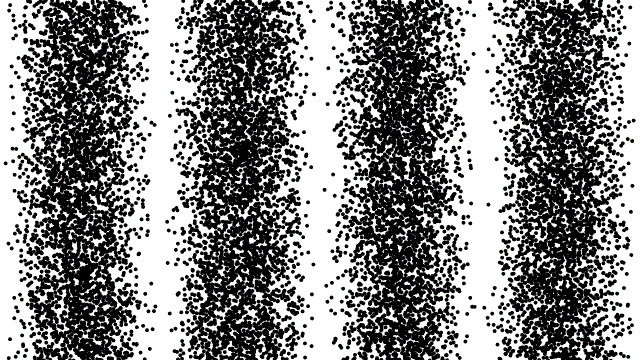}
		\caption{}
	\end{subfigure}
	\caption{Para part\'iculas qu\^anticas, o padr\~ao de interfer\^encia \'e inicialmente impercept\'ivel, mas emerge paulatinamente \`a medida que mais estat\'istica \'e coletada pelo experimento. Fonte: Wikimedia Commons/Dom\'inio p\'ublico.}
	\label{fig:ds_particle_quantum}
\end{figure}
	
	A discuss\~ao do experimento da fenda dupla em sala de aula cria, tamb\'em, uma excelente oportunidade de desmistificar a mec\^anica qu\^antica, o que \'e especialmente importante levando-se em conta a quantidade de charlat\~aes que usam os resultados n\~ao-intuitivos desse experimento para propor que a F\'isica corrobora suas interpreta\c{c}\~oes esot\'ericas do mundo.
	
	Mais especificamente, o surgimento de um padr\~ao de interfer\^encia para um feixe de part\'iculas pode ser explorado para se discutir a necessidade de se abandonar o conceito de trajet\'orias bem definidas (que deveria levar ao padr\~ao dado pela figura~\ref{fig:ds_particle}), e, consequentemente, a necessidade de se reformular a mec\^anica. Pode-se usar esse gancho para discutir o \textit{princ\'ipio da incerteza} de Heisenberg, e como esse princ\'ipio explica o fen\^omeno de destrui\c{c}\~ao do padr\~ao de interfer\^encia quando um observador detecta por qual fenda a part\'icula passou (cf.~ref.~\cite{Feynman}, Vol.~III, Se\c{c}\~ao 1.8). Tamb\'em \'e essencial esclarecer aos estudantes que n\~ao se deve pensar em part\'iculas qu\^anticas como ``bolinhas'' movendo-se classicamente. A prop\'osito, fazer uso de analogias com a mec\^anica cl\'assica geralmente leva a uma confus\~ao de conceitos~\cite{2019PassonETAL}. Os f\'otons devem ser vistos como pacotes de energia e momento. O mesmo vale para todas as part\'iculas elementares. Isso ficar\'a mais claro quando se apresentar a defini\c{c}\~ao de part\'icula no paradigma de Teorias Qu\^anticas de Campos, como a Eletrodin\^amica Qu\^antica, que ser\'a apresentada na se\c{c}\~ao seguinte.
	
	\item Outra evid\^encia em favor do car\'ater corpuscular da luz \'e o efeito fotoel\'etrico/fotovoltaico. Por constituir o princ\'ipio b\'asico do funcionamento de c\'elulas fotovoltaicas, essa tem\'atica pode ser abordada em um contexto de aula investigativa~\cite{FreitasFotoeletrico} ou, ainda, em di\'alogo mais amplo com outras disciplinas escolares em contexto CTSA (Ci\^encia, Tecnologia, Sociedade e Ambiente)~\cite{Lima}. Uma proposta de tal atividade encontra-se delineada no ap\^endice~\ref{sec:fotovoltaico}.
	
	Ademais, a discuss\~ao do efeito fotoel\'etrico permite a apresenta\c{c}\~ao da rela\c{c}\~ao de Einstein para a energia dos ``pacotes de onda eletromagn\'etica'',
	\begin{equation}
		\left(\parbox{2cm}{\centering energia\\ da part\'icula}\right) = \text{(constante) }\times \left(\parbox{2cm}{\centering frequ\^encia\\ da onda\\ associada}\right),
		\label{eq:Ehf}
	\end{equation} 
	rela\c{c}\~ao que quantifica a dualidade ao associar uma grandeza tipicamente ligada a part\'iculas a uma propriedade ondulat\'oria. 
\end{itemize}

Seja qual for o caminho escolhido, o objetivo a se atingir \'e introduzir o conceito de \textbf{f\'otons} enquanto part\'iculas da radia\c{c}\~ao eletromagn\'etica.

\subsection{Eletrodin\^amica Qu\^antica}
\label{sec:qed}

\subsubsection{O campo eletromagn\'etico qu\^antico}
\label{sec:qft}

A introdu\c{c}\~ao do conceito de f\'oton implica, em um primeiro momento, em uma ambiguidade na descri\c{c}\~ao da radia\c{c}\~ao eletromagn\'etica, que passa a ser ao mesmo tempo tratada como oscila\c{c}\~ao de um campo cl\'assico e como ente com propriedades qu\^anticas\footnote{Note que essa ambiguidade n\~ao est\'a associada \`a dualidade onda-part\'icula. O problema acima exposto diz respeito \`a impossibilidade de um mesmo objeto possuir um car\'ater cl\'assico e qu\^antico, simultaneamente. J\'a a dualidade onda-part\'icula \'e introduzida justamente como supera\c{c}\~ao dos conceitos cl\'assicos de ``onda'' e ``part\'icula'', que n\~ao se aplicam a objetos qu\^anticos.}. Como o car\'ater qu\^antico dos f\'otons poderia emergir da propaga\c{c}\~ao de um campo cl\'assico? 
Essa inconsist\^encia aponta para a necessidade de se quantizar o campo eletromagn\'etico, i.e. descrev\^e-lo inteiramente de acordo com o paradigma qu\^antico. 

A teoria qu\^antica que descreve a intera\c{c}\~ao entre part\'iculas carregadas (por ex. el\'etrons) e o campo eletromagn\'etico quantizado \'e chamada de Eletrodin\^amica Qu\^antica (\emph{Quantum Electrodynamics} ou QED em ingl\^es).

Na QED, o campo eletromagn\'etico passa a ser visto como uma ``superf\'icie'' tridimensional\footnote{Mais precisamente, uma hipersuperf\'icie.} vibrat\'oria que preenche todo o volume do espa\c{c}o, an\'aloga \`a superf\'icie de um instrumento musical de percuss\~ao, como a pele de um tambor. \`A parte da dimensionalidade --- o campo preenche todo o espa\c{c}o, enquanto a pele percussiva \'e plana, bidimensional ---, a \'unica (e crucial) distin\c{c}\~ao se deve ao car\'ater cl\'assico das vibra\c{c}\~oes do instrumento musical e ao car\'ater qu\^antico do campo quantizado\footnote{Tecnicamente, tanto a superf\'icie sonora quanto o campo da QED podem ser descritos como uma cole\c{c}\~ao de osciladores harm\^onicos, mas no primeiro caso os osciladores s\~ao cl\'assicos, e no segundo s\~ao osciladores qu\^anticos.}. A partir dessa analogia, e da compreens\~ao de suas limita\c{c}\~oes, pode-se entender conceitos fundamentais da QED (e de Teorias Qu\^anticas de Campos em geral).

Primeiramente, assim como a superf\'icie percussiva possui um estado em que n\~ao h\'a quaisquer vibra\c{c}\~oes e nenhum som \'e emitido, tamb\'em o campo qu\^antico possui um \emph{estado fundamental}, que \'e chamado de \emph{estado de v\'acuo} (a motiva\c{c}\~ao para essa nomenclatura ficar\'a clara a seguir). No entanto, note que, por se tratar de um objeto qu\^antico, n\~ao \'e poss\'ivel afirmar que o campo no estado fundamental est\'a em repouso. O princ\'ipio da incerteza garante que, mesmo no estado fundamental, h\'a vibra\c{c}\~oes do campo, associadas \`a chamada ``energia de ponto-zero''. Isso significa que o v\'acuo qu\^antico tem energia, fato que tem consequ\^encias observ\'aveis e aplica\c{c}\~oes pr\'aticas/tecnol\'ogicas, tais como:
\begin{itemize}
	\item o \textit{efeito Casimir}, que consiste na atra\c{c}\~ao de duas placas met\'alicas, eletricamente neutras, imersas no v\'acuo. A atra\c{c}\~ao se deve \`a modifica\c{c}\~ao da energia do v\'acuo qu\^antico na presen\c{c}a das placas. Mais especificamente, quanto mais pr\'oximas as placas, menor a energia do v\'acuo na regi\~ao por elas limitadas;
	\item o processo de \textit{emiss\~ao espont\^anea} de um f\'oton por um \'atomo no estado excitado, que d\'a origem aos espectros at\^omicos. No formalismo de Schr\"odinger da mec\^anica qu\^antica, todas as camadas eletr\^onicas do \'atomo correspondem a estados estacion\'arios, e a probabilidade de transi\c{c}\~ao entre eles \'e nula. A transi\c{c}\~ao do el\'etron entre camadas, e a consequente emiss\~ao de um f\'oton, s\'o ocorre devido \`a intera\c{c}\~ao do el\'etron com o v\'acuo eletromagn\'etico qu\^antico;
	\item por possuir energia, a relatividade geral de Einstein garante que \textbf{o v\'acuo qu\^antico gravita}. Essa energia do v\'acuo age como uma \emph{constante cosmol\'ogica}, contribuindo para a expans\~ao acelerada do Universo. Entretanto, quando se compara a previs\~ao te\'orica para a energia do v\'acuo qu\^antico com o valor da constante cosmol\'ogica medido com base em observa\c{c}\~oes da expans\~ao do Universo, h\'a uma discrep\^ancia de 120 ordens de magnitude! Ou seja, trata-se da pior previs\~ao da hist\'oria da F\'isica! Esse \'e o chamado \textit{problema da constante cosmol\'ogica}, ainda sem solu\c{c}\~ao~\cite{Burgess:2013ara, Adler:1995vd}.
\end{itemize}

\begin{figure*}[h!]
	\centering
	\includegraphics[page=14, scale=.6]{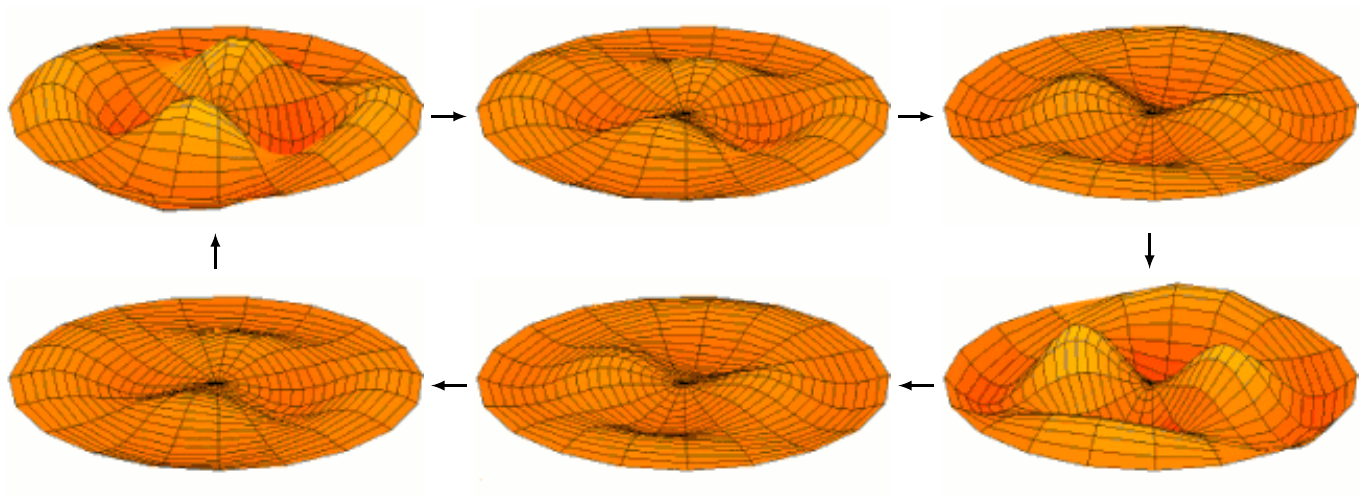}
	\caption{Momentos distintos de um ciclo oscilat\'orio associado a um modo normal de vibra\c{c}\~ao para uma superf\'icie bidimensional. No caso de um instrumento de percuss\~ao, esse padr\~ao corresponderia a uma nota musical (i.e. uma frequ\^encia sonora) espec\'ifica. Para um campo qu\^antico, corresponde a uma part\'icula com energia e momento espec\'ificos. Fonte: adaptado de Wikimedia Commons/Dom\'inio p\'ublico.}
	\label{fig:drum}
\end{figure*}

Al\'em do estado fundamental, uma superf\'icie vibrat\'oria pode, obviamente, oscilar de diversas maneiras. Dentre elas, destacam-se os chamados \emph{modos normais de vibra\c{c}\~ao}, associados a ondas estacion\'arias, em que todos os pontos da superf\'icie oscilam com a mesma frequ\^encia, como exemplificado ilustrativamente na figura~\ref{fig:drum}. Em um instrumento musical, um modo normal est\'a associado a uma nota musical espec\'ifica: quando se ataca a pele de um tambor, ela vibra em um modo normal e emite som da frequ\^encia para a qual o instrumento foi previamente afinado. No caso de um campo qu\^antico, um modo normal de vibra\c{c}\~ao corresponde a uma \emph{part\'icula} de energia e momento definidos. Note o car\'ater explicitamente dual dessa descri\c{c}\~ao, que tra\c{c}a paralelos diretos entre part\'iculas com energia definida e modos de oscila\c{c}\~ao de determinada frequ\^encia. O fato de a vibra\c{c}\~ao n\~ao estar localizada em nenhuma regi\~ao, mas ocorrer em todo o espa\c{c}o ocupado pelo campo, como ilustrado na figura~\ref{fig:drum}, est\'a de acordo com o esperado pelo princ\'ipio da incerteza de Heisenberg quando aplicado a uma part\'icula cujo momento linear \'e conhecido precisamente.

Essa interpreta\c{c}\~ao de part\'iculas como excita\c{c}\~oes do campo em modos normais de vibra\c{c}\~ao justifica, ainda, a nomenclatura de \emph{v\'acuo} para o estado fundamental, por corresponder ao caso em que o n\'umero de part\'iculas se anula.

Por fim, enfatiza-se que a mesma interpreta\c{c}\~ao se aplica a part\'iculas materiais: o el\'etron tamb\'em \'e um modo normal de vibra\c{c}\~ao de um campo fundamental, que podemos chamar de campo eletr\^onico ou campo do el\'etron. Isso explica, ali\'as, por que todos el\'etrons do Universo s\~ao id\^enticos: todos s\~ao vibra\c{c}\~oes de um \'unico campo fundamental. O Modelo Padr\~ao da F\'isica de Part\'iculas, embasado no formalismo de Teoria Qu\^antica de Campos, \'e ao mesmo tempo uma teoria de part\'iculas e uma teoria de intera\c{c}\~ao entre campos.

\subsubsection{F\'otons virtuais como mediadores da intera\c{c}\~ao eletromagn\'etica}

Em discuss\~oes anteriores (vide se\c{c}\~oes~\ref{sec:contato} e \ref{sec:luz}), foi apresentada a ideia de que a mat\'eria n\~ao interage por contato direto entre seus constituintes, mas \`a dist\^ancia, mediada por campos de intera\c{c}\~ao, como o campo eletromagn\'etico. Sob a perspectiva da QED, isso se traduz como a aus\^encia de autointera\c{c}\~ao do campo de el\'etrons. Ao inv\'es disso, a  intera\c{c}\~ao direta do campo eletr\^onico se d\'a com outro(s) campo(s), e a repuls\~ao entre dois el\'etrons nada mais \'e do que um efeito colateral da intera\c{c}\~ao de ambos com um campo intermedi\'ario.

\begin{figure*}[h!]
	\centering
	\begin{subfigure}[b]{.3\textwidth}
		\includegraphics[width=\textwidth]{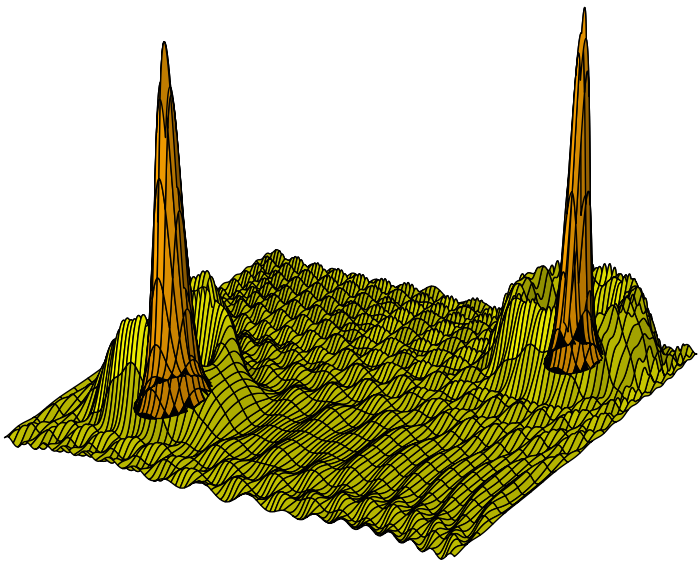}
		\caption{}
	\end{subfigure}
	\begin{subfigure}[b]{.3\textwidth}
		\includegraphics[width=\textwidth]{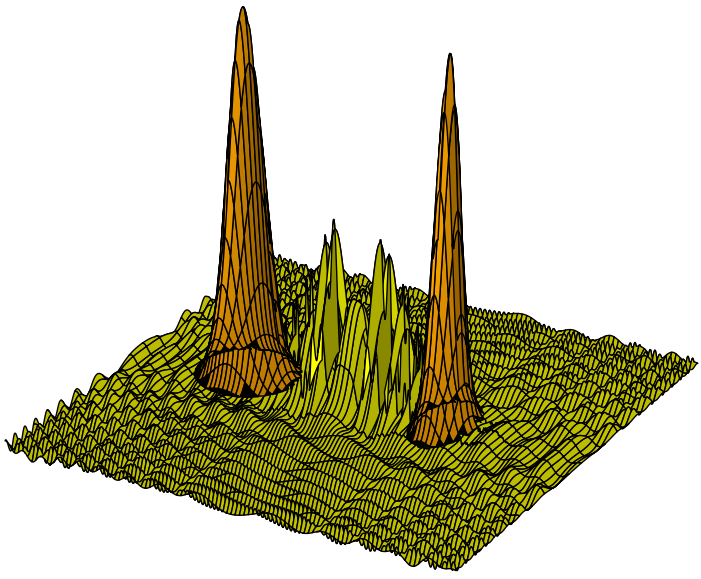}
		\caption{}
	\end{subfigure}
	\begin{subfigure}[b]{.3\textwidth}
		\includegraphics[width=\textwidth]{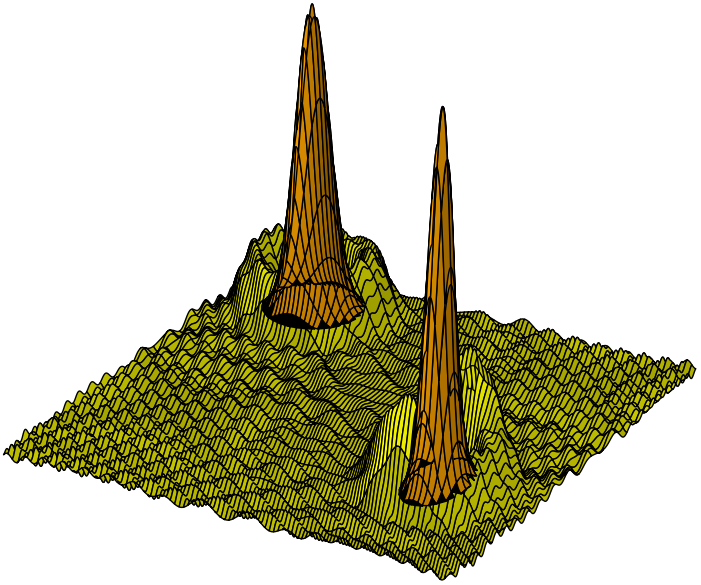}
		\caption{}
	\end{subfigure}
	\caption{(a) Dois el\'etrons (picos mais protuberantes, de cor laranja) afastados um do outro. Devido \`a intera\c{c}\~ao entre o campo de el\'etrons e o campo eletromagn\'etico, um el\'etron nunca existe isoladamente: sua presen\c{c}a induz excita\c{c}\~oes do campo fot\^onico ao seu redor. (b) \`A medida que os el\'etrons se aproximam, intensifica-se a troca de energia-momento entre ambos mediada por excita\c{c}\~oes do campo eletromagn\'etico, causando a m\'utua repuls\~ao. (c) El\'etrons novamente distantes ap\'os a intera\c{c}\~ao. }
	\label{fig:ee_field}	
\end{figure*}

Esse processo de intera\c{c}\~ao na QED est\'a representado esquematicamente na figura~\ref{fig:ee_field}. Os dois picos mais protuberantes, de cor laranja, representam el\'etrons, i.e. excita\c{c}\~oes do campo eletr\^onico\footnote{Caso esses el\'etrons possu\'issem momento bem definido, essas distribui\c{c}\~oes deveriam se espalhar por todo o plano da figura. Aqui, no entanto, tratam-se de pacotes de onda localizados em uma regi\~ao do espa\c{c}o, o que implica em uma incerteza no momento.}, e o campo de fundo, de cor verde, corresponde ao campo eletromagn\'etico. Devido \`a intera\c{c}\~ao entre os dois campos, um el\'etron nunca existe isoladamente: sua presen\c{c}a induz uma ``nuvem'' de excita\c{c}\~oes do campo eletromagn\'etico ao seu redor, como na figura~\ref{fig:ee_field} (a). \`A medida que os el\'etrons se aproximam, intensifica-se a troca de energia-momento entre ambos mediada por excita\c{c}\~oes do campo eletromagn\'etico (b), causando a m\'utua repuls\~ao (c). 

D\'a-se o nome de \emph{f\'oton virtual} a qualquer excita\c{c}\~ao do campo eletromagn\'etico que \textbf{n\~ao} corresponde a um modo normal de vibra\c{c}\~ao, tais como as excita\c{c}\~oes do campo eletromagn\'etico ilustradas na figura~\ref{fig:ee_field}. Um f\'oton virtual \'e, portanto, um estado n\~ao-estacion\'ario, n\~ao necessariamente associado \`a propaga\c{c}\~ao de uma onda, e portanto bastante diferente de um f\'oton real, como ficar\'a ainda mais claro a seguir. Entretanto, justifica-se cham\'a-la de f\'oton por se tratar de uma vibra\c{c}\~ao do campo eletromagn\'etico, e, para distingu\'i-la de um f\'oton observ\'avel, usa-se essa adjetiva\c{c}\~ao de \emph{virtual}\footnote{Existe um acalorado debate na literatura a respeito da exist\^encia ou inexist\^encia de part\'iculas virtuais. Para uma discuss\~ao a esse respeito, cf. ref.~\cite{Jaeger:2019sfp}.}. 

Note que h\'a f\'otons virtuais sendo trocados em todas as tr\^es etapas da intera\c{c}\~ao, representados pelas pequenas perturba\c{c}\~oes do campo eletromagn\'etico entre os dois el\'etrons nas  figuras~\ref{fig:ee_field} (a), (b) e (c). Mas a contribui\c{c}\~ao do f\'oton virtual ao espalhamento \'e mais significativa quando os el\'etrons est\~ao mais pr\'oximos um do outro, como representado na figura~\ref{fig:ee_field} (b). Essa discuss\~ao ser\'a retomada em mais detalhes na se\c{c}\~ao~\ref{sec:feynman} a seguir.

Cabe ressaltar que, apesar de termos definido anteriormente um el\'etron como uma excita\c{c}\~ao do campo eletr\^onico, o que realmente se observa \'e o conjunto de excita\c{c}\~oes do campo eletr\^onico e do campo eletromagn\'etico ao seu redor: diz-se que o el\'etron nunca \'e visto ``nu'', mas est\'a sempre ``vestido'' de uma nuvem de part\'iculas virtuais. Esse fato ser\'a novamente discutido e melhor explorado na se\c{c}\~ao~\ref{sec:effcharge} a seguir.

Ressalta-se, ainda, que a figura~\ref{fig:ee_field} \'e apenas uma representa\c{c}\~ao esquem\'atica da intera\c{c}\~ao, uma alus\~ao semicl\'assica a um processo que \'e fundamentalmente qu\^antico. O simples fato de desenhar as oscila\c{c}\~oes do campo pressup\~oe que saibamos a posi\c{c}\~ao de cada ponto dessa superf\'icie no espa\c{c}o, em desacordo com o princ\'ipio da incerteza. Entretanto, ilustrar um campo qu\^antico \'e imposs\'ivel, e a representa\c{c}\~ao acima, apesar de incompleta, fornece uma boa intui\c{c}\~ao sobre o processo de intera\c{c}\~ao entre (excita\c{c}\~oes de) campos.

Ainda relacionado \`as limita\c{c}\~oes da figura, note que, assim como a presen\c{c}a do el\'etron causa excita\c{c}\~oes do campo eletromagn\'etico, estes f\'otons virtuais tamb\'em induzem excita\c{c}\~oes do campo eletr\^onico, que induzem outras oscila\c{c}\~oes fot\^onicas, e assim sucessivamente. Ou seja, em torno das oscila\c{c}\~oes do campo eletromagn\'etico dever-se-ia tamb\'em ilustrar oscila\c{c}\~oes do campo eletr\^onico, e em torno dessas ilustrar outras excita\c{c}\~oes do campo eletromagn\'etico, etc. A figura~\ref{fig:ee_field} representa, portanto, apenas a primeira ordem de uma s\'erie de outros processos que poderiam ocorrer na presen\c{c}a de dois el\'etrons. Como se trata de um sistema qu\^antico, n\~ao \'e poss\'ivel dizer qual desses processos realmente ocorre: o resultado da intera\c{c}\~ao \'e uma superposi\c{c}\~ao qu\^antica de todas essas possibilidades. Aqui o(a) docente pode fazer uma alus\~ao direta ao experimento da fenda dupla, em que n\~ao \'e poss\'ivel afirmar por qual fenda a part\'icula passa: a interfer\^encia \'e resultado da superposi\c{c}\~ao das duas possibilidades de trajet\'orias, cada uma passando por uma fenda.

\subsubsection{Diagramas de Feynman}
\label{sec:feynman}

\begin{figure}[h!]
	\centering
	\includegraphics[page=3]{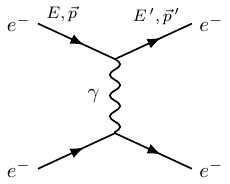}
	\caption{Diagrama de Feynman ilustrando a intera\c{c}\~ao entre dois el\'etrons mediada pela troca de um f\'oton virtual. Os pontos que conectam os el\'etrons ao f\'oton s\~ao chamados de \textit{v\'ertices} do diagrama.
	}
	\label{fig:ee}
\end{figure}

\begin{figure*}[h!]
	\centering
	\includegraphics[page=4]{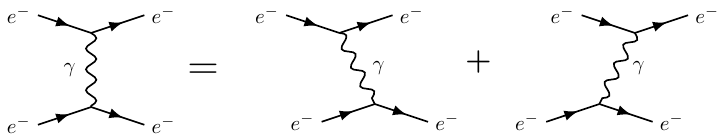}
	\caption{A intera\c{c}\~ao entre el\'etrons pela troca de um f\'oton virtual deve levar em conta as duas possibilidades de ordenamento temporal, em que um el\'etron emite e o outro absorve o f\'oton trocado.}
	\label{fig:feyn_sum}
\end{figure*}

Outra maneira de representar a intera\c{c}\~ao entre el\'etrons \'e atrav\'es de um \emph{diagrama de Feynman}, como na figura~\ref{fig:ee}. Aqui, o eixo vertical representa a dimens\~ao espacial, e o eixo horizontal \'e a passagem do tempo. Representa-se, aqui, o mesmo processo ilustrado na figura~\ref{fig:ee_field} anterior: dois el\'etrons interagindo pela troca de um f\'oton virtual. Entretanto, \emph{n\~ao \'e correto} ler esse diagrama como se um dos el\'etrons (digamos, o el\'etron da parte superior da figura) tivesse emitido o f\'oton, e o outro (da parte inferior) o tivesse absorvido. Pois o processo oposto, em que o el\'etron inferior emitiu e o superior absorveu o f\'oton, \'e igualmente poss\'ivel. E, se h\'a dois processos intermedi\'arios poss\'iveis, a mec\^anica qu\^antica garante que a contribui\c{c}\~ao total aos processos f\'isicos \'e a soma das contribui\c{c}\~oes dos processos intermedi\'arios. A figura~\ref{fig:feyn_sum} mostra que o diagrama de Feynman da figura~\ref{fig:ee} representa a \emph{soma} (i.e. a superposi\c{c}\~ao qu\^antica) das duas ordena\c{c}\~oes temporais poss\'iveis para a emiss\~ao e absor\c{c}\~ao do f\'oton intermedi\'ario~\cite{FeynmanQED, Thomson, HalzenMartin}. Ou seja, a figura~\ref{fig:ee} ilustra apenas que, em primeira ordem, esse processo ocorre mediante a \emph{troca} de um f\'oton virtual, sem representar como esse processo de fato ocorre no espa\c{c}o-tempo.

Para compreender melhor a diferen\c{c}a entre f\'otons reais e virtuais, consideremos o balan\c{c}o de energia-momento em um ponto de ``emiss\~ao/absor\c{c}\~ao'' do f\'oton no diagrama, i.e. em um de seus v\'ertices. Sejam $(E, \vec{p})$ e $(E^{\,\prime}, \vec{p}^{\ \prime})$ a energia e momento linear de um el\'etron da figura~\ref{fig:ee} (por ex., o el\'etron da parte superior do diagrama) antes e depois da intera\c{c}\~ao (i.e. \`a esquerda e \`a direita do v\'ertice, respectivamente), e sejam $(E_\gamma, \vec{p}_\gamma)$ a energia e momento do f\'oton virtual. A conserva\c{c}\~ao de energia e momento na absor\c{c}\~ao/emiss\~ao do f\'oton imp\~oe que 
\begin{eqnarray}
	E &=& E_\gamma + E^{\,\prime},\\
	\vec{p} &=& \vec{p}_\gamma + \vec{p}^{\ \prime},
\end{eqnarray} 
logo
\begin{equation}
	E^2_\gamma- |\vec{p}_\gamma|^2c^2 = (E - E^{\,\prime})^2 - |\vec{p} - \vec{p}^{\ \prime}|^2c^2.
	\label{eq:foton_virtual}
\end{equation}
Como um f\'oton tem massa $m_\gamma=0$, esperar-se-ia, da rela\c{c}\~ao relativ\'istica de energia-momento\footnote{N\~ao \'e necess\'ario discutir extensamente sobre a teoria da relatividade especial para introduzir essa rela\c{c}\~ao aos estudantes. Para os prop\'ositos desta discuss\~ao, \'e suficiente dizer que a rela\c{c}\~ao Newtoniana de energia-momento, $E=p^2/2m=mv^2/2$, \'e modificada para $E^2 = p^2 c^2 + m^2 c^4$, de modo a incluir a energia de repouso. De fato, quando n\~ao h\'a energia cin\'etica, $p=0$, a rela\c{c}\~ao relativ\'istica reduz-se \`a conhecida f\'ormula de Einstein $E=mc^2$.}, que
\begin{equation}
	E_\gamma^2 - |\vec{p}_\gamma|^2c^2 = m_\gamma^2 c^4 = 0\quad \boldsymbol{?}
	\label{eq:onshell}
\end{equation}
No entanto, em um espalhamento de el\'etrons n\~ao h\'a nada que vincule $E$ a $E^{\,\prime}$ ou $\vec{p}$ a $\vec{p}^{\ \prime}$ a priori\footnote{A rela\c{c}\~ao relativ\'istica de energia-momento relaciona $E$ a $\vec{p}$ e $E^\prime$ a $\vec{p}^{\ \prime}$ atrav\'es de 
	\[ E^2 - |\vec{p}|^2c^2 = m_e^2 c^4 = E^{\,\prime\,2} - |\vec{p}^{\ \prime}|^2c^2, \]
que podemos usar para reescrever o lado direito da equa\c{c}\~ao~(\ref{eq:foton_virtual}) como \[ E_\gamma^2 - |\vec{p_\gamma}|^2 c^2 = 2m_e^2c^4 - 2(E E^\prime - \vec{p}\cdot\vec{p}^{\ \prime}),\]
que em geral n\~ao \'e igual a zero.} (tratam-se de vari\'aveis livres do problema), de maneira que, em geral, o lado direito da equa\c{c}\~ao~(\ref{eq:foton_virtual}) \emph{n\~ao} se anula. Isso significa que \emph{um f\'oton virtual em geral n\~ao satisfaz a rela\c{c}\~ao relativ\'istica de energia-momento}. Ou, dito de outra forma, \'e como se a part\'icula virtual pudesse ter massa diferente da massa observada para excita\c{c}\~oes normais do campo em quest\~ao. Essa \'e outra importante distin\c{c}\~ao entre part\'iculas reais e virtuais. Essa viola\c{c}\~ao da equa\c{c}\~ao~(\ref{eq:onshell}) n\~ao \'e um problema, contanto que esteja dentro da margem de incerteza associada \`a energia e momento do f\'oton virtual. Pelo princ\'ipio da incerteza, a precis\~ao com que se pode determinar a energia de um sistema ($\Delta E$) \'e limitada pelo tempo dispon\'ivel para ele interagir com um poss\'ivel aparato medidor ($\Delta t$), assim como a incerteza em seu momento ($\Delta p$) cresce \`a medida que se reduz a regi\~ao do espa\c{c}o a que ele esteja restrito ($\Delta x$). Isso pode ser expresso matematicamente por meio das famosas rela\c{c}\~oes de incerteza
\begin{equation}
    \Delta E\,\Delta t\geq \frac{\hbar}{2},\quad
    \Delta p\,\Delta x\geq \frac{\hbar}{2},
\end{equation}
com $\hbar \approx 1.05\times 10^{-34}~\text{kg\,m}^2/\text{s}$ a constante de Planck reduzida\footnote{Essa constante \'e caracter\'istica da teoria qu\^antica. Sempre que ela aparece, \'e ind\'icio de que h\'a algum fen\^omeno qu\^antico subjacente.}.
Assim, quanto menor for o tempo de vida de uma excita\c{c}\~ao de um campo, e/ou quanto menor for seu alcance espacial, maior ser\'a a incerteza associada \`a sua energia-momento, e portanto maior ser\'a a margem para uma aparente viola\c{c}\~ao da equa\c{c}\~ao~(\ref{eq:onshell}).

Mais especificamente, o espalhamento descrito pelo diagrama da figura~\ref{fig:ee} \'e mediado por f\'otons de \emph{todas} as poss\'iveis virtualidades, e a intera\c{c}\~ao final resulta da soma de todas as contribui\c{c}\~oes. Entretanto, para grandes dist\^ancias entre as part\'iculas interagentes, a sobreposi\c{c}\~ao dos f\'otons virtuais resulta em uma interfer\^encia destrutiva (exatamente no sentido de interfer\^encia de ondas), de modo que quanto maior a virtualidade dos f\'otons mediadores, menores ser\~ao seus impactos no resultado final da intera\c{c}\~ao. Assim, a grandes dist\^ancias, apenas f\'otons de baixa virtualidade contribuem significativamente. Reciprocamente, a menores dist\^ancias, a contribui\c{c}\~ao dos f\'otons de maior virtualidade tamb\'em se torna relevante. Uma discuss\~ao muito detalhada desse efeito, ao mesmo tempo em que \'e acess\'ivel ao p\'ublico geral, pode ser encontrada na ref.~\cite{FeynmanQED}, cuja leitura \'e fortemente recomendada para uma maior compreens\~ao da QED e um complemento da discuss\~ao aqui apresentada.

\begin{figure}[h!]
	\centering
	\includegraphics[page=5]{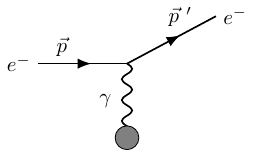}
	\caption{Diagrama de Feynman para um espalhamento el\'astico por uma carga est\'atica. Define-se o \^angulo $\theta$ como o \^angulo entre o momento final ($\vec{p}^{\ \prime}$) e inicial do el\'etron ($\vec{p}$), medindo o desvio relativamente \`a dire\c{c}\~ao do feixe incidente. Ele n\~ao aparece na figura, pois um diagrama de Feynman representa apenas uma esquematiza\c{c}\~ao do processo que contribui para a intera\c{c}\~ao, e n\~ao ilustra como o espalhamento ocorre, de fato, no espa\c{c}o-tempo. Veja coment\'ario no in\'icio da se\c{c}\~ao~\ref{sec:feynman}, bem como a figura~\ref{fig:feyn_sum}.}
	\label{fig:Coulomb_scattering}
\end{figure}

O argumento exposto acima nos fornece uma maneira de compreender o comportamento da Lei de Coulomb sob uma perspectiva microsc\'opica da intera\c{c}\~ao eletromagn\'etica. Consideremos o caso de um el\'etron espalhado elasticamente por uma carga muito mais pesada, e que portanto permanece aproximadamente est\'atica, como esquematizado na figura~\ref{fig:Coulomb_scattering}. Um exemplo \'e um experimento de difra\c{c}\~ao de el\'etrons, ao estilo dos experimentos de Rutherford-Geiger-Marsden que levaram \`a descoberta do n\'ucleo at\^omico\footnote{O experimento de Rutherford-Geiger-Marsden foi realizado com part\'iculas $\alpha$, ao inv\'es de el\'etrons, mas a argumenta\c{c}\~ao independe da natureza da carga teste, contanto que o espalhamento seja el\'astico.}. Nesse caso, a energia cin\'etica do el\'etron n\~ao varia (por se tratar de espalhamento el\'astico) e a equa\c{c}\~ao~(\ref{eq:foton_virtual}) fica
\begin{equation}\begin{split}
	\text{virtualidade}&\equiv E^2_\gamma- |\vec{p}_\gamma|^2c^2 \\
	&= - |\vec{p}-\vec{p}^{\ \prime}|^2c^2 = -4|\vec{p}|^2c^2\sin^2\frac{\theta}{2},
	\label{eq:elastic}
\end{split}\end{equation}
que define o ``grau de virtualidade'' do f\'oton atrav\'es da medida de viola\c{c}\~ao da rela\c{c}\~ao relativ\'istica de energia-momento. Nessa equa\c{c}\~ao, $\theta$ \'e o \^angulo entre o feixe espalhado e a dire\c{c}\~ao do feixe incidente (vide legenda da figura~\ref{fig:Coulomb_scattering}), que varia entre $\theta=0$ (quando n\~ao h\'a desvio) e $\theta=\pi$ (quando o feixe volta na mesma dire\c{c}\~ao e sentido oposto ao incidente).

Nota-se que, quanto maior o \^angulo $\theta$ de espalhamento relativo ao feixe incidente (i.e. quanto maior for a altera\c{c}\~ao do estado de movimento do el\'etron devido \`a intera\c{c}\~ao), maior \'e a virtualidade, e portanto, pelo argumento anterior, menor deve ser o alcance dessa intera\c{c}\~ao. Ou seja, o el\'etron s\'o \'e desviado significativamente de sua trajet\'oria incidente se ele se aproximar muito da outra carga. Por outro lado, um espalhamento em que o momento linear do el\'etron pouco varia, $\vec{p}\approx \vec{p}^{\ \prime}$ e $\theta\approx 0$, est\'a associado a f\'otons que satisfazem a rela\c{c}\~ao energia-momento em boa aproxima\c{c}\~ao, e portanto podem mediar intera\c{c}\~oes a longas dist\^ancias. Posto de outra forma: quanto mais o el\'etron se aproxima de outra carga, maior poder\'a ser a virtualidade dos f\'otons trocados entre elas, e maior ser\'a o desvio da trajet\'oria do el\'etron; ao passo que, se as cargas est\~ao distantes, os f\'otons devem satisfazer a equa\c{c}\~ao~(\ref{eq:onshell}) em boa aproxima\c{c}\~ao, o que corresponde, pela equa\c{c}\~ao~(\ref{eq:elastic}), a um desvio pequeno de trajet\'oria. Esse \'e justamente o comportamento cl\'assico da intera\c{c}\~ao eletrost\'atica: quanto mais distantes as cargas, menor \'e a for\c{c}a e, consequentemente, menor a varia\c{c}\~ao do momento linear da part\'icula teste. O argumento acima apresenta uma explica\c{c}\~ao para esse fen\^omeno em n\'ivel microsc\'opico. \'E poss\'ivel mostrar que a rela\c{c}\~ao de virtualidade dada pela equa\c{c}\~ao~(\ref{eq:elastic}) d\'a origem a uma energia potencial de intera\c{c}\~ao do tipo $V(r)\sim 1/r$, que \'e justamente a energia potencial associada \`a lei de Coulomb. Note, ainda, que todos os f\'otons que contribuem para a Lei de Coulomb s\~ao virtuais.

Uma outra maneira de refrasear a argumenta\c{c}\~ao acima \'e a seguinte. Um f\'oton real, n\~ao massivo, cont\'em apenas duas poss\'iveis polariza\c{c}\~oes, que s\~ao transversais \`a dire\c{c}\~ao de propaga\c{c}\~ao\footnote{Ou seja, o fato de a luz ser uma onda transversal adv\'em do fato de o f\'oton n\~ao ter massa.}. Por outro lado, a viola\c{c}\~ao da equa\c{c}\~ao~(\ref{eq:onshell}) implica que um f\'oton virtual possui uma terceira polariza\c{c}\~ao, dita longitudinal, que n\~ao est\'a associada a um grau de liberdade que se propaga. De fato, \'e poss\'ivel mostrar que esse grau de liberdade longitudinal atua como mensageiro da Lei de Coulomb~\cite{HalzenMartin}. Segue-se que quanto maior for a virtualidade do f\'oton, maior \'e a contribui\c{c}\~ao da polariza\c{c}\~ao longitudinal, e mais intensa a intera\c{c}\~ao Coulombiana, como conclu\'imos acima.

\subsection{P\'ositrons: os antiel\'etrons}
\label{sec:positrons}

Na se\c{c}\~ao~\ref{sec:qft} acima vimos que, em uma teoria qu\^antica relativ\'istica, o el\'etron \'e interpretado como uma excita\c{c}\~ao de um campo fundamental, que chamamos de ``campo do el\'etron'' ou ``campo eletr\^onico''. Contudo, \'e poss\'ivel mostrar que os postulados da relatividade restrita s\'o s\~ao satisfeitos quando se considera que o campo eletr\^onico possui tamb\'em um outro tipo de excita\c{c}\~ao, com a \textbf{mesma massa} do el\'etron mas carga oposta. Em outras palavras, uma teoria qu\^antica relativ\'istica do el\'etron prev\^e, tamb\'em, a exist\^encia de um \textbf{antiel\'etron}! Como a carga dessa part\'icula \'e positiva, ela \'e chamada de \textbf{p\'ositron}.

Pode-se entender a afirma\c{c}\~ao acima da seguinte maneira. Considere dois eventos que ocorrem nos pontos $x=(t,\vec{x})$ e $x+\Delta x=(t+\Delta t,\vec{x}+\Delta\vec{x})$ do espa\c{c}o-tempo, e suponhamos que $|\Delta\vec{x}|>c\Delta t$, ou seja, a dist\^ancia espacial $\Delta\vec{x}$ \'e maior do que a dist\^ancia que a luz pode percorrer no intervalo $\Delta t$. Isso significa que esses dois eventos s\~ao causalmente desconexos, pois s\'o poderiam ser relacionados por um sinal superluminal, o que violaria os postulados da relatividade restrita. 
No entanto, quando calculamos a probabilidade de que um el\'etron se propague de $x$ at\'e $x+\Delta x$, constata-se que essa probabilidade \emph{n\~ao se anula}! Assim, se o campo eletr\^onico propagasse informa\c{c}\~ao somente via el\'etrons, haveria viola\c{c}\~ao da causalidade: ver\'iamos efeitos precedendo suas causas, o que \'e absurdo! Para preservar a causalidade,  o campo eletr\^onico deve possuir, tamb\'em, um segundo tipo de excita\c{c}\~ao, que s\~ao os chamados ``p\'ositrons'': as antipart\'iculas dos el\'etrons.
A transmiss\~ao de informa\c{c}\~ao entre dois pontos espa\c{c}o-temporais envolve a propaga\c{c}\~ao de um el\'etron de $x$ a $x+\Delta x$ bem como a propaga\c{c}\~ao de um p\'ositron de $x+\Delta x$ a $x$ e, quando essas duas part\'iculas t\^em exatamente a mesma massa, esses dois processos interferem destrutivamente\footnote{Vale enfatizar que essa superposi\c{c}\~ao entre propaga\c{c}\~ao de el\'etrons e p\'ositrons s\'o se anula quando $x$ e $x+\Delta x$ est\~ao fora do cone de luz um do outro. Quando $\Delta \vec{x}\leq c\Delta t$, a interfer\^encia entre el\'etron e p\'ositron n\~ao \'e totalmente destrutiva, de modo a (felizmente) ser poss\'ivel a transmiss\~ao de informa\c{c}\~ao nesse caso.}, garantindo a causalidade em n\'ivel qu\^antico. Note que, nesse caso, a propaga\c{c}\~ao de um el\'etron \'e exatamente cancelada pela de um p\'ositron se propagando no sentido oposto no espa\c{c}o e no tempo. Esse \'e um dos aspectos da interpreta\c{c}\~ao de Feynman-St\"uckelberg para as antipart\'iculas, que garante que, sob todos os aspectos, \emph{as antipart\'iculas se comportam como part\'iculas viajando para tr\'as no tempo}. Isso explica porque part\'iculas e antipart\'iculas possuem a mesma massa: s\~ao, no fundo, a mesma part\'icula, ou, dito de outra forma, s\~ao excita\c{c}\~oes de um mesmo campo fundamental.

\begin{figure*}[h!]
    \centering
    \includegraphics[page=11]{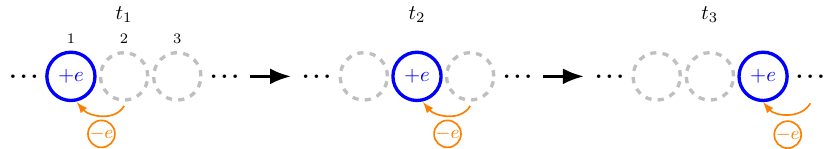}
    \caption{Representa\c{c}\~ao esquem\'atica do movimento de um p\'ositron, interpretado como o movimento de um el\'etron movendo-se ``para tr\'as no tempo''. O p\'ositron pode ser visto como uma ``aus\^encia de el\'etron'' ou como um buraco nesta p\'agina em branco: \`a medida que o ``buraco'' se move para a direita, o espa\c{c}o que ele antes ocupava passa a ser novamente branco, como o restante da p\'agina. \emph{Reduzir a aus\^encia} de carga negativa em uma determinada posi\c{c}\~ao \'e equivalente a \emph{acrescentar uma carga negativa nesse ponto}. Isso porque $-(+e)=-e$. Assim, o p\'ositron movendo-se de $1$ para $2$ em $t_1\to t_2$ corresponde a um el\'etron movendo-se de $2$ para $1$. Em $t_2\to t_3$ o el\'etron se desloca de $3$ para $2$. Note que o el\'etron que saiu de $2$ para $1$ precisou, primeiro, sair de $3$ para $2$.  Assim, efetivamente, o movimento do p\'ositron de $1\to 2\to 3$ em $t_1\to t_2\to t_3$ corresponde ao movimento de um el\'etron de $3\to 2 \to 1$ em $t_3\to t_2\to t_1$.}
    \label{fig:anti}
\end{figure*}

Essa interpreta\c{c}\~ao de antipart\'iculas pode ser visualizada da seguinte maneira. Imagine um p\'ositron se propagando da esquerda para a direita, como ilustrado na figura~\ref{fig:anti}. Vamos denominar as tr\^es posi\c{c}\~oes que o p\'ositron ocupa na figura como posi\c{c}\~oes $1$, $2$ e $3$, respectivamente. No instante $t_1$ h\'a uma carga el\'etrica positiva ocupando a posi\c{c}\~ao $1$, mais \`a esquerda da figura, enquanto as posi\c{c}\~oes vizinhas correspondem a espa\c{c}os vazios, portanto com carga nula. Na passagem do instante $t_1$ para um instante $t_2$ posterior, a carga $+e$ \'e subtra\'ida da posi\c{c}\~ao $1$ e acrescida na posi\c{c}\~ao $2$. Mas esse processo \'e totalmente an\'alogo a subtrairmos uma carga $-e$ de $2$ (pois $0-(-e)=+e$) e adicionarmos $-e$ em $1$ (pois $-e+e=0$). Dito de outra forma, \'e como se a exist\^encia de um p\'ositron pudesse ser vista como a ``aus\^encia de um el\'etron'', ou como um ``buraco'': \`a medida que o p\'ositron se move da esquerda para a direita, os espa\c{c}os mais \`a direita v\~ao se movendo para a esquerda, para preencher a lacuna do p\'ositron. A figura~\ref{fig:anti} mostra, assim, como o movimento do p\'ositron da esquerda para a direita ao longo de $t_1 \to t_2 \to t_3$ corresponde ao movimento de um el\'etron da direita para a esquerda ao longo de $t_3\to t_2\to t_1$, portanto andando ``para tr\'as no tempo''.

Esse mesmo argumento se aplica a qualquer campo qu\^antico, de modo que \emph{toda} part\'icula elementar possui uma antipart\'icula associada, e ambas part\'icula e antipart\'icula possuem a mesma massa, mas cargas opostas. 

Refraseando em poucas palavras: antipart\'iculas emergem inevitavelmente dos postulados da mec\^anica qu\^antica e da relatividade restrita. Ou, ainda de outra forma: \'e imposs\'ivel formular uma teoria qu\^antica relativ\'istica de apenas \emph{uma} part\'icula. Isso n\~ao \'e inesperado, visto que a rela\c{c}\~ao relativ\'istica $E=mc^2$ j\'a aponta para a possibilidade de se converter energia em massa e, portanto, criar novas part\'iculas a partir de uma energia inicial ou, reciprocamente, converter massa em energia atrav\'es da aniquila\c{c}\~ao de part\'iculas.

\begin{figure}[h!]
    \centering
    \includegraphics[page=6]{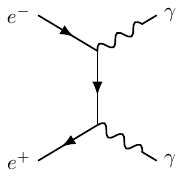}
    \caption{Diagrama representando a aniquila\c{c}\~ao de um par el\'etron-p\'ositron em dois f\'otons. Note como a part\'icula mediadora \'e uma excita\c{c}\~ao do campo eletr\^onico. Esse caso ilustra como todas as part\'iculas podem ser virtuais, e n\~ao apenas as mediadoras das intera\c{c}\~oes.}
    \label{fig:ep_ann}
\end{figure}

Na QED, em que el\'etrons e p\'ositrons interagem com o campo eletromagn\'etico, essa energia que cria part\'iculas ou resulta de suas aniquila\c{c}\~oes est\'a na forma de f\'otons. Ou seja, um par el\'etron-p\'ositron pode se aniquilar mutuamente, resultando em (dois ou mais) f\'otons no produto final, ou, reciprocamente, f\'otons podem produzir um par el\'etron-p\'ositron\footnote{\'E f\'acil mostrar, por conserva\c{c}\~ao de energia-momento, que um par el\'etron-p\'ositron n\~ao pode resultar de um \'unico f\'oton n\~ao-massivo, devendo portanto existir pelo menos dois f\'otons no estado inicial de um processo de cria\c{c}\~ao de pares. Para um f\'oton virtual, o processo $\gamma \leftrightarrow e^+ + e^-$ \'e permitido.}. Esse processo est\'a representado diagramaticamente pela figura~\ref{fig:ep_ann}. 

Historicamente, a exist\^encia dos p\'ositrons foi prevista por Dirac em 1928, mesmo antes de essas part\'iculas terem sido observadas. Ap\'os ter formulado sua equa\c{c}\~ao relativ\'istica para o el\'etron, Dirac notou que as solu\c{c}\~oes inevitavelmente continham, tamb\'em, el\'etrons de energia negativa. O que parecia um problema foi transformado em um triunfo da teoria quando Dirac interpretou essas solu\c{c}\~oes como fenomenologicamente equivalentes a antiel\'etrons, ou ``el\'etrons'' de carga positiva\footnote{Inicialmente Dirac relutou em propor a exist\^encia de uma nova part\'icula, e sup\^os que os ``antiel\'etrons'' que apareciam como solu\c{c}\~oes de sua equa\c{c}\~ao eram os pr\'otons. A exorbitante diferen\c{c}a entre as massas do el\'etron e do pr\'oton logo desmentiu essa hip\'otese, e, para a sorte de Dirac, Anderson detectou o p\'ositron poucos anos depois, corroborando o sucesso da descri\c{c}\~ao de Dirac do el\'etron.}. A corrobora\c{c}\~ao experimental dessa predi\c{c}\~ao foi feita por Anderson em 1932, ao observar part\'iculas que, sob um campo magn\'etico, curvavam-se como se tivessem a mesma massa do el\'etron, por\'em carga oposta. A hist\'oria do p\'ositron \'e, portanto, um dos muitos exemplos de part\'iculas preditas teoricamente antes de serem detectadas experimentalmente, o que ilustra o car\'ater altamente preditivo dessas teorias.

\begin{figure}[h!]
	\centering
	\includegraphics[width=.45\textwidth]{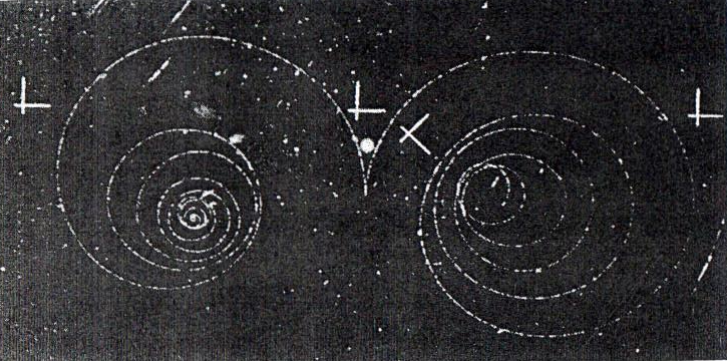}
	\caption{Cria\c{c}\~ao de um par el\'etron-p\'ositron visualizado em uma c\^amara de bolhas. Fonte: CERN.}
	\label{fig:pair}
\end{figure}

A figura~\ref{fig:pair} ilustra a cria\c{c}\~ao de um par el\'etron-p\'ositron observada em uma c\^amara de bolhas. O experimento \'e desenhado para detectar o rastro de part\'iculas carregadas. Por isso, o rastro do f\'oton n\~ao aparece, embora possa ser deduzido pelo v\'ertice de cria\c{c}\~ao do par. As part\'iculas est\~ao submetidas a um campo magn\'etico perpendicular ao plano da figura, e, por possu\'irem mesma massa e cargas opostas, s\~ao curvadas em dire\c{c}\~oes opostas, mas com trajet\'orias de aproximadamente mesmo raio.

\begin{figure}[h!]
	\centering
	\includegraphics[width=.45\textwidth]{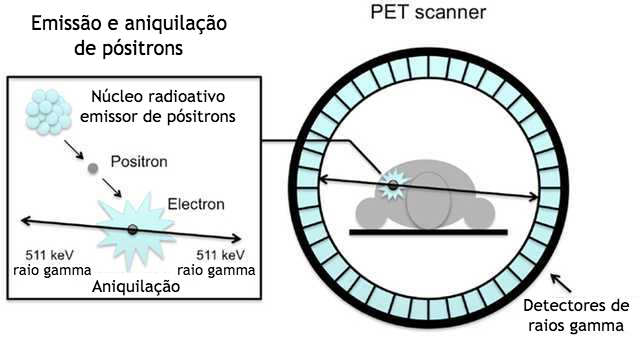}
	\caption{Esquema de funcionamento de um aparelho de Tomografia por Emiss\~ao de P\'ositrons (\emph{PET scan}). Fonte: adaptado de~\cite{physicsforums}.}
	\label{fig:PET}
\end{figure}

Esse fen\^omeno de aniquila\c{c}\~ao de pares resultando em radia\c{c}\~ao eletromagn\'etica tem aplica\c{c}\~ao pr\'atica na medicina, nos chamados \emph{PET scans} (PET \'e a sigla em ingl\^es para \textit{Tomografia por Emiss\~ao de P\'ositrons}). Trata-se de uma modalidade de obten\c{c}\~ao de imagens e dados sobre o funcionamento metab\'olico do organismo, que funciona da seguinte maneira~\cite{PET}. Primeiramente, modifica-se uma mol\'ecula usualmente metabolizada pelo organismo (por exemplo, glucose) substituindo-se um de seus \'atomos por um elemento radioativo que decai por emiss\~ao de p\'ositrons (usualmente $^{18}$F). Essa subst\^ancia \'e inserida intravenosamente no paciente, e o organismo, confundindo-a com glucose, redireciona-a a \'org\~ao vitais, onde ser\'a metabolizada. Ali, o elemento radioativo emitir\'a um p\'ositron, que se aniquilar\'a com um el\'etron do corpo do paciente, emitindo dois f\'otons de radia\c{c}\~ao $\gamma$ que ser\~ao observados pelo aparelho detector, como na figura~\ref{fig:PET}. Por conserva\c{c}\~ao de momento, os f\'otons sempre s\~ao emitidos em dire\c{c}\~oes opostas. O aparelho calcula o intervalo de tempo entre a detec\c{c}\~ao de cada f\'oton, e determina, assim, o local de onde o p\'ositron foi emitido. Faz-se, assim, um mapeamento da atividade metab\'olica do organismo, podendo-se determinar a integridade do tecido cerebral, a exist\^encia de tumores, o funcionamento cardiovascular, dentre outras aplica\c{c}\~oes.

\subsection{A carga efetiva do el\'etron}
\label{sec:effcharge}

Outra consequ\^encia da convers\~ao de f\'otons em pares el\'etron-p\'ositron (e vice-versa) \'e a chamada ``polariza\c{c}\~ao do v\'acuo'', que tem uma importante consequ\^encia fenomenol\'ogica: a constata\c{c}\~ao de que a carga el\'etrica do el\'etron \emph{n\~ao \'e constante}!

\begin{figure}[h!]
	\centering
	\includegraphics[scale=1.3, page=7]{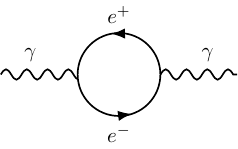}
	\caption{Diagrama de Feynman ilustrando um poss\'ivel processo ocorrendo durante a propaga\c{c}\~ao de um f\'oton, temporariamente interrompida pela cria\c{c}\~ao e posterior aniquila\c{c}\~ao de um par el\'etron-p\'ositron. Este \'e um exemplo de um diagrama com um la\c{c}o.
	}
	\label{fig:props}
\end{figure}

Para entender melhor essa afirma\c{c}\~ao, considere o diagrama ilustrado na figura~\ref{fig:props}, em que a propaga\c{c}\~ao de um f\'oton \'e temporariamente interrompida pela cria\c{c}\~ao de um par el\'etron-p\'ositron, que posteriormente se aniquila e d\'a origem novamente a um f\'oton que continua a se propagar. Esse processo ilustra que, devido \`a intera\c{c}\~ao com o campo eletr\^onico, um f\'oton n\~ao pode ser visto apenas como uma excita\c{c}\~ao do campo eletromagn\'etico. Sua propaga\c{c}\~ao \'e afetada por sua intera\c{c}\~ao com el\'etrons e p\'ositrons. E isso, por sua vez, tem impacto sobre a maneira como esse f\'oton intermedeia a intera\c{c}\~ao eletromagn\'etica entre cargas el\'etricas.

\begin{figure}[h!]
    \centering
    \begin{subfigure}[t]{.23\textwidth}
        \includegraphics[scale=1.15, page=8]{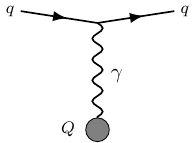}
        \caption{}
    \end{subfigure}
    \begin{subfigure}[t]{.23\textwidth}
       \includegraphics[scale=1.15, page=9]{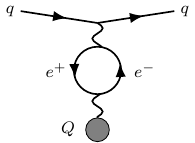}
        \caption{}
    \end{subfigure}
    \caption{Dois processos poss\'iveis de media\c{c}\~ao da intera\c{c}\~ao entre as cargas $q$ e $Q$. (a) Diagrama respons\'avel pela intera\c{c}\~ao descrita no eletromagnetismo cl\'assico. (b) Uma corre\c{c}\~ao qu\^antica oriunda da intera\c{c}\~ao m\'utua entre o f\'oton e as excita\c{c}\~oes do campo eletr\^onico, que altera a propaga\c{c}\~ao do f\'oton. Por incluir uma modifica\c{c}\~ao na propaga\c{c}\~ao do f\'oton, esse diagrama induz uma modifica\c{c}\~ao da lei de Coulomb para a intera\c{c}\~ao entre part\'iculas carregadas. Isso tamb\'em pode ser interpretado como uma altera\c{c}\~ao na ``carga el\'etrica efetiva'' das part\'iculas interagentes.}
    \label{fig:charge_renorm}
\end{figure}

De fato, considere a intera\c{c}\~ao entre uma carga $q$ incidindo contra uma carga $Q$, que por simplicidade suporemos ser muito mais pesada, de modo que ela permanece essencialmente im\'ovel ao longo do processo. O eletromagnetismo cl\'assico diz que essa intera\c{c}\~ao \'e regida pela lei de Coulomb, ou seja, a energia potencial de intera\c{c}\~ao \'e
\begin{equation}
    V_\text{Coulomb}(r) = \frac{kQq}{r},
\end{equation}
com $k\approx 9\times 10^9~\text{N\,m}^2/\text{C}^2$ a constante de Coulomb. Esse comportamento \'e exatamente o que se obt\'em na eletrodin\^amica qu\^antica a partir do diagrama da figura~\ref{fig:charge_renorm} (a), como j\'a discutido na se\c{c}\~ao~\ref{sec:feynman}. Ou seja, a teoria eletromagn\'etica cl\'assica est\'a contida na QED como primeira aproxima\c{c}\~ao, ao ignorarmos todos diagramas que contenham la\c{c}os.

Mas h\'a v\'arios outros processos tamb\'em presentes na QED, como o ilustrado na figura~\ref{fig:charge_renorm} (b), que  contribuem para a media\c{c}\~ao da intera\c{c}\~ao entre as cargas, e que portanto fornecem ``corre\c{c}\~oes qu\^anticas'' \`a predi\c{c}\~ao da teoria cl\'assica. Esses processos est\~ao \emph{sempre} presentes, independentemente de a intera\c{c}\~ao ser macrosc\'opica ou n\~ao. Quer dizer, \'e imposs\'ivel ``desligarmos'' esses processos tipicamente qu\^anticos e observarmos apenas os processos cl\'assicos. Sendo assim, por que esses efeitos qu\^anticos n\~ao s\~ao comumente observados em nosso dia a dia? Por que a lei de Coulomb \'e uma boa aproxima\c{c}\~ao para intera\c{c}\~oes macrosc\'opicas? E, ainda mais importante: o qu\~ao pequenas t\^em que ser as dist\^ancias envolvidas para que esses efeitos qu\^anticos se manifestem mais perceptivelmente? 

Podemos responder essa quest\~ao recorrendo novamente ao conceito de virtualidade das part\'iculas intermedi\'arias, elaborado na se\c{c}\~ao~\ref{sec:feynman}. Naquela ocasi\~ao, definimos o ``grau de virtualidade'' de uma part\'icula de energia $E$, momento $\vec{p}$ e massa $m$ como uma medida do quanto ela viola a rela\c{c}\~ao relativ\'istica de massa-energia-momento, i.e.
\begin{equation}
\label{eq:virtualidade}
    \text{virtualidade}\equiv E^2 - |\vec{p}|^2 c^2 - m^2 c^4.
\end{equation}
Assim, tudo se passa como se a part\'icula virtual tivesse uma massa diferente da correspondente part\'icula real. Por exemplo, \'e como se um el\'etron virtual tivesse massa diferente de $m_e\approx 9.1\times 10^{-31}$~kg. Uma part\'icula real \'e aquela para a qual o lado direito da equa\c{c}\~ao~(\ref{eq:virtualidade}) se anula. 

Intera\c{c}\~oes de longa dist\^ancia s\~ao mediadas predominantemente por f\'otons de baixa virtualidade, o que por sua vez implica em uma alta virtualidade para o par el\'etron-p\'ositron da figura~\ref{fig:charge_renorm} (b)\footnote{Um exerc\'icio interessante consiste em aplicar as leis de conserva\c{c}\~ao de energia e momento aos v\'ertices do diagrama da figura~\ref{fig:props} para mostrar que o par el\'etron-p\'ositron s\'o pode ser real se
\[ E_\gamma^2 - |\vec{p}_\gamma|^2 c^2 \geq (2m_e c^2)^2.\]}. 
Isso faz com que os efeitos desse diagrama sejam atenuados em compara\c{c}\~ao com o da figura~\ref{fig:charge_renorm} (a). Assim, para grandes dist\^ancias, a intera\c{c}\~ao \'e regida essencialmente pela lei de Coulomb, como esperado. A que dist\^ancias a contribui\c{c}\~ao do diagrama com la\c{c}o se torna relevante? Ora, s\'o existe uma escala de energia caracter\'istica desse processo, que \'e a energia de repouso do par el\'etron-p\'ositron, $2m_e c^2$. Essa energia est\'a naturalmente associada, na teoria qu\^antica, a uma escala de dist\^ancia da ordem\footnote{Veja o ap\^endice~\ref{sec:unidades} para uma atividade elucidando a rela\c{c}\~ao entre energia e dist\^ancias que emerge naturalmente no paradigma qu\^antico.}
\begin{equation}
    \small
    \widetilde{r}= \frac{\hbar c}{2m_e c^2}
                =\frac{\left(\parbox{13mm}{\scriptsize\centering constante\\ de Planck}\right)\times \left(\parbox{13mm}{\scriptsize\centering velocidade\\ da luz}\right)}{\left(\parbox{24mm}{\centering\scriptsize energia de repouso\\ do par $e^+ e^-$}\right)}
                \sim 2\times 10^{-13}~\text{m}.
\end{equation}
O fato de essa escala de dist\^ancia, caracter\'istica do processo da figura~\ref{fig:charge_renorm} (b), envolver a constante de Planck e a velocidade da luz mostra que se trata de um efeito caracteristicamente qu\^antico e relativ\'istico. Como j\'a poder\'iamos antecipar, trata-se de uma escala de dist\^ancia associada ao comprimento de onda de um f\'oton com energia $2m_e c^2$. Nota-se que $\widetilde{r}$ \'e cerca de duas ordens de magnitude menor do que o t\'ipico raio at\^omico, portanto as part\'iculas interagentes devem estar de fato muito pr\'oximas para que a modifica\c{c}\~ao na propaga\c{c}\~ao do f\'oton pela cria\c{c}\~ao de pares intermedi\'arios seja percept\'ivel. 

Essas previs\~oes s\~ao confirmadas quando computamos as corre\c{c}\~oes qu\^anticas ao potencial de Coulomb para $r\gtrsim \widetilde{r}$ devido ao diagrama da figura~\ref{fig:charge_renorm} (b), resultando no chamado \emph{potencial de Uehling},
\begin{equation}
	V(r) = k\frac{Qq}{r}\left( 1 + \frac{\alpha}{\sqrt{2\pi}}\left(~\frac{\widetilde{r}}{r}~\right)^{3/2} e^{-r/\widetilde{r}} + \ldots\right).
	\label{eq:Uehling}
\end{equation}
Aqui, $\alpha\approx 1/137$ \'e uma constante que dita a intensidade da intera\c{c}\~ao eletromagn\'etica\footnote{\'E a  chamada ``constante de estrutura fina'', uma grandeza adimensional definida a partir da carga el\'etrica elementar $e$, da constante de Coulomb $k$, da velocidade da luz $c$ e da constante de Planck reduzida $\hbar$, e \'e dada por $\alpha\equiv k e^2/\hbar c$.}, e $\widetilde{r}$ \'e a dist\^ancia caracter\'istica ao processo qu\^antico da figura~\ref{fig:charge_renorm} (b) que d\'a origem a essa corre\c{c}\~ao.

\'E f\'acil ver que, para dist\^ancias muito maiores do que essa escala caracter\'istica, $r\gg \widetilde{r}$, o potencial de Uehling da equa\c{c}\~ao~\ref{eq:Uehling} reduz-se ao potencial de Coulomb, como esperado. As retic\^encias indicam que existem ainda outras corre\c{c}\~oes advindas de processos ainda mais complicados, e que s\'o ser\~ao not\'aveis a dist\^ancias ainda menores.

\begin{figure}[h!]
    \centering
	\includegraphics[scale=.5]{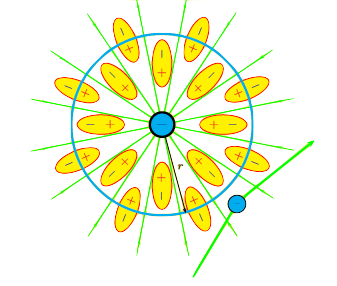}
	\caption{Efeito de blindagem da carga do el\'etron devido aos pares el\'etron-p\'ositron virtuais. Fonte: cortesia de~\copyright~INFN, Laboratori Nazionali di Frascati~\cite{IFNF_Frascati}.}
	\label{fig:screening}
\end{figure}

Essas corre\c{c}\~oes qu\^anticas devido \`a cria\c{c}\~ao de pares el\'etron-p\'ositron virtuais em torno de uma carga el\'etrica podem ser interpretadas fisicamente como oriundas de uma \emph{blindagem} da carga devido a um efeito de \emph{polariza\c{c}\~ao do v\'acuo}, como ilustrado na figura~\ref{fig:screening}. 
Suponhamos, por exemplo, que a part\'icula ao centro da figura seja um el\'etron. Devido a processos como os da figura~\ref{fig:charge_renorm} (b), em torno desse el\'etron existe uma multitude de pares el\'etron-p\'ositron virtuais que atuam como min\'usculos dipolos, efetivamente reduzindo a densidade de carga do el\'etron e, portanto, a intensidade de sua intera\c{c}\~ao eletromagn\'etica com outras cargas. Diz-se que o el\'etron est\'a sempre ``vestido'', de modo que, quando dois el\'etrons interagem, a densidade de carga efetiva da intera\c{c}\~ao \'e a da nuvem ilustrada na figura~\ref{fig:screening}, e n\~ao do el\'etron ``nu'' (ilustrado no centro da figura). Por\'em, \`a medida que os el\'etrons s\~ao lan\c{c}ados um contra o outro a mais altas energias, a intera\c{c}\~ao se d\'a a dist\^ancias cada vez menores, at\'e que eventualmente (quando a energia do centro de massa for $E_{\rm c.m.}\gtrsim 2m_e c^2$) um el\'etron come\c{c}a a penetrar a nuvem de polariza\c{c}\~ao do outro e enxerga uma maior carga efetiva, o que faz com que o comportamento da intera\c{c}\~ao seja diferente do Coulombiano. O ``raio efetivo'' dessa nuvem de polariza\c{c}\~ao em torno do el\'etron \'e a escala caracter\'istica $\widetilde{r}$ associada ao diagrama~\ref{fig:charge_renorm} (b), como j\'a discutido anteriormente.

Essa varia\c{c}\~ao da carga efetiva do el\'etron com a dist\^ancia \'e, de fato, observada. Em processos que envolvem intera\c{c}\~oes entre el\'etrons a altas energias, tudo se passa como se a carga do el\'etron fosse de fato maior do que a que observamos em nosso mundo de baixas energias. Uma pergunta plaus\'ivel, neste momento, seria: e qual \'e, ent\~ao, a carga do el\'etron ``nu''? A resposta \'e que a carga do el\'etron \emph{tende a infinito} \`a medida que a dist\^ancia entre as cargas interagentes diminui. \'E claro que uma carga infinita n\~ao faz sentido fisicamente, e esse comportamento j\'a nos indica que, a partir de uma certa escala de energia, essa descri\c{c}\~ao do el\'etron e suas intera\c{c}\~oes em termos de part\'iculas e campos j\'a n\~ao \'e mais aplic\'avel. Isso \'e an\'alogo ao caso em que a descri\c{c}\~ao cont\'inua de um s\'olido ou fluido deixa de ser v\'alida a partir de certas escalas de dist\^ancia, quando a estrutura at\^omico-molecular da mat\'eria se torna relevante, e  novos graus de liberdade precisam ser inclu\'idos na descri\c{c}\~ao do sistema. Quest\~oes sobre o que \'e e quais s\~ao as propriedades de um el\'etron ``nu'' n\~ao podem ainda ser resolvidas conclusivamente, e \'e mais prov\'avel que, a partir de uma determinada escala de energia, a pergunta sequer fa\c{c}a sentido, pois o pr\'oprio conceito de ``el\'etron'' como hoje o entendemos deve ser modificado\footnote{\'E esperado que, a dist\^ancias da ordem do comprimento de Planck, $\ell_P\sim 10^{-35}$~m, a intera\c{c}\~ao gravitacional se torne relevante em n\'ivel qu\^antico. A descri\c{c}\~ao desses efeitos tipicamente exige reformula\c{c}\~oes de nossos conceitos de part\'iculas e campos, como nas chamadas teorias de cordas, ou uma discretiza\c{c}\~ao do pr\'oprio espa\c{c}o-tempo, como no formalismo de \emph{spin foams} e gravita\c{c}\~ao qu\^antica de la\c{c}os.}.

Uma importante consequ\^encia desse efeito de polariza\c{c}\~ao do v\'acuo \'e o deslocamento de Lamb entre os n\'iveis de energia dos orbitais $2S_{1/2}$ e $2P_{1/2}$ do \'atomo de hidrog\^enio. Se a intera\c{c}\~ao el\'etron-n\'ucleo ocorresse apenas via diagrama da figura~\ref{fig:charge_renorm} (a), esses orbitais seriam degenerados (i.e. possuiriam a mesma energia). No entanto, experimentos cuidadosos mostraram que h\'a uma transi\c{c}\~ao entre eles, correspondendo \`a emiss\~ao de um f\'oton de frequ\^encia $\sim 1057$~MHz. Na QED, esse valor \'e predito corretamente com grande precis\~ao levando-se em conta que o el\'etron no orbital $S$ \'e capaz de penetrar o n\'ucleo at\^omico\footnote{Para um el\'etron no orbital $S$, a probabilidade de encontr\'a-lo na regi\~ao do n\'ucleo at\^omico \'e n\~ao-nula. Isso n\~ao ocorre para nenhum outro orbital eletr\^onico, e portanto apenas os orbitais $S$ s\~ao afetados.} e, com isso, sentir os efeitos do termo an\^omalo do potencial de Uehling. Isso n\~ao ocorre no orbital $P$, para o qual o potencial eletromagn\'etico \'e simplesmente o potencial de Coulomb. Portanto, o n\'ivel de energia do orbital $S$ fica ligeiramente deslocado do orbital $P$, dando origem a uma poss\'ivel transi\c{c}\~ao eletr\^onica que \'e, de fato, observada. A obten\c{c}\~ao do valor correto para o deslocamento Lamb por meio do formalismo da QED constitui ainda um dos maiores triunfos dessa teoria, e um forte indicativo de seu poder preditivo.

\section{An\'alise das interven\c{c}\~oes}
\label{sec:resultados}

\subsection{O contexto: o local e os sujeitos}
\label{subsec:contexto}

O N\'ucleo Cosmo-UFES, da Universidade Federal do Esp\'irito Santo (UFES), iniciou, no ano de 2017, um projeto de extens\~ao chamado ``Universo na Escola''~\cite{UniversoEscola}, em que professores e pesquisadores da Universidade s\~ao convidados a ministrar palestras a estudantes do ensino m\'edio da rede p\'ublica do estado do Esp\'irito Santo. As palestras versam sobre t\'opicos atuais de pesquisa em F\'isica, visando complementar a forma\c{c}\~ao dos estudantes e despertar-lhes o interesse pelas Ci\^encias em geral e, em particular, pela F\'isica. Elas s\~ao elaboradas como atividade complementar eletiva a estudantes nas escolas participantes e realizadas ap\'os o per\'iodo regular de aulas. 

Os autores elaboraram uma sequ\^encia sobre F\'isica de Part\'iculas para ser inserida no contexto desse projeto e foi ministrada por um dos autores a um grupo de, no máximo, quatorze estudantes do Centro Estadual de Ensino Médio em
Tempo Integral (CEEMTI) Prof$^{\text a}$ Maura Abaurre, no município de Vila Velha, no estado do Espírito Santo. Os(as) estudantes compunham uma turma mista contemplando as 1$^{\text{a}}$, 2$^{\text{a}}$ e 3$^{\text{a}}$ s\'eries do ensino m\'edio. Alguns participantes j\'a integravam um grupo com interesse pr\'evio em F\'isica, que vinha sendo acompanhado e orientado durante o ano pelo professor-coordenador. No entanto, no decorrer da sequ\^encia outros(as) estudantes se juntaram ao grupo, frequentemente a convite dos demais. A organiza\c{c}\~ao local foi realizada pelo professor-coordenador da \'area de F\'isica na escola. Os encontros da sequ\^encia como um todo tiveram duração aproximada de uma hora em aulas extra-classe durante os meses de setembro, outubro, novembro e dezembro de 2019, totalizando uma carga horária de 10 horas de atividades de Física de Partículas desenvolvidas com os(as) discentes. Cabe ressaltar que os(as) estudantes n\~ao tinham qualquer obriga\c{c}\~ao curricular de frequentar as aulas das interven\c{c}\~oes, que constitu\'iam atividade voluntária. Ainda ressaltamos que a sequ\^encia did\'atica foi elaborada em formato de aulas e n\~ao se deu no formato usual de palestras, como geralmente ocorre na perspectiva do projeto de extens\~ao mencionado. 

\subsection{Das aulas}
\label{subsec:aulas}

Na se\c{c}\~ao~\ref{sec:momentos} apresentamos propostas de momentos did\'aticos para introduzir a tem\'atica de F\'isica de Part\'iculas e da Eletrodin\^amica Qu\^antica em particular, organizados de maneira tal que promoveram o engajamento dos(as) estudantes desde o primeiro instante da interven\c{c}\~ao e de modo que os conceitos relevantes \`a F\'isica de Part\'iculas eram constru\'idos conjuntamente com os(as) discentes, conectando-se a seus conhecimentos pr\'evios. 

Em nossa interven\c{c}\~ao na CEEMTI Prof$^{\rm a}$ Maura Abaurre, o conte\'udo apresentado neste trabalho, com foco na Eletrodin\^amica Qu\^antica, serviu como base para a primeira parte da sequ\^encia did\'atica elaborada sobre F\'isica de Part\'iculas. Por sua vez, tal parte foi dividida em tr\^es aulas, ministradas nos dias 16, 26 e 30 de setembro do ano de 2019, cada uma com dura\c{c}\~ao entre 55 e 65 minutos. Os principais t\'opicos abordados em sala, e as respectivas subse\c{c}\~oes discutidas mais detalhadamente acima, encontram-se listadas nas tabelas~\ref{tab:Aula1},~\ref{tab:Aula2} e \ref{tab:Aula3}.

\begin{table*}[h!]
    \centering
	\footnotesize
	\bgroup
	\def\arraystretch{1.1}
	\begin{tabular}{|>{\small} c |l | c |} 
		\hline
		\multicolumn{3}{|c|}{\small\textbf{Aula 1: Introdu\c{c}\~ao \`a F\'isica de Part\'iculas}} \\
		\hline
		Se\c{c}\~ao & \multicolumn{2}{c|}{\small Momento}  \\
		\hline
		\multirow{4}{1cm}{\ref{sec:intro}} &
		Busca por conhecimentos pr\'evios & 1\\
		& N\'os ``enxergamos'' as part\'iculas? Debate sobre o funcionamento da Ci\^encia & 2\\
		& Apresenta\c{c}\~ao do Modelo Padr\~ao da F\'isica de Part\'iculas & 3\\
		& Alguns fen\^omenos e aplica\c{c}\~oes da F\'isica de Part\'iculas & 4\\
		\hline
		\multirow{8}{1cm}{\ref{sec:atom}} &
		    O cont\'inuo e o v\'acuo: embate sobre a estrutura da mat\'eria & 5\\
		& Os ``quatro elementos'' de Emp\'edocles e Arist\'oteles & 6\\
		& Atomismo de Dem\'ocrito, Leucipo e Epicuro & 7\\
        & Sucessos da hip\'otese atomista: & \\
        &   \hspace*{1cm} - a eletr\'olise da \'agua e as leis ponderais & 8\\
        &   \hspace*{1cm} - a lei dos gases ideais & 9\\
        &    \hspace*{1cm} - o movimento browniano & 10\\
        &    Ci\^encia como ferramenta de empoderamento & 11\\
		\hline
		\multirow{2}{1cm}{\ref{sec:eletrons}} &
		    A divisibilidade do ``indivis\'ivel'' & 13\\
		& O experimento de Thomson e a descoberta do el\'etron & 14\\
		\hline
		\multirow{3}{1cm}{\ref{sec:emgrav}} &
		    Atra\c{c}\~ao e repuls\~ao entre el\'etrons& 15\\
		& Lei de Coulomb e Gravita\c{c}\~ao de Newton: semelhan\c{c}as e diferen\c{c}as & 16\\
		& Por que percebemos mais a for\c{c}a gravitacional do que a for\c{c}a el\'etrica? & 17\\
		\hline
	\end{tabular}
	\egroup
	\caption{Os momentos da primeira aula da sequ\^encia.}
	\label{tab:Aula1}
\end{table*}

\begin{table*}[h!]
	\centering
	\footnotesize
	\bgroup
	\def\arraystretch{1.1}
	\begin{tabular}{|>{\small} c | l | c|} 
		\hline
		\multicolumn{3}{|c|}{\small \textbf{Aula 2: A natureza da luz -- Parte I}} \\
		\hline
		Se\c{c}\~ao & \multicolumn{2}{c|}{\small Momento}\\
		\hline
		\multirow{3}{1cm}{\ref{sec:contato}} & 
		    Breve revis\~ao da aula anterior & 1\\
		& Ubiquidade da intera\c{c}\~ao eletromagn\'etica & 2\\
		& Gravita\c{c}\~ao e eletromagnetismo predominam no nosso dia a dia & 3\\
		\hline
		\multirow{12}{1cm}{\ref{sec:luz}} &
    	Conhecimentos pr\'evios sobre o conceito de campo & 4\\
		& O campo el\'etrico e a lei de Coulomb & 5\\
		& Linhas de campo el\'etrico, lei de Gauss e o comportamento $1/r^2$ & 6\\
		& Analogia com fluxo de \'agua de um chafariz (se\c{c}\~ao \ref{sec:emgrav} revisitada) & 7\\
		& A inexist\^encia de monopolos magn\'eticos & 8\\
		& As equa\c{c}\~oes de Maxwell e a unifica\c{c}\~ao do eletromagnetismo & 9\\
		& Ondas eletromagn\'eticas & 10\\
        & Simula\c{c}\~oes: gera\c{c}\~ao e recep\c{c}\~ao de ondas eletromagn\'eticas & 11\\
        & Aplica\c{c}\~oes: ondas de r\'adio, \emph{wifi} e telecomunica\c{c}\~oes em geral & 12\\
        & A luz \'e uma onda eletromagn\'etica & 13\\
	    & O espectro eletromagn\'etico & 14\\
		& Fazendo pipoca no microondas & 15\\
		\hline
		\multirow{3}{1cm}{\ref{sec:solar}} & Espectro de emit\^ancia solar: &\\
		& \hspace*{1cm} evolu\c{c}\~ao da vis\~ao humana & 16 \\
		& \hspace*{1cm} por que as plantas s\~ao verdes? & 17 \\
		\hline
	\end{tabular}
	\egroup
	\caption{Os momentos da segunda aula da sequ\^encia.}
	\label{tab:Aula2}
\end{table*}

\begin{table*}[h!]
	\centering
	\footnotesize
	\bgroup
	\def\arraystretch{1.1}
	\begin{tabular}{| >{\small} c | l | c |} 
		\hline
		\multicolumn{3}{|c|}{\small\textbf{Aula 3: A natureza da luz -- Parte II}} \\
		\hline
		Se\c{c}\~ao &
		\multicolumn{2}{ c| }{\small Momento} \\
		\hline
		\multirow{9}{1cm}{\ref{sec:dualidade}} &
		Breve revis\~ao das aulas anteriores & 1 \\
		& O que s\~ao part\'iculas? O que s\~ao ondas? & 2 \\
		& A dualidade onda-part\'icula e o f\'oton & 3 \\
		& O f\'oton n\~ao tem massa & 4 \\
		& O el\'etron tamb\'em \'e onda-part\'icula & 5 \\
		& O microsc\'opio eletr\^onico & 6 \\
		& Trajet\'orias em Mec\^anica Qu\^antica & 7 \\
		& A rela\c{c}\~ao de de Broglie e a constante de Planck & 8 \\
		& Constantes fundamentais da Natureza & 9 \\
	\hline
	    \ref{sec:feynman} &  Intera\c{c}\~ao eletromagn\'etica: troca de f\'otons e diagramas de Feynman & 10 \\
\hline
        \multirow{7}{1cm}{\ref{sec:positrons}} & O p\'ositron: a antipart\'icula do el\'etron prevista por Dirac & 11 \\
        & O poder preditivo de uma teoria cient\'ifica & 12 \\
        & Aniquila\c{c}\~ao mat\'eria-antimat\'eria & 13 \\
        & Antipart\'iculas como part\'iculas se movendo para tr\'as no tempo & 14 \\
        & A cria\c{c}\~ao de um par el\'etron-p\'ositron & 15 \\
        & A rela\c{c}\~ao massa-energia de Einstein & 16 \\
        & Raios-X e tomografia PET--scan: s\~ao perigosos? & 17 \\
        \hline
    \end{tabular}
    \egroup
    \caption{Os momentos da terceira aula da sequ\^encia.}
    \label{tab:Aula3}
\end{table*}

\'E muito importante ressaltar que as aulas \emph{n\~ao} foram meramente expositivas mas o professor adotou uma postura amplamente dial\'ogica com os(as) alunos(as), a partir da qual diversos momentos citados nas tabelas emergiram organicamente da discuss\~ao com os(as) estudantes e das respostas \`as interroga\c{c}\~oes e propostas de debates. Por exemplo, durante o primeiro momento, em que os(as) estudantes se expressaram sobre seus conhecimentos pr\'evios da tem\'atica, foram citadas ``part\'iculas subat\^omicas'', ao que o professor rebateu com a pergunta ``\textit{E como sabemos que a mat\'eria \'e composta por \'atomos?}''. Essa quest\~ao suscitou uma breve discuss\~ao sobre como o conhecimento cient\'ifico consiste de modelos, que devem fazer previs\~oes test\'aveis e, mais importante, promover um empoderamento daqueles que portam tal conhecimento. Modelos s\~ao acatados ou suplantados de acordo com o n\'ivel de empoderamento que promovem. Essa situa\c{c}\~ao foi, ent\~ao, ilustrada na discuss\~ao dos sucessos emp\'iricos da teoria at\^omica, enfatizando-se que n\~ao precisamos ``enxergar'' um \'atomo para saber que existem, bastando testar alguns ``efeitos colaterais'' preditos pelo modelo. No caso, as leis ponderais constituem forte evid\^encia de que as rea\c{c}\~oes qu\'imicas s\~ao meros rearranjos de \'atomos, e n\~ao metamorfoses de um cont\'inuo.

Outro exemplo s\~ao os momentos 8 e 9 da aula 2. Aqui, ao inv\'es de simplesmente expor as equa\c{c}\~oes de Maxwell, o professor perguntou aos alunos onde se encaixaria o magnetismo, que ainda n\~ao havia se manifestado a n\~ao ser no nome da intera\c{c}\~ao eletro\emph{magn\'etica}. J\'a tendo discutido a lei de Gauss para campos el\'etricos, o professor prop\^os aos(\`as) alunos(as) formularem a lei an\'aloga para campos magn\'eticos, informando-lhes apenas que monopolos magn\'eticos n\~ao existem. Isso estimulou uma interessante discuss\~ao sobre como  equa\c{c}\~oes matem\'aticas s\~ao express\~oes de fatos f\'isicos em uma outra forma de linguagem. Houve certa dificuldade por parte dos(as) estudantes em construir essa formula\c{c}\~ao matem\'atica da lei de Gauss para campos magn\'eticos, mas, guiados pelo professor, foi poss\'ivel obter o resultado desejado. Seguiu-se, ent\~ao, uma discuss\~ao subsequente sobre as outras leis de Maxwell, mas dessa vez expressas verbalmente de maneira qualitativa, conforme discutido na se\c{c}\~ao \ref{sec:em}.

Por fim, outro exemplo \'e o momento sobre ``Trajet\'orias em Mec\^anica Qu\^antica'' na aula 3. Essa discuss\~ao emergiu de uma pergunta sobre a aparente contradi\c{c}\~ao entre a inexist\^encia de trajet\'orias na descri\c{c}\~ao qu\^antica das part\'iculas, e a observa\c{c}\~ao de ``trajet\'orias'' no mundo cl\'assico ou mesmo em experimentos de el\'etrons se movendo em c\^amaras de bolhas. O questionamento tornou-se uma oportunidade para discutir mais aprofundadamente o princ\'ipio da incerteza de Heisenberg nesse contexto, mostrando como a aparente trajet\'oria do el\'etron n\~ao consiste em uma linha, mas em um tubo com espessura finita correspondendo \`a incerteza em sua localiza\c{c}\~ao.

Uma an\'alise detalhada das narrativas presentes em sala de aula, considerando as participa\c{c}\~oes dos estudantes, as trocas com o professor e colegas, e diversas outras observa\c{c}\~oes sobre todas as aulas da sequ\^encia did\'atica ser\'a publicada em trabalho futuro dos autores.

\subsection{Metodologia e coleta de dados}
\label{subsec:metodologia}

O percurso metodol\'ogico adotado em nossa an\'alise possui cunho qualitativo e car\'ater de estudo de caso. A pesquisa qualitativa considera relevante todos os sujeitos envolvidos, os significados e os pontos de vista atribu\'idos \`as situa\c{c}\~oes a partir de dados coletados diretamente no ambiente natural de a\c{c}\~ao~\cite{LUEDKE1986}. Obtivemos autoriza\c{c}\~ao da Secretaria de Educa\c{c}\~ao do Estado do Esp\'irito Santo e da diretoria da escola para tomar notas durante as interven\c{c}\~oes sobre as intera\c{c}\~oes entre os(as) discentes, seus n\'iveis de interesse e de aten\c{c}\~ao, as perguntas realizadas, e demais rea\c{c}\~oes. A coleta de dados ocorreu a partir de observa\c{c}\~oes registradas durante as aulas em um di\'ario de campo, contemplando aspectos descritivos e reflexivos dos sujeitos envolvidos, dos objetos, do espa\c{c}o, das atividades e dos acontecimentos~\cite{LUEDKE1986, OLIVEIRA2014}, bem como reflex\~oes posteriores \`as interven\c{c}\~oes sobre rela\c{c}\~oes entre tais aspectos, sobre atividades para casa e intera\c{c}\~oes imediamente antes e ap\'os as interven\c{c}\~oes. A viv\^encia com o local onde as a\c{c}\~oes se desenvolveram favoreceu uma metodologia de estudo de caso, ideal quando se pretende conhecer as perspectivas de significados dos sujeitos e a identifica\c{c}\~ao de rela\c{c}\~oes causais e padr\~oes em contextos complexos que n\~ao permitem a utiliza\c{c}\~ao de levantamentos e experimentos~\cite{MORAESTAZIRI2019}. Em nosso caso focamos na compreens\~ao de conceitos-chave e suas rela\c{c}\~oes com tecnologia, sociedade e meio-ambiente, avaliando o surgimento de indicadores de alfabetiza\c{c}\~ao cient\'ifica e de engajamento ao longo das aulas, bem como rela\c{c}\~oes entre esses indicadores. 

A fim de manter o sigilo e preservar as identidades dos participantes, adotamos nomes fict\'icios para os estudantes, fazendo refer\^encia a cientistas que atuaram na \'area da F\'isica de Part\'iculas e/ou contribu\'iram para o seu desenvolvimento.

\subsection{Os indicadores de alfabetiza\c{c}\~ao cient\'ifica e de engajamento}
\label{subsec:indicadores}

As observa\c{c}\~oes realizadas durante as interven\c{c}\~oes inclu\'iram a an\'alise dos indicadores de alfabetiza\c{c}\~ao cient\'ifica e de engajamento demonstrado pelos estudantes para avaliar a viabilidade, a efetividade e as potencialidades da inclus\~ao da tem\'atica de F\'isica de Part\'iculas em salas de aula do ensino m\'edio. Novamente ressaltamos que, para al\'em do conte\'udo abordado em sala de aula, as interven\c{c}\~oes incorporaram uma postura amplamente dial\'ogica do professor e acreditamos que essa combina\c{c}\~ao favoreceu a viabilidade e a efetividade de nossa proposta.

\subsubsection*{Indicadores de alfabetiza\c{c}\~ao cient\'ifica}

Os indicadores de alfabetiza\c{c}\~ao cient\'ifica utilizados se alicer\c{c}am na perspectiva freireana de alfabetiza\c{c}\~ao cient\'ifica~\cite{FREIRE2000, FREIRE1989}, que destaca o(a) alfabetizado(a) cientificamente como aquele(a) que: compreende as rela\c{c}\~oes entre Ci\^encia, Tecnologia, Sociedade e Meio-Ambiente; compreende a natureza da Ci\^encia; compreende a \'etica que envolve o trabalho de um(a) cientista; e possui conhecimentos b\'asicos sobre as ci\^encias para atuar no mundo. Em qualquer processo de ensino-aprendizagem que busca incorporar a alfabetiza\c{c}\~ao cient\'ifica identificam-se indicadores que demonstram o desenvolvimento de habilidades e conhecimentos associados ao trabalho de cientistas que, de acordo com a ref.~\cite{SasseronCarvalho2008}, podem ser divididos em tr\^es blocos de a\c{c}\~oes:
\begin{enumerate}[label={\roman*)}]
    \item a\c{c}\~oes que envolvem o trabalho com os dados obtidos em uma investiga\c{c}\~ao;
    \item a\c{c}\~oes que estruturam o pensamento cient\'ifico;
    \item a\c{c}\~oes que buscam o entendimento da situa\c{c}\~ao analisada.
\end{enumerate}
No primeiro bloco de a\c{c}\~oes podem ser identificados os seguintes elementos: a seria\c{c}\~ao de informa\c{c}\~oes, a organiza\c{c}\~ao de informa\c{c}\~oes e a classifica\c{c}\~ao de informa\c{c}\~oes. O primeiro busca estabelecer as bases para uma pesquisa, um conjunto de dados, por exemplo. O segundo elemento aparece quando se discute a metodologia de uma pesquisa, um arranjo das informa\c{c}\~oes. J\'a o terceiro elemento aparece quando se pretende hierarquizar as informa\c{c}\~oes, estabelecendo rela\c{c}\~oes entre elas. O segundo bloco de a\c{c}\~oes envolve as formas de organizar o pensamento, seja atrav\'es do racioc\'inio l\'ogico (relacionando a forma das ideias) e/ou do racioc\'inio proporcional (ideias e vari\'aveis s\~ao relacionadas). J\'a o terceiro bloco contempla o maior n\'umero de elementos: o levantamento de hip\'oteses, o teste de hip\'oteses, a justificativa, a previs\~ao e a explica\c{c}\~ao. O levantamento de hip\'oteses ocorre nos momentos em que suposi\c{c}\~oes sobre determinada ideia ou assunto surgem, e o teste de hip\'oteses ocorre quando essas hip\'oteses s\~ao colocadas em xeque. J\'a a justificativa aparece quando ideias s\~ao incorporadas a uma linha de racioc\'inio que fortalecem uma determinada afirma\c{c}\~ao. A etapa da previs\~ao ocorre quando se sugere que algum evento seja proveniente de alguma ideia e/ou fen\^omeno em an\'alise. O \'ultimo elemento nesse bloco de a\c{c}\~oes, a explica\c{c}\~ao, prop\~oe reunir e relacionar todos os elementos anteriores: informa\c{c}\~oes, ideias, hip\'oteses, previs\~oes. 

Tais elementos do fazer cient\'ifico foram considerados em nossa an\'alise como indicadores de alfabetiza\c{c}\~ao cient\'ifica e s\~ao resumidos na tabela~\ref{tab:Acoes}. Ressaltamos que cada um desses indicadores pode se manifestar independentemente dos outros em sala de aula. Al\'em de serem independentes entre si podem, inclusive, aparecer concomitantemente. 
\begin{table*}[h!]
	\centering
	\scriptsize
	\def\arraystretch{1.75}
	\begin{tabular}{|m{5cm}| m{6cm}|} 
		\hline
		\textbf{A\c{c}\~ao do fazer cient\'ifico} &
		    \textbf{Indicadores de alfabetiza\c{c}\~ao cient\'ifica\newline
		    (Elementos do fazer cient\'ifico)}
		\\ \hline
		Trabalho com os dados obtidos\newline
		    em uma investiga\c{c}\~ao
		& 
		Seria\c{c}\~ao de informa\c{c}\~oes\newline 
		Organiza\c{c}\~ao de informa\c{c}\~oes\newline
		Classifica\c{c}\~ao de informa\c{c}\~oes		
		\\
		\hline
		Estrutura\c{c}\~ao do pensamento cient\'ifico
		& 
		Racioc\'inio l\'ogico\newline
		Racioc\'inio proporcional
		\\ \hline
		Entendimento da situa\c{c}\~ao analisada
		& 
		Levantamento de hip\'oteses\newline
		Teste de hip\'oteses\newline	
		Justificativa\newline		
		Previs\~ao\newline	 
		Explica\c{c}\~ao\\ \hline
	\end{tabular}
	\caption{Os indicadores de alfabetiza\c{c}\~ao cient\'ifica a partir de elementos do fazer cient\'ifico.}
	\label{tab:Acoes}
\end{table*}

\subsubsection*{Indicadores de engajamento}

O engajamento escolar possui natureza multifacetada e \'e fruto da complexa intera\c{c}\~ao social do sujeito com o ambiente escolar no qual est\'a inserido, as situa\c{c}\~oes que presencia e os outros sujeitos desse espa\c{c}o. Al\'em disso, engloba todo um processamento psicol\'ogico do sujeito com rela\c{c}\~ao aos est\'imulos e quaisquer modifica\c{c}\~oes que alterem esse ambiente e suas experi\^encias. Dessa maneira, um engajamento positivo por parte dos estudantes est\'a diretamente ligado \`a exist\^encia de est\'imulos e situa\c{c}\~oes de aprendizagem favor\'aveis ao seu desenvolvimento. Considerar a incid\^encia de indicadores referentes ao engajamento dos estudantes auxilia na compreens\~ao das experi\^encias no ambiente escolar, na prepara\c{c}\~ao de atividades, e na compreens\~ao da efetividade e das potencialidades do processo de ensino-aprendizagem adotado.

A literatura sobre o assunto contempla tr\^es tipos de engajamento que se relacionam de forma din\^amica e n\~ao podem ser pensados como processos isolados: o comportamental, o emocional e o cognitivo~\cite{FREDERICKSetal2004, COELHO2011, BORGESetal2005, SASSERONSOUZA2019, FARIAVAZ2019, Finn1993, Voelkl1997, STIPEK2002, CONNELLWELLBORN1991, BROPHY1987, AMES1992, DWECKLEGGETT1988, HARTER1981, CORNOMADINACH1983}. Listamos na tabela~\ref{tab:Engajamento} os indicadores de engajamento que consideramos abaixo na apresenta\c{c}\~ao de nossos resultados quanto \`a viabilidade, \`a efetividade e \`as potencialidades da inser\c{c}\~ao da tem\'atica de F\'isica de Part\'iculas em sala de aula do ensino m\'edio.
\bgroup
\def\arraystretch{1.4}
\begin{table*}[h!]
	\scriptsize
	\centering
	\begin{tabular}{|>{\centering}m{4cm}| >{\centering}m{4cm} | c|} 
		\hline
		\multicolumn{3}{|c|}{\textbf{Indicadores de engajamento}} \\
		\hline
		\textbf{Comportamental} &
		\textbf{Emocional} &
		\textbf{Cognitivo} \\
		\hline
		Participa\c{c}\~ao nas aulas &
		Emo\c{c}\~ao &
		Investimento no aprendizado \\
		\hline
		Participa\c{c}\~ao/execu\c{c}\~ao\newline
		   nas/das tarefas de sala &
		Identifica\c{c}\~ao com\newline 
		    a escola e/ou colegas &
		Autonomia \\
		\hline
		\multirow{3}{4cm}{\centering Participa\c{c}\~ao/execu\c{c}\~ao\newline
		    nas/das tarefas de casa} &
		Identifica\c{c}\~ao com o professor &
		Desejo de ir al\'em do b\'asico \\\cline{2-3}
		&
		Atribui\c{c}\~ao de valores \`a F\'isica\newline
		   e/ou \`as Ci\^encias em geral &
		Uso de estrat\'egias \\		\hline
	\end{tabular}
	\caption{Os indicadores de engajamento.}
	\label{tab:Engajamento}
\end{table*}
\egroup

\subsection{Resultados}
\label{subsec:resultados}

As incid\^encias dos indicadores de alfabetiza\c{c}\~ao cient\'ifica e de engajamento ser\~ao apresentadas apenas para as tr\^es primeiras aulas da sequ\^encia, foco deste trabalho, relacionando-as com algumas observa\c{c}\~oes da sequ\^encia como um todo. Uma elabora\c{c}\~ao detalhada e a an\'alise aprofundada da incid\^encia desses indicadores ao longo de toda a sequ\^encia ser\'a publicada separadamente.

\begin{figure*}[h!]
	\centering
	\textbf{Incid\^encia das a\c{c}\~oes do fazer cient\'ifico -- Aulas 1, 2 e 3}\\
	\includegraphics[trim=0 0 180 70, clip, scale=.52]{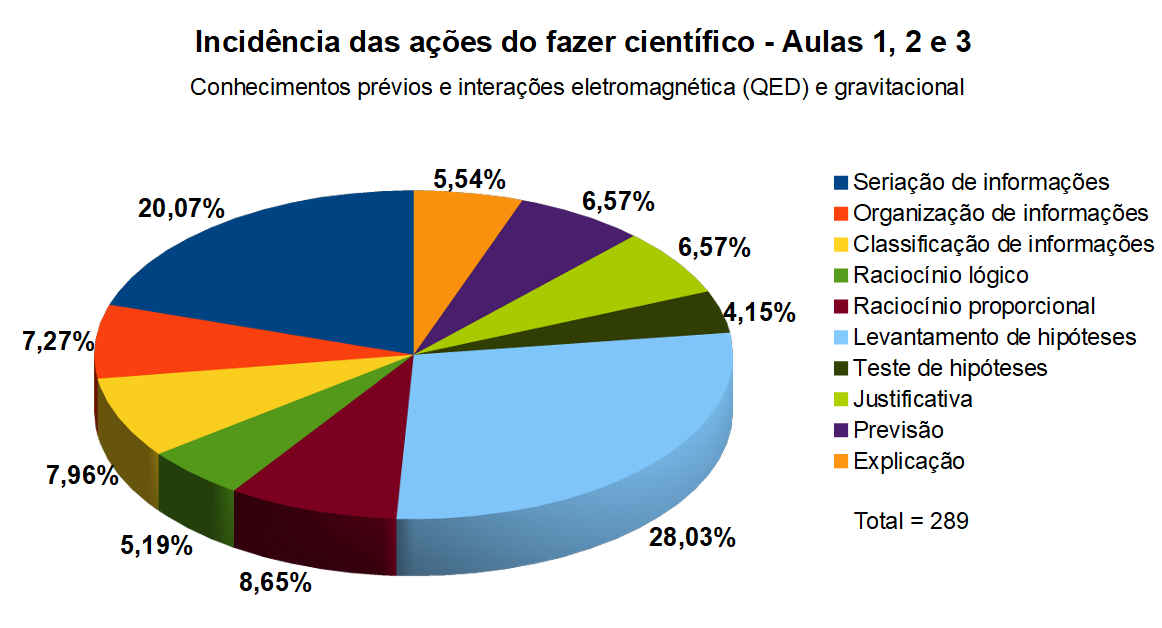}\qquad
	\includegraphics[trim=400 40 0 70, clip, scale=.7]{AC_123_Total}
	\caption{Indicadores de alfabetiza\c{c}\~ao cient\'ifica observados nas interven\c{c}\~oes realizadas. A legenda est\'a ordenada no sentido anti-hor\'ario do gr\'afico a partir de ``seria\c{c}\~ao de informa\c{c}\~oes'' (azul escuro, 20,07\%).} 
	\label{fig:AC_123}
\end{figure*}

Os resultados referentes aos indicadores de alfabetiza\c{c}\~ao cient\'ifica mais not\'aveis nas tr\^es primeiras aulas da sequ\^encia s\~ao apresentados na figura~\ref{fig:AC_123}.

Entre as a\c{c}\~oes que envolvem o trabalho com os dados obtidos em uma investiga\c{c}\~ao, a mais presente foi a de seria\c{c}\~ao de informa\c{c}\~oes. Esse resultado \'e bastante natural considerando que o bloco de aulas em quest\~ao se refere \`as primeiras aulas da sequ\^encia, com diversos momentos em que os(as) alunos(as) buscavam estabelecer as informa\c{c}\~oes bases, pois ou estabeleceram seus primeiros contatos com alguns dos t\'opicos ou relembraram conhecimentos pr\'evios, ressignificando-os no contexto das discuss\~oes ocorridas em sala de aula.
		
Com rela\c{c}\~ao \`as a\c{c}\~oes que estruturam o pensamento cient\'ifico, percebemos uma maior incid\^encia do racioc\'inio proporcional. Isso se deu principalmente por conta do est\'imulo dado para o uso do racioc\'inio proporcional por parte dos(as) alunos(as) quando uma abordagem qualitativa sustentou a discuss\~ao sobre as semelhan\c{c}as e diferen\c{c}as da lei de Coulomb com a lei da gravita\c{c}\~ao universal de Newton na primeira aula. Tamb\'em houve forte ocorr\^encia desse indicador durante a discuss\~ao da segunda aula sobre a express\~ao da lei de Coulomb, quando foram analisados casos limites, como o de duas cargas el\'etricas colocadas muito pr\'oximas uma da outra. 
		
J\'a com rela\c{c}\~ao \`as a\c{c}\~oes que buscam o entendimento da situa\c{c}\~ao analisada, a maior incid\^encia ficou por conta do levantamento de hip\'oteses, principalmente devido \`a primeira e \`a segunda aulas. Na primeira aula, esse indicador apareceu mais no momento de busca pelos conhecimentos pr\'evios dos(as) estudantes, quando eles ficaram bem \`a vontade para responder algumas quest\~oes-chave propostas pelo docente, com o intuito de entender o que os(as) alunos(as) j\'a conheciam sobre F\'isica de Part\'iculas, e tamb\'em para discutirem entre si, complementando as falas dos(as) colegas com sugest\~oes que avan\c{c}avam a conversa inicial. J\'a na segunda aula, esse indicador apareceu com maior frequ\^encia nos momentos em que o professor estimulou os(as) alunos(as) a elaborarem uma lei de Gauss para o magnetismo ap\'os a discuss\~ao sobre a inexist\^encia de monopolos magn\'eticos. Tamb\'em houve alta incid\^encia desse indicador no momento em que a simula\c{c}\~ao interativa com as ondas eletromagn\'eticas foi utilizada~\cite{PhET_antena}, bem como no momento em que conversaram sobre o que caracterizava uma onda.
		
\begin{figure*}[h!]
	\centering
	\includegraphics[scale=.7]{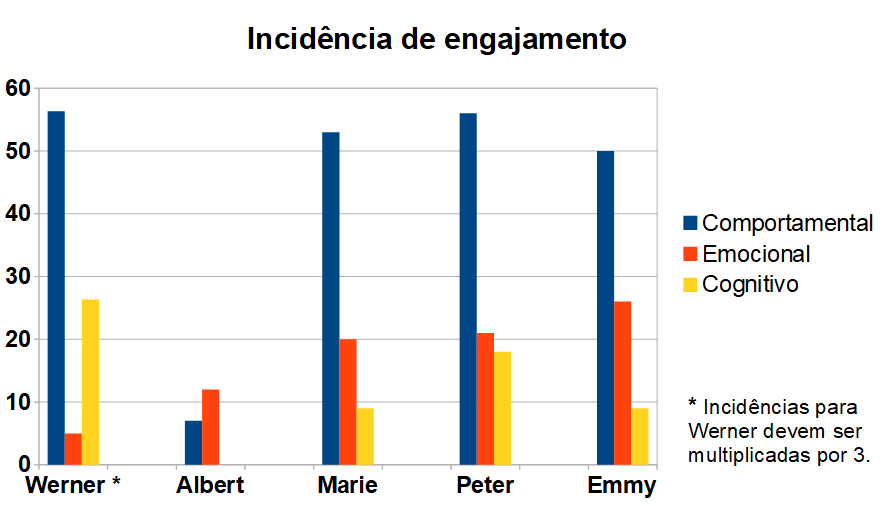}
	\caption{Incid\^encia dos indicadores de engajamento comportamental, emocional e cognitivo nas tr\^es primeiras aulas da sequ\^encia did\'atica proposta para os cinco estudantes mais presentes nas interven\c{c}\~oes. Note que o estudante Werner se destacou apresentando altas incid\^encias em todos os indicadores. Optamos por reescalonar essas incid\^encias no gr\'afico e, portanto, os n\'umeros totais para esse estudante s\~ao obtidos a partir dos n\'umeros apresentados multiplicados por um fator 3. }
	\label{fig:Resultados_Engajamento_Bloco1}
\end{figure*}

Na figura~\ref{fig:Resultados_Engajamento_Bloco1} apresentamos a alta incid\^encia dos indicadores de engajamento  manifestados durante as interven\c{c}\~oes, para cada um dos tipos de engajamento, e para os cinco estudantes mais presentes nas aulas da sequ\^encia de F\'isica de Part\'iculas como um todo. Estes ser\~ao denominados  Werner, Albert, Marie, Peter e Emmy, e estiveram presentes em 10, 10, 8, 7 e 6 das 10 aulas da sequ\^encia, respectivamente.

As maiores incid\^encias de engajamento comportamental e cognitivo partiram do estudante Werner. Marie, Peter e Emmy demonstraram incid\^encias similares para cada um dos tipos de engajamento. J\'a Albert foi o que menos participou das tr\^es primeiras aulas entre os cinco alunos. No entanto, a partir da quarta aula (sobre f\'isica nuclear, que ser\'a abordada em futura publica\c{c}\~ao) este aluno revelar\'a um enorme salto na incid\^encia de engajamento comportamental. Al\'em disso, a partir da quinta aula da sequ\^encia, ele tamb\'em revelar\'a um aumento significativo dos indicadores de engajamento emocional e, principalmente, de engajamento cognitivo. Considerando todas as aulas da sequ\^encia que ser\~ao abordadas em futuras publica\c{c}\~oes, observamos em Albert uma significativa evolu\c{c}\~ao nas incid\^encias de todos os tipos de engajamento e, junto a Werner, que demonstrou engajamento ao longo de todas as aulas da sequ\^encia, ele ir\'a se configurar com um dos estudantes com maiores n\'iveis de engajamento da sequ\^encia. Esse crescimento progressivo em seus \'indices de engajamento \'e demonstrativo do sucesso da sequ\^encia, que,  aliada \`as interven\c{c}\~oes dial\'ogicas do docente, mostra-se capaz de trazer os(as) estudantes para as aulas, mesmo aqueles(as) mais inibidos, t\'imidos e/ou inseguros em um primeiro momento.


\section{Conclus\~oes}
\label{sec:conclusoes}

Neste trabalho apresentamos material referente \`a primeira parte de uma proposta de sequ\^encia did\'atica sobre F\'isica de Part\'iculas para o ensino m\'edio. A sequ\^encia did\'atica de 10 aulas contemplou tamb\'em outras tem\'aticas da F\'isica de Part\'iculas: a F\'isica Nuclear; a intera\c{c}\~ao forte (QCD, da sigla em ingl\^es \textit{Quantum Cromodynamics}); e a intera\c{c}\~ao fraca, o b\'oson de Higgs e t\'opicos al\'em do Modelo Padr\~ao da F\'isica de Part\'iculas. Est\~ao em prepara\c{c}\~ao outras publica\c{c}\~oes com material similar ao apresentado neste trabalho para essas outras tem\'aticas.

A primeira parte dessa sequ\^encia, foco do presente trabalho, lidou com a Eletrodin\^amica Qu\^antica. Na se\c{c}\~ao~\ref{sec:momentos} apresentamos os momentos did\'aticos que podem ser explorados em sala de aula dentro desse contexto. Na se\c{c}\~ao~\ref{sec:resultados} avaliamos a viabilidade e a efetividade de nossa proposta a partir de aplica\c{c}\~ao pr\'opria deste material a um grupo misto de 14 alunos das tr\^es s\'eries do ensino m\'edio de uma escola p\'ublica estadual do Esp\'irito Santo. O material apresentado neste trabalho serviu como base para tr\^es aulas com dura\c{c}\~ao aproximada de uma hora cada, e foi utilizado em sala de aula aliado a uma postura dial\'ogica do docente e a uma perspectiva de ensino-aprendizagem que contemplou a alfabetiza\c{c}\~ao cient\'ifica dos estudantes. As aulas incorporaram elementos para fomentar uma discuss\~ao sistematizada sobre o assunto e seus conceitos-chave, sobre a natureza e a \'etica das ci\^encias e os fatores recorrentes na pr\'atica cient\'ifica, bem como sobre as rela\c{c}\~oes da tem\'atica com Tecnologia, Sociedade e Meio-Ambiente. Percebemos um bom entendimento dos conceitos-chave referentes ao material aqui apresentado ao avaliarmos as intera\c{c}\~oes em sala de aula, que ser\~ao publicadas \`a parte. Os resultados da se\c{c}\~ao~\ref{sec:resultados} revelam a presen\c{c}a de diversos indicadores de alfabetiza\c{c}\~ao cient\'ifica bem como de engajamento dos estudantes. 
Tais resultados indicam a viabilidade de se incorporar tem\'aticas da f\'isica moderna e contempor\^anea no ensino m\'edio, satisfazendo um longo desejo dos pr\'oprios estudantes que, curiosos, acabam se envolvendo em um processo de ensino-aprendizagem que os empodera para participarem do processo coletivo de constru\c{c}\~ao hist\'orico-social da Ci\^encia e os insere em discuss\~oes relevantes \`a realidade concreta do mundo contempor\^aneo. Acreditamos que a efetividade das aulas foi fruto da combina\c{c}\~ao da postura dial\'ogica do docente aliada \`a perspectiva voltada \`a alfabetiza\c{c}\~ao cient\'ifica dos estudantes.

Nota-se que o material apresentado neste artigo \'e um pouco mais extenso do que aquele que ministramos em sala de aula. Isso se deve a dois fatores. Por um lado, nossas aulas ocorreram em um contexto de um projeto de extens\~ao, envolvendo um grupo heterog\^eneo de estudantes, com o prop\'osito de servir-lhes como introdu\c{c}\~ao \`a F\'isica de Part\'iculas. Nesse formato, n\~ao julgamos cab\'ivel aprofundar discuss\~oes sobre o eletromagnetismo cl\'assico, mas tampouco se poderia pressupor que o grupo j\'a estivesse familiarizado com esse conte\'udo. Ao mesmo tempo, por limita\c{c}\~oes de tempo dispon\'ivel, julgamos inadequado discutir a QED enquanto teoria de campo (se\c{c}\~ao~\ref{sec:qft}) ou aprofundar a discuss\~ao sobre o conceito de f\'oton virtual, apesar de tal discuss\~ao fornecer uma rica ferramenta para compreens\~ao da intera\c{c}\~ao eletromagn\'etica em termos de microf\'isica. Em suma, foi necess\'ario encontrar um meio termo, de acordo com nossas circunst\^ancias concretas. Por outro lado, a outros(as) docentes, trabalhando em situa\c{c}\~oes diferentes, pode ser mais conveniente expandir a discuss\~ao sobre algumas se\c{c}\~oes, e por isso disponibilizamos material que auxilie nesse prop\'osito. Estamos, com isso, aderindo ao princ\'ipio norteador deste trabalho, que \'e elaborar um material que sirva n\~ao como receita a ser seguida linear e mecanicamente, mas como fonte auxiliar ao planejamento de aulas sobre esse assunto, um produto org\^anico que sirva de mat\'eria prima a ser trabalhada pelo(a) docente, ao inv\'es de f\'ormulas r\'igidas, engessadas. Uma valiosa sugest\~ao de material complementar sobre a QED voltada ao p\'ublico leigo, extremamente instrutivo e educativo, pode ser lido no livro de um dos pioneiros dessa teoria: Richard Feynman, ``A estranha teoria da luz e da mat\'eria''~\cite{FeynmanQED}. Al\'em do material e dos resultados explorat\'orios relativos aos indicadores de alfabetiza\c{c}\~ao cient\'ifica e de engajamento apresentados neste trabalho, inclu\'imos um ap\^endice com propostas de atividades que podem ser incorporadas em sala de aula dentro da tem\'atica de Eletrodin\^amica Qu\^antica.

Nossos resultados mostram conclusivamente que, aliando-se a utiliza\c{c}\~ao (e adapta\c{c}\~ao) do presente material a uma postura dial\'ogica por parte do(a) docente, \'e poss\'ivel engajar os(as) estudantes no processo de ensino-aprendizagem e promover a alfabetiza\c{c}\~ao cient\'ifica em seu sentido pleno.  


\section*{Agradecimentos}

Os autores agradecem J\'ulio C\'esar Fabris, organizador do projeto ``Universo na Escola'', e Thiago Pereira da Silva, professor de F\'isica da CEEMTI Prof$^{\text{a}}$ Maura Abaurre (Vila Velha -- ES) no per\'iodo das interven\c{c}\~oes, por viabilizarem o presente estudo. Agradecemos tamb\'em Geide Rosa Coelho, por sugest\~oes e coment\'arios relevantes \`a an\'alise das interven\c{c}\~oes.


\appendix

\section{Propostas de Atividades}
\label{apend:propostadeatividades}

\subsection{Espectro de emit\^ancia solar} 
\label{sec:solar}

\begin{figure}[h!]
	\centering
	\includegraphics[width=.5\textwidth]{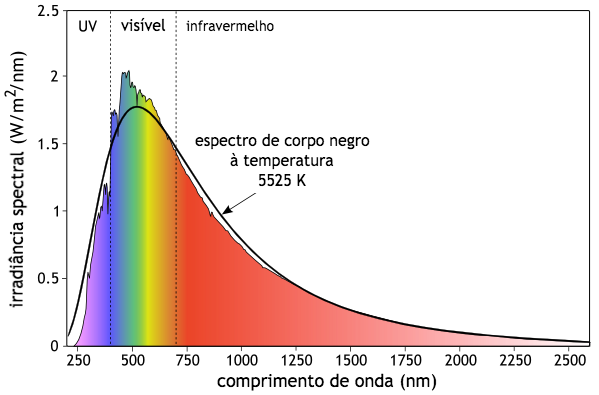}
	\caption{Espectro de emiss\~ao de radia\c{c}\~ao pelo Sol. Note o pico exatamente na regi\~ao do vis\'ivel, com m\'aximo na faixa da cor verde. Fonte: Dom\'inio p\'ublico.}
	\label{fig:solar}
\end{figure}

Uma interessante atividade em di\'alogo com a Biologia consiste em apresentar o espectro de emiss\~ao solar (i.e. a pot\^encia de ondas eletromagn\'eticas emitidas pelo Sol em fun\c{c}\~ao da frequ\^encia da onda), figura~\ref{fig:solar}, que tem pico de m\'aximo precisamente na regi\~ao do vis\'ivel, e lan\c{c}ar \`a turma o desafio de tentar explicar a ``coincid\^encia'' de o Sol ser mais luminoso justamente na faixa em que n\'os enxergamos o espectro eletromagn\'etico. Pela experi\^encia dos autores, n\~ao haver\'a grandes dificuldades da turma em invocar o mecanismo de sele\c{c}\~ao natural. 
	
Ainda mais intrigante, entretanto, \'e que o pico do espectro solar corresponde \`a radia\c{c}\~ao na faixa da cor verde, que \'e justamente a cor das folhas das plantas. Isso significa que, embora usem absor\c{c}\~ao de luz para realizarem fotoss\'intese, as folhas \emph{refletem} a faixa de frequ\^encia em que h\'a mais luminosidade dispon\'ivel. Como explicar esse aparente paradoxo? Tamb\'em essa quest\~ao, cuja solu\c{c}\~ao \'e bastante mais elaborada, pode ser lan\c{c}ada \`a turma, para que, individualmente ou em grupos, tentem formular hip\'oteses plaus\'iveis que expliquem tal fato, exercitando assim o racioc\'inio e criatividade no \^ambito cient\'ifico. Uma proposta recente de se explicar o aparente paradoxo \'e a chamada ``hip\'otese da Terra roxa''~\cite{PurpleEarth}. Uma discuss\~ao sobre o assunto pode ser articulada conjuntamente com o(a) professor(a) de Biologia da institui\c{c}\~ao de ensino, pois envolve, em igual medida, conhecimentos de F\'isica (espectro eletromagn\'etico, espectro de emiss\~ao solar, a cor dos objetos) e Biologia (mecanismos de sele\c{c}\~ao natural e especia\c{c}\~ao, fotoss\'intese), fornecendo uma excelente oportunidade para se superar a compartimentaliza\c{c}\~ao do ensino cient\'ifico na escola, al\'em de promover atividades diferenciadas com um vi\'es de ensino investigativo.

\subsection{Efeito fotovoltaico e energia solar}
\label{sec:fotovoltaico}

\begin{enumerate}
	\item Ao inv\'es de apresentar em aula a explica\c{c}\~ao de Einstein para o efeito fotoel\'etrico, o docente pode descrever brevemente o fen\^omeno da eje\c{c}\~ao de el\'etrons de uma superf\'icie met\'alica quando iluminada por uma fonte de luz, junto dos seguintes fatos experimentais:
	\begin{itemize}
		\item existe uma frequ\^encia m\'inima da luz incidente abaixo da qual nenhum el\'etron \'e ejetado. Essa frequ\^encia m\'inima depende do material de que \'e feita a superf\'icie;
		\item a energia cin\'etica dos el\'etrons ejetados aumenta com a frequ\^encia da luz incidente, mas n\~ao depende de sua intensidade;
		\item o n\'umero de el\'etrons ejetados n\~ao depende da frequ\^encia da luz incidente, mas aumenta com sua intensidade.
	\end{itemize}
	Sugere-se, ent\~ao, que os(as) estudantes discutam poss\'iveis explica\c{c}\~oes para essas observa\c{c}\~oes. Pode-se direcion\'a-los(as) a usar o conceito de f\'oton introduzido por outras vias, ou deix\'a-los(as) concluir sobre a necessidade de se tratar a luz como constitu\'ida de corp\'usculos. O objetivo central \'e que sejam capazes de construir a rela\c{c}\~ao Einsteiniana dada pela equa\c{c}\~ao~(\ref{eq:Ehf}), e identificar que a intensidade da luz est\'a relacionada ao n\'umero de f\'otons. Pode ser uma atividade desafiadora aos estudantes, mas o(a) docente pode e deve intervir para ajud\'a-los a construir o racioc\'inio que leva \`a impossibilidade de explicar esse comportamento pela teoria ondulat\'oria da luz, e formular o conceito de f\'oton.

	\item Complementar \`a discuss\~ao do efeito fotoel\'etrico, pode-se propor que os(as) alunos(as) pesquisem e redijam um texto, ou apresentem curtos semin\'arios sobre assuntos relativos a energia solar e c\'elulas fotovoltaicas. Como essas c\'elulas funcionam? Quais s\~ao as vantagens e desvantagens perante outras formas de gera\c{c}\~ao de energia el\'etrica? Quais s\~ao os potenciais impactos ambientais? Qual \'e a efici\^encia t\'ipica de tais c\'elulas? Onde elas s\~ao utilizadas? Pode-se indicar as refs.~\cite{IEEE, RVQ, ANEEL} (e demais refer\^encias citadas por essas) como alguns textos base \`a pesquisa. Caso o(a) docente queira preparar uma sequ\^encia de aulas voltada a essa tem\'atica, pode se basear na ref.~\cite{Lima}.
\end{enumerate}

\subsection{Dualidade onda-part\'icula e o laborat\'orio SIRIUS}
\label{sec:sirius}

\begin{figure}[h!]
	\centering
	\includegraphics[width=.43\textwidth]{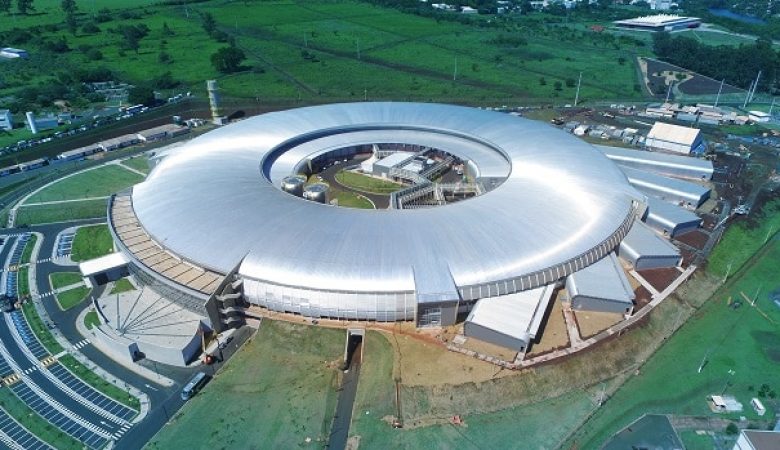}
	\caption{Sirius, o novo laborat\'orio de luz s\'incrotron brasileiro, localizado no Laborat\'orio Nacional de Luz S\'incroton (LNLS) em Campinas. Fonte: LNLS/CNPEM.}
	\label{fig:Sirius}
\end{figure}

O SIRIUS (figura~\ref{fig:Sirius}) \'e o novo acelerador de part\'iculas brasileiro, localizado no Laborat\'orio Nacional de Luz S\'incrotron (LNLS) em Campinas, e que tem por objetivo contribuir para o estudo da estrutura de materiais e at\'e mesmo de mol\'eculas. Ao produzir f\'otons altamente energ\'eticos, ele pode ser usado essencialmente como um microsc\'opio extremamente potente, capaz de discernir dist\^ancias muito inferiores \`as acess\'iveis por um microsc\'opio usual, que funcione \`a base de luz vis\'ivel. Isso porque a menor dist\^ancia $d$ que um instrumento \'otico pode discernir est\'a associada ao comprimento da onda $\lambda$ incidente sobre o objeto que se deseja observar, que por sua vez est\'a associado \`a energia do f\'oton $E$ via
\begin{equation}
	d\sim \lambda = \frac{hc}{E}\approx \frac{1.24\times 10^{-6} \text{ eV}\cdot \text{m}}{E}.
	\label{eq:resolucao}
\end{equation}

No Sirius, um feixe de el\'etrons \'e acelerado em torno de uma circunfer\^encia de 518,4~m, e, por se tratarem de cargas el\'etricas aceleradas, emitem radia\c{c}\~ao eletromagn\'etica de diversas frequ\^encias --- chamada radia\c{c}\~ao s\'incrotron. Essa radia\c{c}\~ao \'e separada e coletada em 13 diferentes canais (chamados \textit{linhas de luz}), cada um com intensidades e energias caracter\'isticas, e \'e usada para ``fotografar'' objetos em alta resolu\c{c}\~ao. 

Para se ter uma ideia das capacidades desse laborat\'orio, podemos considerar a linha de luz Sapucaia, em que o feixe de luz \'e otimizado para energias de $12$~keV. O(a) docente pode propor que os(as) estudantes estimem a escala de comprimento m\'inima que pode ser discernida por essa fonte, usando a rela\c{c}\~ao~(\ref{eq:resolucao}), e citem alguns tipos de objetos que podem ser estudados nessa linha de luz. 

\'E f\'acil ver que a resolu\c{c}\~ao estimada \'e suficiente para enxergar subestruturas da ordem de $10^{-10}$~m, portanto essa linha de luz \'e apropriada ao estudo de mol\'eculas ou pol\'imeros. 

Essa discuss\~ao tamb\'em ajuda a construir a compreens\~ao de que experimentos a mais altas energias est\~ao associados \`a ``sondagem'' de dist\^ancias cada vez menores, o que fica claro da rela\c{c}\~ao~(\ref{eq:resolucao}). \'E por isso que, se queremos estudar a estrutura elementar da mat\'eria, precisamos de aceleradores de part\'iculas cada vez mais potentes.

\subsection{Unidades naturais}
\label{sec:unidades}

A F\'isica de Part\'iculas lida com fen\^omenos em escalas subat\^omicas, envolvendo ordens de grandeza muito d\'ispares daquelas com que estamos familiarizados no dia a dia. Por isso, \'e dif\'icil adquirir uma intui\c{c}\~ao acurada das quantidades f\'isicas relevantes em termos das unidades do Sistema Internacional (metro, segundo, grama, etc.). Nesse caso \'e mais conveniente usar um outro sistema de unidades, chamado ``sistema natural de unidades'', que ser\'a objeto de discuss\~ao nessa atividade. 

Mais especificamente, a proposta principal da atividade a seguir \'e discutir o papel das constantes universais na Natureza como \emph{fatores de convers\~ao} entre grandezas f\'isicas que possuem uma rela\c{c}\~ao \'intima entre si, mas que foram inicialmente consideradas distintas no desenvolvimento hist\'orico das teorias cient\'ificas. Um sistema natural de unidades \'e aquele em que essas grandezas intimamente relacionadas s\~ao tratadas como tal, em p\'e de igualdade.

A discuss\~ao pode se iniciar com a constata\c{c}\~ao de que, por se tratar de uma teoria qu\^antica relativ\'istica, duas constantes fundamentais aparecem frequentemente nas express\~oes das grandezas f\'isicas computadas em F\'isica de Part\'iculas:

\bgroup
\footnotesize
\begin{eqnarray*}
	\hbar &\simeq& 1.055\times 10^{-34}~\text{J} \cdot \text{s} \quad \left(\dfrac{\text{constante de Planck}}{2\pi}\right), \\
	c &\simeq& 2.998\times 10^{8}~\text{m/s} \quad~~~~ \left(\parbox{30mm}{\centering velocidade da luz\\ no v\'acuo}\right).
\end{eqnarray*}
\egroup
O(A) professor(a) pode, ent\~ao, questionar os(as) estudantes por que essas constantes possuem esses valores espec\'ificos. O que esses n\'umeros significam, de fato? A discuss\~ao \'e longa e profunda, mas um ponto que deve ser tocado \'e que esses n\'umeros s\~ao espec\'ificos a um sistema de unidades em particular, e que, em outro sistema, os n\'umeros seriam outros. Deve ficar claro aos(\`as) estudantes que isso \emph{n\~ao} significa que o valor da constante muda. O que muda \'e a sua \emph{representa\c{c}\~ao}, que vai depender no sistema de unidades: expressar a velocidade da luz em milhas por hora resultar\'a em um valor num\'erico diferente do citado acima, mas isso certamente n\~ao significa que a velocidade da luz foi alterada de alguma maneira.

Essa discuss\~ao inicial abre margem para a quest\~ao: seria poss\'ivel escolher um sistema de unidades em que $\hbar=1$ e $c=1$? Deve ficar claro, aqui, que esses valores s\~ao \emph{adimensionais}. Isso pode introduzir alguma confus\~ao na turma: como pode uma grandeza com dimens\~ao de ``metro/segundo'' converter-se em uma constante adimensional? Podemos igualar unidades de comprimento e unidades de tempo? Se sim, o que isso significa?

As respostas a essas perguntas s\~ao bastante profundas e, em \'ultima inst\^ancia, essa discuss\~ao pode at\'e mesmo ser utilizada para introduzir \`a turma alguns princ\'ipios b\'asicos da relatividade especial. Para iniciar o debate, o(a) docente pode propor o seguinte problema.

\begin{figure}[h!]
    \centering
    \includegraphics[scale=.08]{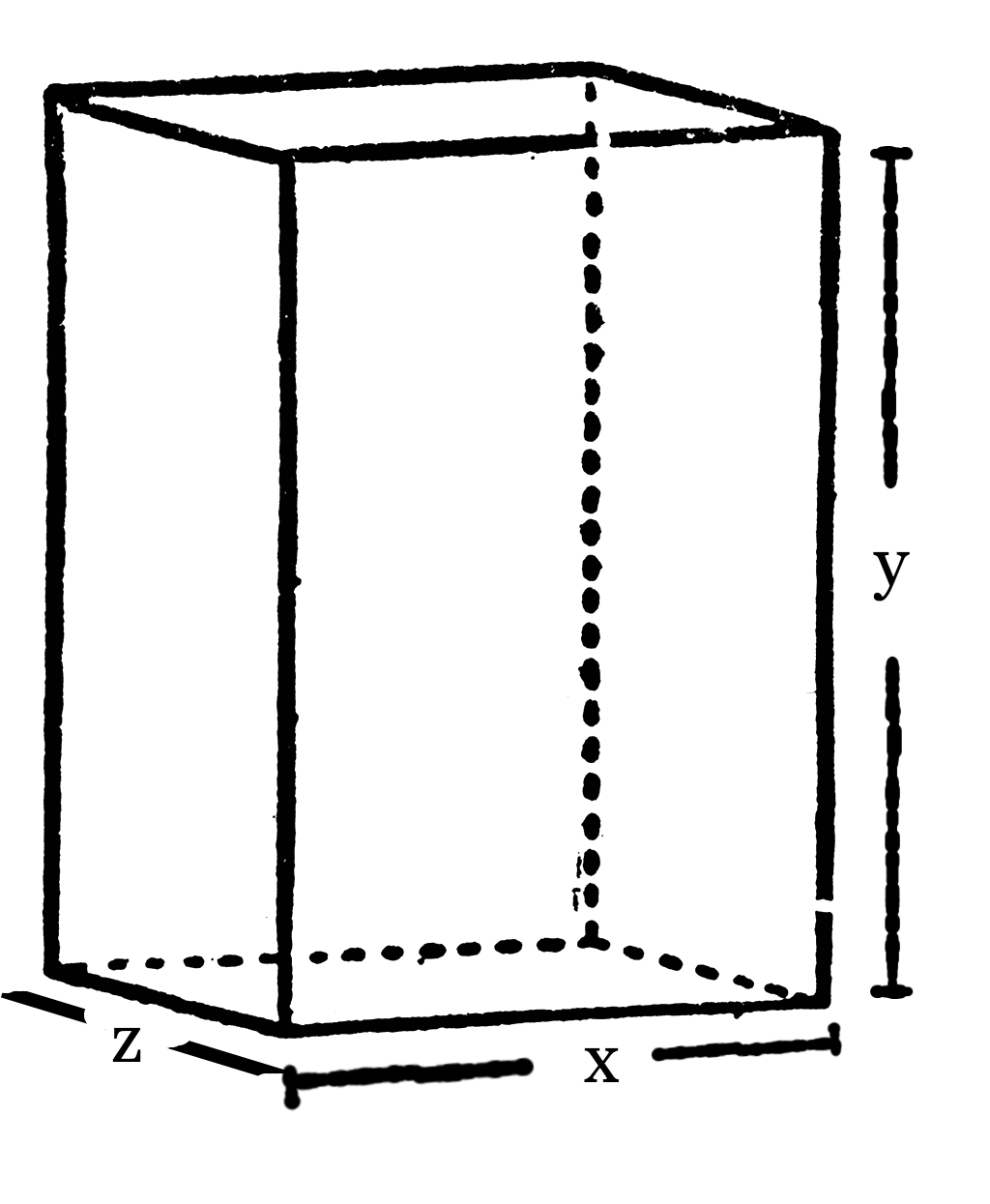}\qquad
    \includegraphics[scale=.08]{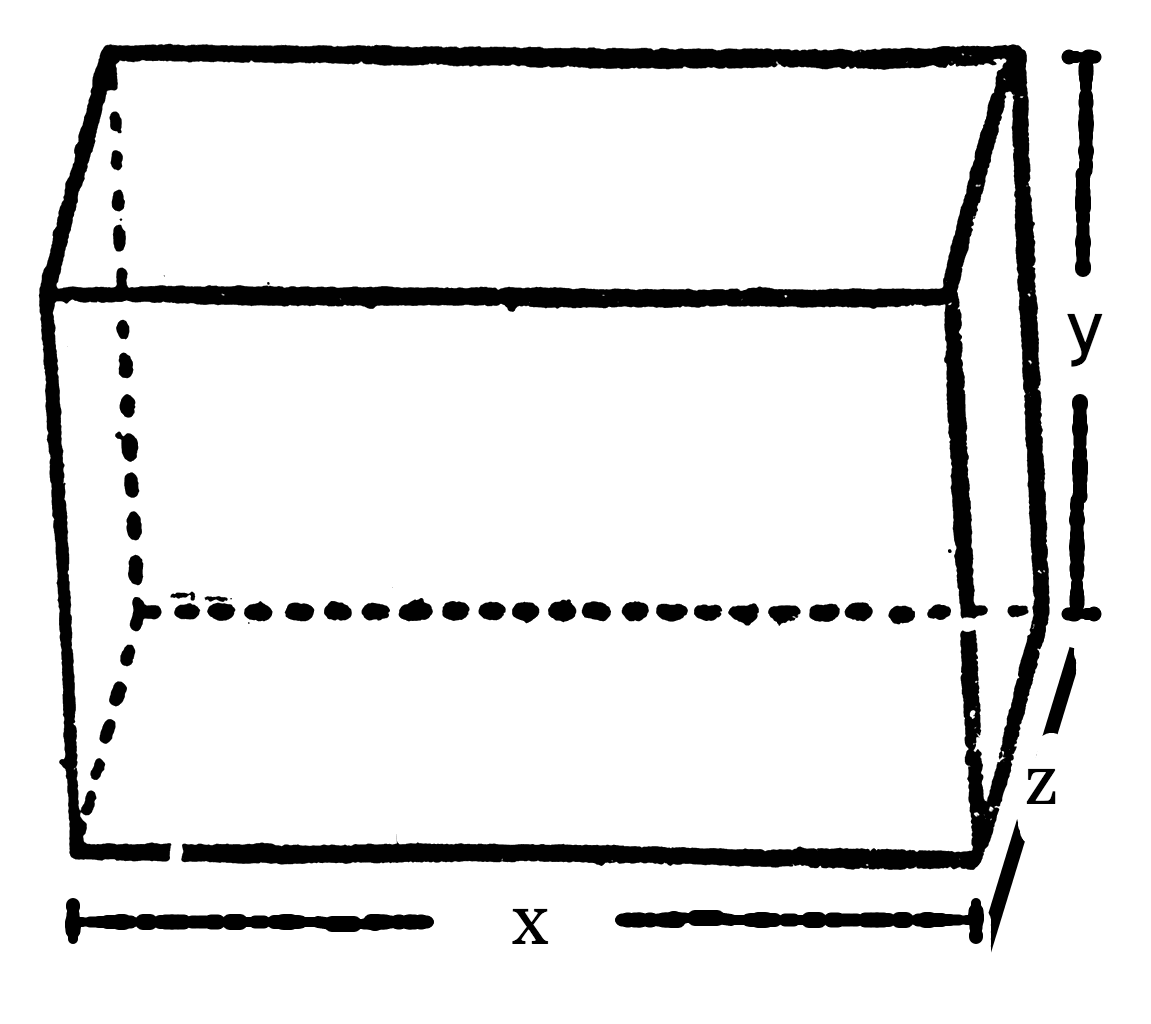}
    \caption{Se utiliz\'assemos unidades e dimens\~oes distintas para medir altura, largura, e comprimento, sempre que fiz\'essemos a rota\c{c}\~ao de um objeto ter\'iamos que recalcular suas dimens\~oes. Como rota\c{c}\~oes n\~ao alteram o tamanho do objeto, esse c\'alculo envolveria fatores de convers\~ao universais. A escolha mais conveniente \'e aquela em que esses fatores s\~ao todos unit\'arios, ou seja, medimos todos os lados dos objetos usando as mesmas unidades. A universalidade da velocidade da luz nos indica, de igual modo, que espa\c{c}o e tempo s\~ao dimens\~oes distintas de um \'unico ente: o espa\c{c}o-tempo. E a convers\~ao entre elas \'e feita pela propaga\c{c}\~ao da luz, que oferece, aqui, o an\'alogo \`a rota\c{c}\~ao espacial do objeto no exemplo da presente figura.}
    \label{fig:box_dims}
\end{figure}
~\\
\emph{``Imagine, por um momento, que decidamos medir comprimentos, larguras e alturas usando unidades (e dimens\~oes!) diferentes para cada uma dessas dire\c{c}\~oes espaciais. Explique por que devem existir fatores de convers\~ao entre essas unidades, e discuta suas propriedades. Discuta por que essa decis\~ao, mais do que artificial, \'e antinatural.''}
~\\

O(a) docente pode pedir que os(as) estudantes criem nomes para as diferentes unidades de comprimento, largura e altura a serem usadas no exemplo, bem como fatores de convers\~ao entre elas. O importante \'e concluir que usar unidades distintas n\~ao faz sentido, porque sempre podemos fazer uma rota\c{c}\~ao do objeto e converter largura em altura, altura em comprimento, e assim por diante (vide figura~\ref{fig:box_dims}). Se usamos unidades diferentes para cada dimens\~ao espacial, sempre que rotacion\'assemos um objeto ter\'iamos que usar fatores de convers\~ao para reescrever suas dimens\~oes nas unidades apropriadas. Como rota\c{c}\~oes n\~ao alteram o tamanho dos objetos, esses fatores de convers\~ao seriam ``constantes fundamentais da Natureza''. A escolha mais conveniente seria, portanto, fazer com que tais fatores de convers\~ao sejam iguais \`a unidade.

Emendando essa conclus\~ao, o professor pode ent\~ao lan\c{c}ar a seguinte provoca\c{c}\~ao:

~\\
\emph{``A velocidade da luz no v\'acuo, $c$, \'e uma constante universal, com cujo valor quaisquer observadores sob quaisquer circunst\^ancias no Universo concordam. Podemos, portanto, considerar $c$ como um fator de convers\~ao, an\'alogo ao caso da discuss\~ao anterior. Repetindo a argumenta\c{c}\~ao acima, a que conclus\~oes podemos chegar baseado na const\^ancia da velocidade da luz?''}
~\\

Em analogia direta \`a discuss\~ao anterior, a conclus\~ao \'e que a propaga\c{c}\~ao da luz constitui um processo f\'isico universal que conecta tempo e espa\c{c}o (an\'alogo \`a rota\c{c}\~ao dos objetos no exemplo anterior, figura~\ref{fig:box_dims}). A const\^ancia da velocidade da luz indica que tempo e espa\c{c}o s\~ao dimens\~oes distintas de um mesmo objeto: o espa\c{c}o-tempo. Portanto, \'e mais natural lidarmos com unidades em que $c=1$ e espa\c{c}o e tempo s\~ao tratados indistintamente.

Por fim, o(a) docente pode estender a analogia \`a outra constante universal: a constante de Planck.

~\\
\emph{``Analogamente, que conclus\~oes f\'isicas podemos extrair da const\^ancia de $\hbar$?''}
~\\

A conclus\~ao importante, aqui, \'e que $\hbar$ conecta o conceito de energia a espa\c{c}o e tempo. Essa conex\~ao foi utilizada na se\c{c}\~ao~\ref{sec:effcharge} quando convertemos uma energia caracter\'istica de um processo na escala de comprimento \`a qual ele est\'a tipicamente associado. Essa rela\c{c}\~ao tamb\'em pode ser explorada, junto \`a atividade proposta anteriormente, na se\c{c}\~ao~\ref{sec:sirius}, para discutir por que precisamos de aceleradores cada vez mais potentes, que realizem colis\~oes a energias cada vez mais altas, para explorarmos a f\'isica microsc\'opica, i.e. para entendermos como a mat\'eria se comporta a dist\^ancias cada vez menores.

\end{document}